\documentclass[fleqn,usenatbib]{mnras}

\usepackage{newtxtext,newtxmath}

\usepackage[T1]{fontenc}

\DeclareRobustCommand{\VAN}[3]{#2}
\let\VANthebibliography\thebibliography
\def\thebibliography{\DeclareRobustCommand{\VAN}[3]{##3}\VANthebibliography}

\usepackage{graphicx}	
\usepackage{amsmath}	
\usepackage{xcolor}   
\usepackage{lineno} 
\usepackage{comment} 
\usepackage{subfiles} 
\usepackage{upgreek}

\newcommand{\Rein}{\theta_{\rm E}}
\newcommand{\Reff}{R_{\rm eff}}

\newcommand{\cyone}[1]{{{{{#1}}}}}
\newcommand{\cytwo}[1]{{{{#1}}}}
\newcommand{\cyref}[1]{{#1}}
\newcommand{\ajs}[1]{{{{#1}}}}
\newcommand{\ajstwo}[1]{{{{#1}}}}
\newcommand{\ajsref}[1]{{#1}}

\title[Deviation from the power-law mass model]{Project Dinos I: A joint lensing--dynamics constraint on the deviation from the power law in the mass profile of massive ellipticals\thanks{\ajsref{The lens modelling products and the table of model parameters from this paper are released at \url{https://projectdinos.com}.}}}

\author[C. Y. Tan et al.]{Chin Yi Tan$^{1,2}$\thanks{E-mail: chinyi@uchicago.edu},
Anowar J. Shajib$^{3,2}$\thanks{NFHP Einstein Fellow, E-mail: ajshajib@uchicago.edu},  Simon Birrer$^{4}$,  Alessandro Sonnenfeld$^{5,6}$,  Tommaso Treu$^{7}$, \newauthor Patrick Wells$^{8}$, Devon Williams$^{7}$, Elizabeth J. Buckley-Geer$^{9,2}$, Alex Drlica-Wagner$^{9,2,3}$ and Joshua Frieman$^{2,3,9}$ 
\\
$^{1}$Department of Physics, University of Chicago, Chicago, IL 60637, USA\\
$^{2}$Kavli Institute for Cosmological Physics, University of Chicago, Chicago, IL 60637, USA\\
$^{3}$Department of Astronomy \& Astrophysics, The University of Chicago, Chicago, IL 60637, USA\\
$^{4}$Stony Brook University, Department of Physics and Astronomy, Stony Brook, New York, 11794, U.S.A.\\
$^{5}$Department of Astronomy, School of Physics and Astronomy, Shanghai Jiao Tong University, Shanghai 200240, China \\
$^{6}$Leiden Observatory, Leiden University, P.O. Box 9513, 2300 RA Leiden, The Netherlands \\
$^{7}$Department of Physics and Astronomy, University of California, Los Angeles, CA 90095 \\
$^{8}$Department of Physics and Astronomy, University of California, Davis, CA 95616, USA \\
$^{9}$Fermi National Accelerator Laboratory, P. O. Box 500, Batavia, IL 60510, USA
}

\date{Accepted XXX. Received YYY; in original form ZZZ}

\pubyear{2023}

\begin{document}
\label{firstpage}
\pagerange{\pageref{firstpage}--\pageref{lastpage}}
\maketitle


\begin{abstract}
\ajstwo{The mass distribution in massive elliptical galaxies encodes their evolutionary history, thus providing an avenue to constrain the baryonic astrophysics in their evolution. The power-law assumption for the radial mass profile in ellipticals has been sufficient to describe several observables to the noise level, including strong lensing and stellar dynamics. 
In this paper, we quantitatively constrained any deviation, or the lack thereof, from the power-law mass profile in massive ellipticals through joint lensing--dynamics analysis of a large statistical sample with 77 galaxy--galaxy lens systems. We performed an improved and uniform lens modelling of these systems from archival \textit{Hubble Space Telescope} imaging using the automated lens modelling pipeline \textsc{dolphin}. We combined the lens model posteriors with the stellar dynamics to constrain the deviation from the power law after accounting for the line-of-sight lensing effects, a first for analyses on galaxy--galaxy lenses.
We find that the Sloan Lens ACS Survey (SLACS) lens galaxies with a mean redshift of 0.2 are consistent with the power-law profile within 1.1$\sigma$ (2.8$\sigma$) and the Strong Lensing Legacy Survey (SL2S) lens galaxies with a mean redshift of 0.6 are consistent within 0.8$\sigma$ (2.1$\sigma$), for a spatially constant (Osipkov--Merritt) stellar anisotropy profile.
We adopted the spatially constant anisotropy profile as our baseline choice based on previous dynamical observables of local ellipticals. However, spatially resolved stellar kinematics of lens galaxies are necessary to differentiate between the two anisotropy models. Future studies will use our lens models to constrain the mass distribution individually in the dark matter and baryonic components.}
\end{abstract}



\section{Introduction}
\label{Sec:1}

Strong gravitational lensing is the phenomenon where multiple images of a background object appear due to the deflection of light by the gravitational effect of a massive foreground object. The foreground object is typically a massive galaxy or a galaxy cluster. Strong lensing at the galaxy scale is a valuable probe of the mass distribution of galaxies \cytwo{yielding} numerous applications in astrophysics and cosmology \citep[for a review, see][]{Shajib22b}.

For galaxy-scale deflectors, \cytwo{a} background galaxy can be imaged into multiple lensed arcs or a full Einstein ring. \cyone{Furthermore,} multiple images of a point-like source -- e.g., of a supernova \citep{Goobar17}  or an active galactic nucleus \citep[AGN;][]{Walsh79} -- can also appear. Galaxy--galaxy lens systems, i.e., systems that do not contain a prominent resolvable point source in the background galaxy, are more commonly discovered than multiply-imaged point source systems, as supernovae and AGNs are rarer than galaxies. Some of the largest samples of galaxy--galaxy lens systems to date are the Sloan Lens ACS\footnote{Advanced Camera for Surveys} (SLACS) survey \citep{Bolton06}, the Strong Lensing Legacy Survey \citep[SL2S;][]{Gavazzi12}, the SLACS for the Masses \citep[S4TM;][]{Shu15}, the BOSS\footnote{Baryon Oscillation Spectroscopic Survey} Emission-Line Lens Survey \citep[BELLS;][]{Brownstein11}, and the BELLS for the GALaxy-Ly$\alpha$ EmitteR sYstems \citep[BELLS GALLERY;][]{Shu16}. These samples of galaxy--galaxy strong lenses have been used to study the internal structure of elliptical galaxies by modelling the imaging data from the \textit{Hubble Space Telescope} (\textit{HST}) and ground-based telescopes \citep[e.g.,][]{Auger09, Auger10, Treu10, Sonnenfeld13b, Sonnenfeld15}.

\subsection{Radial form of the mass profile in massive ellipticals}
\label{Section:Intro_Evo}

In the current paradigm, massive elliptical galaxies are considered to be the end product of hierarchical mergers of dark matter haloes along with the stars and gas contained within them. In this \ajstwo{formation} scenario, \cytwo{the elliptical galaxies undergo two processes\ajstwo{, which can sometimes co-occur.} }

\ajstwo{For one process, the} baryonic gas dissipatively cools within the dark matter halo, starting from an initial Navarro--Frenk--White (NFW) mass distribution \ajstwo{at $z \gg 2$} \citep{Navarro96, Navarro97}. The dark matter halo also contracts in response to the baryonic infall, becoming steeper than the $\rho(r) \propto r^{-1}$ profile at the inner region \citep{Blumenthal86, Gnedin04, Abadi10}. 
\cytwo{For the other process}, galaxies undergo multiple mergers and accretion events to eventually reach a dynamically pressure-supported state. \cytwo{These events cause the} individual mass density profiles of the dark matter and the baryons, and thus the total density profile, to change in various ways. \cytwo{While} dissipationless mergers do not lead to a change in the dark matter density profile \citep{Gnedin04, Ma04}, dissipational mergers lead to a steeper slope at the inner region \citep{Sonnenfeld14}. Furthermore, baryonic feedback mechanisms -- for example, feedback from stellar wind, supernovae, \cyone{and} AGN -- can cause an outflow of gas, making the dark matter halo expand in response, which leads to a shallower total density profile \citep{Naab07, Duffy10, Johansson12, Dubois13}. \ajstwo{These feedback mechanisms can also counteract the initial dark matter contraction in the first place \citep{Governato12, Pontzen12}}. Indeed, recent cosmological hydrodynamical simulations, such as the Magneticum and the IllustrisTNG, predict a shallowing trend in the density profile slope with decreasing redshift at $z < 1$, thus positing gas-poor mergers being the dominant channel for the growth of massive ellipticals \citep{Remus17, Wang19}. Therefore, \cytwo{ observationally constraining the total density profile of elliptical galaxies at various redshifts would be a valuable approach to studying their formation history and mechanisms. By fine-tuning the baryonic physics prescriptions used in the cosmological hydrodynamical simulations to reproduce the density profile trends of these galaxies, we can gain insights into the underlying physical processes that drive their evolution.}

The total density profile in elliptical galaxies is found to be well approximated by a power-law mass distribution (i.e., $\rho(r) \propto r^{-\gamma}$) using observations from the combination of strong lensing and stellar dynamics \citep[e.g.,][]{Treu06, Shajib21}, from stellar dynamics only \citep[e.g.,][]{Cappellari16, Derkenne21}, from X-ray observations \citep[e.g.,][]{Humphrey10}, and from a combination of strong lensing, weak lensing, and dynamics \citep[e.g.,][]{Gavazzi07}. Although neither the dark matter nor the baryons individually follow the power law, the total density profile following the power law is called the `bulge--halo conspiracy' \citep{Dutton14, Treu06}. The total density profile in elliptical galaxies is typically observed to be slightly steeper than isothermal \citep[i.e., $\gamma \sim 2.1$;][]{Auger10b, Ritondale19, Shajib21}. 

More recently, however, \citet{Etherington23} suggest that the actual mass distribution may potentially deviate from the power-law profile based on their observation that the surface mass density correlates differently with the logarithmic slope $\gamma$ of the same systems measured from lensing-only and joint lensing--dynamics analyses. However, such a potential deviation from the power-law profile is yet to be quantitatively constrained with a high confidence level, and unaccounted systematic uncertainties pertaining to the lens or dynamical modelling are not yet ruled out as the source of the discrepancy mentioned above. Nonetheless, it is essential to use an accurate model of the mass distribution in the lensing galaxies before extracting observed structural properties to compare with the predictions from simulations. Otherwise, modelling systematics can bias the observed properties of the lensing galaxies, invalidating their use to confirm or rule out simulation predictions.

\ajstwo{Furthermore, any significant deviation from the power-law mass profile has important implications for time-delay cosmography \citep[for recent reviews, see][]{Treu22, Birrer22}. The 2.4 per cent measurement of the Hubble constant from the $H_0$ Lenses in the COSMOGRAIL's Wellspring (H0LiCOW) collaboration assumes power-law mass profile in the lensing galaxies \citep[e.g.,][]{Suyu13, Wong20}. If the power-law mass profile is accurate, then this measurement rules out systematics in the local $H_0$ measurement based on the cosmic distance ladder of type Ia supernova calibrated with Cepheids \citep[e.g.,][]{Riess22} and confirms the existence of new physics beyond the $\Lambda$ cold dark matter ($\Lambda$CDM) cosmology \citep[for a review, see][]{DiValentino21}. However, any potential deviation from the power-law profile in elliptical galaxies can be a source of major systematic in the time-delay cosmography. Therefore, the Time-Delay Cosmography (TDCOSMO) collaboration simultaneously constrains the Hubble constant and the galaxy mass profile by combining stellar kinematics with the lensing information \citep{Birrer20, Shajib23}.}




\subsection{Project Dinos}

In this paper, we introduce Project Dinos (PI: Shajib)\footnote{\textit{HST} archival program AR-16149}, whose primary goal is to study elliptical galaxy evolution using strong-lensing galaxies. This project aims to constrain the mass distribution in the elliptical galaxies at various redshifts, both in the total mass and individually in dark matter and baryons, and compare them with cosmological hydrodynamical simulations to understand the impact of baryonic feedback and mergers on the elliptical galaxy evolution at $z \lesssim 1$.

In this first paper of the Project Dinos series, we \cytwo{quantified any deviation from the power-law profile in elliptical galaxies}. We constrained the logarithmic slope of the mass profiles from a large statistical sample collated from multiple archival samples: SLACS, SL2S, and BELLS. The principal criteria to adopt these specific surveys are large sample size, diversity in the covered redshift range, and availability of the \textit{HST} archival imaging and stellar kinematic measurements. 

In most of the previous analyses of these lens samples, the imaging data were only used to measure the Einstein radius by modelling the mass profile of the system with a singular isothermal ellipsoid (SIE). The logarithmic slope of the mass distribution was then constrained by combining the stellar dynamics with the lensing-inferred enclosed projected mass. However, if the data quality is sufficient, it is also possible to constrain the local logarithmic slope of the mass profile directly from the imaging data \citep{Birrer21}. Then, combining the stellar dynamics with the lensing constraints can provide us with more detailed information on the mass structure, for example, by breaking the degeneracy between the normalizations of the stellar and dark matter components in the lensing galaxy \citep[e.g.,][]{Shajib21}. \ajsref{Furthermore, the power-law mass models for a large sample of galaxy-galaxy lens systems with available kinematics are required to provide a prior on the mass profile shape to improve the Hubble constant measurement from time-delay cosmography \citep{Birrer20, Birrer21b}.} In this paper, we \cytwo{modelled} \textit{HST} archival images of galaxy--galaxy lens systems with more flexible models than the simple SIE model adopted by previous studies on these same datasets. In this way, we \cytwo{extracted} \ajstwo{more} lensing information \ajstwo{from} the imaging data beyond \ajstwo{only constraining the Einstein radius}. \ajstwo{Furthermore, our sample is the largest to date with state-of-the-art power-law models of strong lenses.} Future papers in this series will investigate the dark matter and baryonic distribution in elliptical galaxies and their evolution across redshift using the sample presented in this paper.

This paper is organized as follows. In Section \ref{Sec2:Lens Sample}, we present our lens sample, the \textit{HST} imaging data used for lens modelling, and the available ground-based stellar velocity dispersion measurements. In Section \ref{Sec3:Methods}, we describe the lens modelling procedure using an automated and uniform modelling framework. \cyone{We then compare our lens modelling results with previous studies in Section \ref{Section 5}. In Section \ref{Section4:Model_paramters}, we describe \ajstwo{the hierarchical Bayesian framework for combining the lens model posteriors with the stellar dynamics to quantify} any deviation from the power-law mass model. We discuss the implications of our results in Section \ref{sec:discussion} and \cytwo{conclude the paper in Section \ref{Section6}.}} Throughout the paper, we adopt the flat $\Lambda$CDM model as our fiducial cosmology with $H_0 = 70$ km s$^{-1}$ Mpc$^{-1}$ and $\Omega_{\rm m} = 0.3$.

\section{Lens samples and data}\label{Sec2:Lens Sample}

We chose an initial lens sample that consists of lens systems from the SLACS, SL2S, and BELLS samples. We used all the available archival \textit{HST} images of these systems in the ultraviolet and visible (UVIS) bands. The specific \textit{HST} filters used to model each lens system are provided in Appendix \ref{Appendix B}. We also obtained the redshift and velocity dispersion measurements for these systems from the literature.

\subsection{Lens sample selection}
\label{subsec: lens_sel}

In this section, we describe the parent lens survey for each of our chosen subsamples and how we selected these initial subsamples for lens modelling. The lenses in the SLACS survey were found by selecting targets with higher-redshift emission lines and lower-redshift absorption features measured using the spectrum obtained from the Sloan Digital Sky Survey \citep[SDSS;][]{Bolton06}. The sample is then further refined by selecting systems showing clear evidence of multiple imaging of the background source by the lensing galaxy. 
Following \citet{Shajib21}, we chose a subsample of 50 systems to model out of the full sample of 85 `Grade A' lenses \citep[described in][]{Auger09}. This subsample was selected by visually inspecting the lens images and excluding systems where (i) there are nearby satellites or line-of-sight galaxies, (ii) there is highly complex source morphology, (iii) the deflector galaxy is disc-like, or (iv) there is no archival \textit{HST} imaging data for the system in the visible band (F555W or F606W). Requirements (i) and (ii) were imposed due to simpler lens systems being more suitable for uniform modelling with a higher probability of success, whereas requirement (iii) was set so that we only have elliptical galaxies in the sample. We imposed requirement (iv) so that the lens light profile of the galaxies can be measured uniformly across the SLACS sample.

Another sample analysed in this paper is the SL2S sample. The sample was identified using images from the Canada--France--Hawaii Telescope Legacy Survey \citep{Gavazzi12, Gavazzi14, More12}. The candidates were identified by finding blue features around elliptical galaxies consistent with the presence of gravitationally lensed arcs. The samples were confirmed using follow-up \textit{HST} imaging and Very Large Telescope (VLT), Keck, or Gemini spectroscopy. Out of the 56 lenses in the sample, we selected 31 lenses primarily based on the availability of visible band \textit{HST} images for that particular system.  However, we also excluded two systems (SL2SJ0213$-$0743, SL2SJ1405$+$5502), which have $V$-band \textit{HST} images, but the nearby foreground objects are too complex to mask effectively.

The last sample used in the analysis is the BELLS sample. The sample consists of candidates discovered by higher redshift emission lines within the BOSS galaxies \citep{Brownstein11} in a similar method as the SLACS lenses. Our sample consists of the 22 `Grade A' systems for the BELLS sample. `Grade B' lenses and below were excluded as they either lack the counter images or have an insufficient signal-to-noise ratio ($S/N$) for modelling.

\subsection{Processing of the \textit{HST} imaging data}
The imaging data for the selected lenses are from \textit{HST}'s Advanced Camera for Surveys (ACS), Wide Field and Planetary Camera 2 (WFPC2), and Wide Field  Camera 3 (WFC3). Table \ref{table:filters} provides the \textit{HST} program IDs, instruments, and filters for the imaging data of our lens sample.

\begin{table}
\centering
\caption{ \label{table:filters}
Description of the \textit{HST}  programs from which we assemble our archival lens sample.  
\ajstwo{Programs marked with an asterisk are snapshot programs that acquired only single-exposure images.}
}
\begin{tabular}{lllll}
\hline
Survey & \textit{HST} program & Program PI  & Instrument   & Filter  \\
\hline
SLACS  & 10174* & Koopmans & ACS        & F435W, F814W     \\
       & 10494  & Koopmans & ACS        & F555W, F814W     \\
       & 10587* & Bolton & ACS        & F435W, F814W     \\
       & 10798  & Koopmans & ACS        & F555W, F814W          \\
       &        &          & WFPC2      & F606W          \\
       & 10886  & Bolton   & ACS        & F814W    \\
       &        &          & WFPC2      & F606W          \\
       & 11202  & Koopmans & WFPC2      & F606W            \\
       & 12898  & Koopmans & WFC3       & F390W            \\
\hline
SL2S   & 10876  & Kneib    & ACS        & F606W, F814W     \\
       & 11289  & Kneib    & WFPC2      & F606W            \\
       & 11588  & Gavazzi  & WFC3       & F475X, F600LP    \\
\hline
BELLS  & 12209  & Bolton   & ACS        & F814W            \\  
\hline
\end{tabular}
\end{table}

We combined images with multiple exposures using the \textsc{astrodrizzle} software package \citep{Avila15}. The drizzling process with \textsc{astrodrizzle} also allowed us to remove cosmic rays from the single exposure image. The drizzling pixel scale was chosen based on the camera used: 0.04 arcsec for WFC3, 0.05 arcsec for ACS, and 0.10 arcsec for WFPC2. 
An exception was made for the SLACS WFPC2 images, where we used the data reduced by \cite{Auger09} with the pixel scale set to 0.05 arcsec. \cite{Auger09} employed a custom data-reduction pipeline where the sub-pixel shifts between exposures were used to allow the drizzling pixel scale to be significantly smaller than the detector pixel scale. To achieve a higher image quality after drizzling, we left the orientation of the final output image with respect to the RA-Dec coordinate system to be the same as the individual raw exposures. We then used \textsc{SourceExtractor} to estimate the mean background light and subtract it from each image \citep{Bertin96}.

We find \cyone{ up to a 0.5 arcsec} global offset in the world coordinate system (WCS) between images from different bands due to differences in their absolute astrometric reference. This often occurs when comparing images from different \textit{HST} cameras. Thus, to ensure that the coordinate systems of the images are correctly aligned across all filters, we constrained the centroid of the deflector galaxy in each filter individually by fitting a single S\'ersic profile to the light distribution within the central 0.6 arcsec radius of the deflector galaxy using \textsc{lenstronomy} \citep{Birrer18,Birrer21b}. We then corrected the relative offsets between the WCSs using the difference between the fitted centroids. We find this method sufficient in reducing the modelling residuals caused by image misalignment between the filters.

\begin{figure}
    \centering
    \includegraphics[width=0.95\columnwidth]{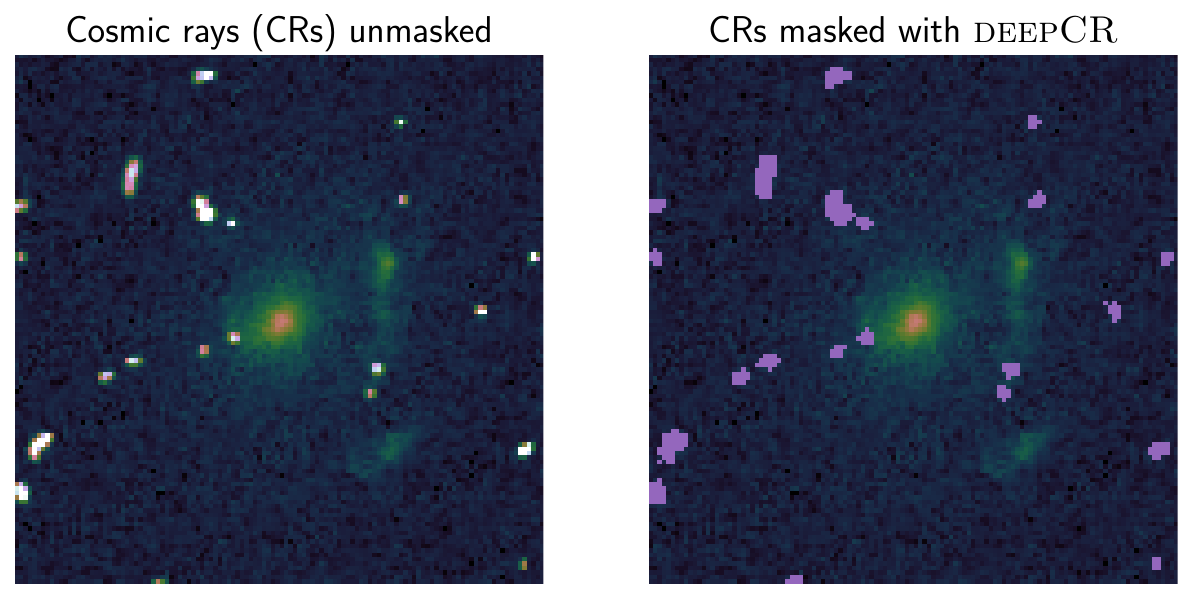}
    \caption{An example demonstration of \textsc{deepCR} in masking cosmic rays for a single-exposure image (in F435W band) of SDSSJ0819$+$4534. The left-hand panel shows the unmodified \textit{HST} image, and the right-hand panel shows the \textit{HST} image with the masked pixels \cyone{shown in purple} }
    \label{deepCR_example}
\end{figure}

As shown in Table \ref{table:filters}, the \textit{HST} images for some SLACS lenses only have a single exposure. Therefore, \textsc{astrodrizzle} cannot remove cosmic rays from these images. We used the \textsc{deepCR} software program for these images to automatically detect and mask the cosmic rays \citep{Zhang20}. \textsc{deepCR} uses a neural network trained on \textit{HST} images to detect and mask cosmic rays. Figure \ref{deepCR_example} shows an example of the mask produced by \textsc{deepCR} for the system SDSSJ0819$+$4534. 

 Since images from different cameras and filters have different point spread functions (PSF), we simulated the PSFs for each combination of camera and filter with  
\textsc{tinytim} \citep{Krist11} using a G2V star's spectral energy distribution. When producing the PSF, we also accounted for zero-point offsets produced by the time-dependent aberration changes on the mirror \citep{Krist11}. The pixelized PSFs were set to have the same pixel sizes as the corresponding drizzled \textit{HST} images. \cite{Shajib21} find that different \textsc{tinytim} configurations and various empirical methods to estimate the PSF \citep[such as ePSF from stars within the corresponding image;][]{Anderson00} produce negligible differences in sample mean of the model parameters.

\subsection{Measured stellar velocity dispersions}
\label{Sec2:Spectra}
We obtained the measured redshift of the source and deflector galaxies and the single-aperture stellar velocity dispersion of the deflector galaxy from the literature. \cyone{For the SLACS systems, we used the same SDSS velocity dispersion measurements as \citet{Birrer20}. The circular fibre used by the SDSS program had a diameter of 3 arcsec, and the nominal full width at half-maximum (FWHM) of the seeing was 1.4 arcsec in the $r$-band \citep{Bolton08, Birrer20}. The velocity dispersion for the SLACS systems was then measured using templates described in \cite{Shu15}. \citet{Birrer20} find that the uncertainty in the SDSS velocity dispersion measurements was underestimated with an unaccounted systematic uncertainty of 6 per cent. We also accounted for such underestimated uncertainties as described in Section \ref{Section4:Measure_lambda}. For the SL2S lenses, we used \ajstwo{the velocity dispersions measured by previous studies using long-slit spectra} obtained from the Keck Observatory, the VLT, and the Gemini North Telescope \citep[][see therein for the slit dimensions and seeing]{Sonnenfeld13b, Sonnenfeld15}. The velocity dispersions for the BELLS systems were obtained from BOSS spectroscopy as part of the SDSS-III program. The SDSS-III program used an upgraded optical spectrograph having a smaller fibre diameter of 2 arcsec \ajstwo{and the observations had a 1.3 arcsec nominal seeing in the $r$-band}} \citep{Brownstein11}.

\section{Lens modelling} \label{Sec3:Methods}
We used the automated lens modelling pipeline \textsc{dolphin}\footnote{\url{https://github.com/ajshajib/dolphin}} to perform uniform lens modelling for all the lenses in our sample \citep{Shajib21}. \textsc{dolphin} uses the publicly available software package \textsc{lenstronomy}\footnote{\url{https://github.com/lenstronomy/lenstronomy}} as its modelling engine \citep{Birrer18,Birrer21b}. The effectiveness of the \textsc{lenstronomy} package was demonstrated with the Time-Delay Lens Modelling Challenge, where multiple participating teams used the software package to successfully recover the hidden lens model parameters \citep{Ding21b}. \textsc{lenstronomy} has also been used in many other applications related to strong gravitational lenses, such as in time-delay cosmography \citep{Birrer19b, Shajib20, Shajib22} and in studying dark matter substructures \citep{Gilman20}. Whereas \textsc{lenstronomy} provides a powerful and feature-rich API to allow fine-tuning of the lens model settings for individual systems, \textsc{dolphin} vastly simplifies the modelling procedure for the user through a semi-automated decision-making process without requiring a large amount of investigator time spent for fine-tuning. 

In Section \ref{Lensing Formalism}, we briefly review the theory of strong gravitational lensing and introduce the relevant terminologies. Then, in Section \ref{ModelComp}, we discuss the model components used to describe the deflector and source galaxy's mass and light distributions. Section \ref{Section: Modelling procedure} presents the methods used to obtain the model posteriors from the data. Finally, in Section \ref{Section:Final_Sample_Selection}, we present our final lens model sample and discuss our quality assurance procedure for the lens models.

\subsection{Introduction to strong lensing formalism}
\label{Lensing Formalism}
\cyone{In the single-lens-plane regime,} the effect of gravitational lensing is described with the lens equation, which maps the source plane vector $\boldsymbol{\beta}$  to the image plane vector $\boldsymbol{\theta}$ as
\begin{equation}
    \boldsymbol{\beta} = \boldsymbol{\theta} - \boldsymbol{\alpha}(\boldsymbol{\theta}),
\label{Eqn:Lens_Equation}
\end{equation}
where $\boldsymbol{\alpha}(\boldsymbol{\theta})$ is the deflection angle. 
The lensing potential $\psi$ relates to the deflection angle as 
\begin{equation}
    \boldsymbol\alpha(\boldsymbol{\theta}) = \nabla \psi(\boldsymbol{\theta}) .
\end{equation}

For a 2D projected surface mass density $\Sigma(\boldsymbol{\theta})$, we can define the dimensionless convergence quantity as  
\begin{equation}
    \kappa(\boldsymbol{\theta}) = \frac{\Sigma(\boldsymbol{\theta})}{\Sigma_\text{crit}} , 
\end{equation}
where  $\Sigma_\text{crit}$ is the lensing critical surface mass density, defined as
\begin{equation}
    \Sigma_\text{crit} \equiv \frac{c^2}{4 \uppi G} \frac{D_{\rm s}}{D_{\rm l} D_{\rm ds }} ,
\end{equation}
where $c$ is the speed of light, $G$ is the gravitational constant, and $D_{\rm l}$, $D_{\rm s}$, $D_{\rm ls}$ are the angular diameter distances from the observer to the lens, from the observer to the source, and from the lens to the source, respectively. The lensing potential is then connected to the convergence via the Poisson equation
\begin{equation}
    \begin{aligned}
         &\nabla^2 \psi(\boldsymbol{\theta}) = 2 \kappa(\boldsymbol{\theta}) \\
         \Rightarrow &\nabla \cdot \boldsymbol{\alpha} (\boldsymbol{\theta}) = 2 \kappa(\boldsymbol{\theta}).
    \end{aligned}
\end{equation}
Therefore, the distortion of the source galaxy's light in a lensing system, described by $\boldsymbol{\alpha}(\boldsymbol{\theta})$, can be used to probe the surface mass density $\Sigma(\boldsymbol{\theta})$.

\subsection{Definitions of model components}
\label{ModelComp}
The lens model consists of three main components: the mass profile of the deflector galaxy, the light profile of the deflector galaxy, and the light profile of the source galaxy. 
\cytwo{We \ajstwo{modelled} the} mass profile of the deflector galaxy with a power-law ellipsoidal mass distribution \citep[PEMD;][]{Barkana98}. The convergence of this profile is given by
\begin{equation}
    \kappa(\theta_1,\theta_2) = \frac{3-\gamma}{2}\left(\frac{\Rein}{\sqrt{q_{\rm m}\theta_1^2+\theta_2^2/q_{\rm m}}}\right)^{\gamma-1},
\label{Eqn:PEMD}
\end{equation}
where $\Rein$ is the Einstein Radius, $\gamma$ is the logarithmic slope of the mass distribution, and  $q_\text{m}$ is the axis ratio. The coordinates $(\theta_1,\theta_2)$ are defined by rotating the (RA, Dec) coordinate system by the major axis's position angle ${\phi}_\text{m}$. We also added a `residual shear' field (commonly referred to as the `external shear' in the literature) in the lens model parametrized by a shear magnitude $\gamma_{\rm shear}$ and angle $\phi_{\rm shear}$. This residual shear field accounts for additional shear contributed by the line-of-sight structures and any unaccounted angular structure in the central deflector  \citep[e.g.,][]{Etherington23b}. 

For the light profile of the deflector galaxy, we used the elliptical S\'ersic profile given by
\begin{equation}
\label{Eq:Sersic}
I(\theta_1,\theta_2)=I_0\exp\left[-b_n\left(\frac{\sqrt{q_{\rm L}\theta_1^2+\theta_2^2/q_{\rm L}}}{\Reff}\right)^{1/n_{\rm s}}\right],
\end{equation}
where $\Reff$ is the effective radius of the profile, $n_{\rm s}$ is the S\'ersic index, $q_\text{L}$ is the ratio between semi-major and semi-minor axis, $b_n = 1.999n_s - 0.327$ is the normalizing factor to make $R_{\rm eff}$ the half-light radius, and $I_0$ is the amplitude \citep{Sersic68}. Similar to the mass profile, the coordinates $(\theta_1,\theta_2)$ are defined by rotating the (RA, Dec) coordinate system by the major axis's position angle ${\phi}_\text{L}$.

As a single S\'ersic  profile often cannot accurately fit the light distribution at the galaxy's centre \citep{Claeskens06, Suyu13}, we used two S\'ersic profiles to allow more radial freedom in fitting the galaxy's light distribution. To reduce the degeneracies between the two  S\'ersic profiles, we fixed the S\'ersic indices of the profiles as $n_\text{s}$ = 4 (i.e., the de Vaucouleurs profile) and $n_\text{s}$ = 1 (i.e., the exponential profile), respectively. To minimize the number of unnecessary free parameters in the model, we joined the ellipticity parameters $q_\text{L}$ and ${\phi}_\text{L}$ between the two S\'ersic profiles. 

We find that the two S\'ersic profiles are still inadequate in some cases to reproduce the light profile of the deflector at the very centre (within $\sim$0.4 arcsec). Thus, to avoid a large residual in the centre introducing any bias, we adopt a circular mask at the centre of the deflectors \cyref{to exclude the residuals within the masked region from the computed likelihood.} \cytwo{In our default \ajstwo{setup}, we set the radius of the mask to be 0.4 arcsec. However, }for systems with lensed arcs falling within the central 0.4 arcsec, we chose a smaller radius to avoid masking out the arcs. Nearby galaxies along the line of sight exist in some lens systems, for example, SDSSJ0944$-$0147 or SDSSJ1215$+$0047. For these systems, we manually introduce an appropriately sized circular mask to block out the light from the line-of-sight galaxies. The mask settings for each lens system are given in Appendix \ref{Appendix B}.

For the source galaxy's light distribution, we used the combination of an elliptical S\'ersic profile and a basis set of multiple shapelet components \citep[i.e., weighted 2D Gauss-Hermite polynomials;][]{Refregier03, Birrer15}. The extent of all the shapelet components simultaneously scale with a single parameter $\varsigma$. The order parameter $n_\text{max}$ determines the number of shapelet components $N_\text{shapelets} = (n_\text{max} + 1)(n_\text{max} +2)/2$. Initially, we set $n_\text{max}=6$ as the default setting for all systems. However, for systems with complicated source light profiles, we adopted a higher value of shapelet order parameter, $n_\text{max}$, chosen by trial and error to improve the model's fit. The specific choices made for $n_\text{max}$ in each system are provided in Appendix \ref{Appendix B}.

We simultaneously fit all the available \textit{HST} filters for a given system. In doing so, we allowed the amplitudes and scale radii of all the light profiles to be free across different filters while imposing the same deflector mass profile for all the filters.

The lens model discussed in this section was designed to increase the success rate in uniformly modelling the galaxy--galaxy lenses in our sample. However, several systems contain complex features -- either in the source light or in the deflector environment -- that must be manually modelled on a case-by-case basis. For this paper, we consider this type of lens system as a `failure case' and exclude those from our final lens model sample. A more detailed discussion of how we selected our final lens model sample is given in Section \ref{Section:Final_Sample_Selection}.

\subsection{Modelling procedure}
\label{Section: Modelling procedure}
We sampled from the posterior probability distribution function (PDF) of the model parameters with the Markov chain Monte Carlo (MCMC) method using \textsc{emcee} \citep{Goodman10, Foreman-Mackey13}. To achieve faster convergence in the MCMC sampling, we obtained a point close to the maxima of the posterior PDF using the particle swarm optimization (PSO) method \citep{Kennedy95}.  We then initiated the random walkers for the MCMC sampling within a small region around the best solution from the PSO step. We visually inspected for MCMC convergence by ensuring that the median and standard deviation of the parallel walker positions had reached equilibrium for at least 1000 steps.

We improved the efficiency of finding the posterior probability distribution by following the same optimization sequence or `recipe' \cytwo{as} \citet{Shajib21}. The optimization recipe first adopted a SIE mass model for the deflector with no residual shear (i.e., $\gamma_\text{shear}=0$). The Einstein radius of the system, $\Rein$, was also 
fixed to a value obtained from previous analyses \citep[i.e.,][]{Auger09, Sonnenfeld13, Brownstein11}, while the ellipticity parameters were set to the values for the circular case $(q_{\rm m}=1)$. For the first step of the optimization sequence, the pipeline optimized the deflector galaxy's light profile parameters with all the other parameters fixed. While optimizing for the deflector galaxy's light profile, the pipeline also created a mask to cover the lensed arcs. Then, the source light parameters were optimized after fixing the lens light and deflector mass parameters. The later steps involved freeing the previously fixed parameters and finding the best-fitting values for all the parameters simultaneously.

Since we do not know the order of magnitude of the shapelet scale parameter $\varsigma$, we adopted a log-uniform prior or Jeffreys prior on $\varsigma$, that is, $p(\varsigma)\propto 1/\varsigma$.  We also adopted an empirically motivated \cytwo{\ajstwo{uniform} prior} such that the deflector mass profile's position angle ${\phi}_{\rm m}$ and the light profile's position angle ${\phi}_{\rm L}$ are aligned within 15 deg. \ajstwo{To achieve faster convergence in the optimization, we set the edges of this uniform prior to fall as a one-sided Gaussian function with a small standard deviation ($0.03$ deg) instead of having a sharp cut.} This choice was motivated by the finding of previous studies that the misalignment between the deflector mass and light profiles is less than 10 deg when the residual shear magnitude is small ($\gamma_{\rm ext}<0.1$) \citep[e.g.,][]{Treu09, Sluse12, Shajib19}. The close alignment between the deflector mass and light profiles is also supported by the  Illustris simulation \citep{Xu17}.

\subsection{Quality assurance for lens models}
\label{Section:Final_Sample_Selection}
Once the MCMC chains had converged, we excluded the lens systems that fell into any of these scenarios: (i) the reconstructed lens image missing prominent lens features observed in the \textit{HST} image, (ii) poor source reconstruction indicated by prominent model residuals, (iii) the reconstructed source retaining the distorted shape of the lensed arc, or (iv) the logarithmic slope $\gamma$ converging to extremely atypical values ($\gamma < 1.4$ or $\gamma > 2.8$). The  $\gamma$ values usually diverge in such a way when the imaging data do not contain enough lensing information to constrain it \citep{Shajib21}. Examples include when the lensed arcs are too faint or when no counter-image can be detected in the image. For lens systems without any detectable counter-image above the noise level, our pipeline tends to reconstruct the source to retain the distorted shape of the single arc visible in the system.

\begin{figure*}
    \centering
    \includegraphics[width=\textwidth]{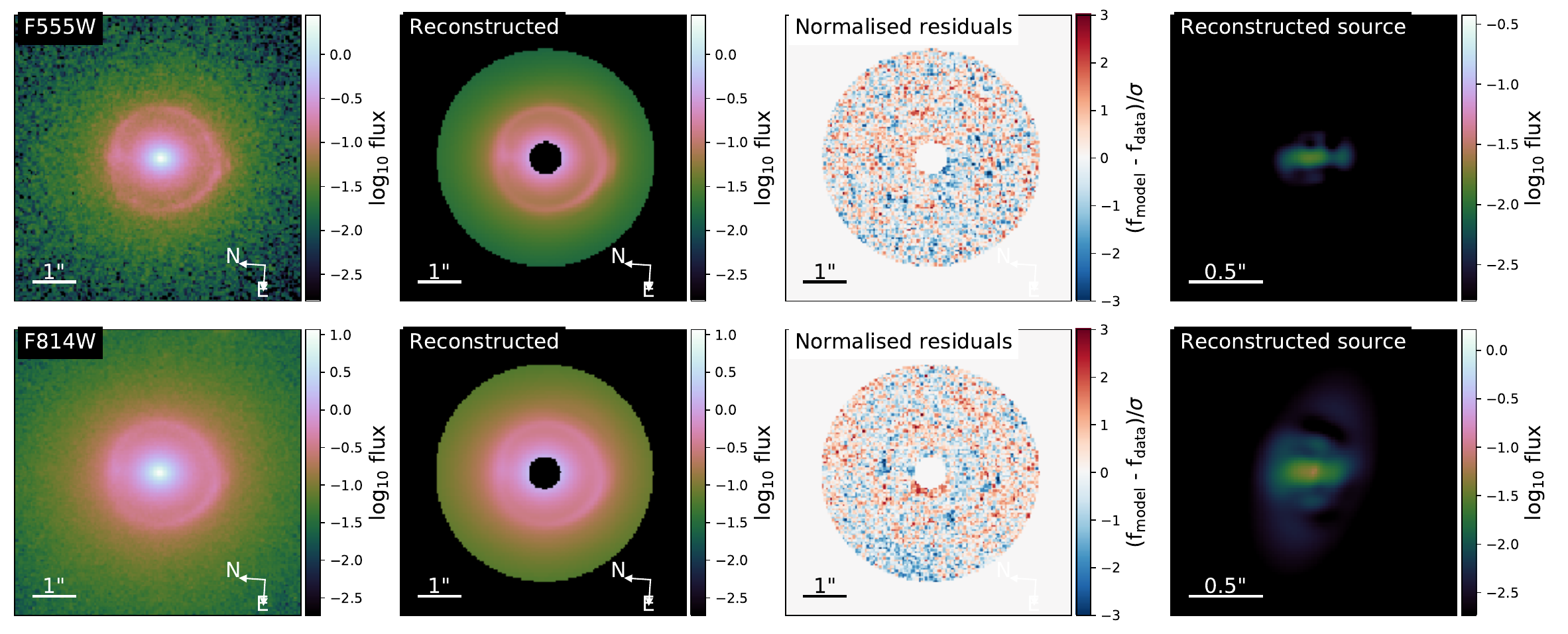}
    \caption{Best-fitting lens model for SDSSJ1627$-$0053. The columns (from left to right) represent the observed \textit{HST} image, the model-based reconstruction, the normalised residual, and the reconstructed source light distribution. The top row corresponds to the F555W filter, and the bottom row corresponds to the F814W filter. The flux values are in units of e$^{-}$ s$^{-1}$. All the scale bars are in units of arcsec.}
\label{fig:multi_band_0219}
\end{figure*}

\begin{figure*}
    \centering
    \includegraphics[width=\textwidth]{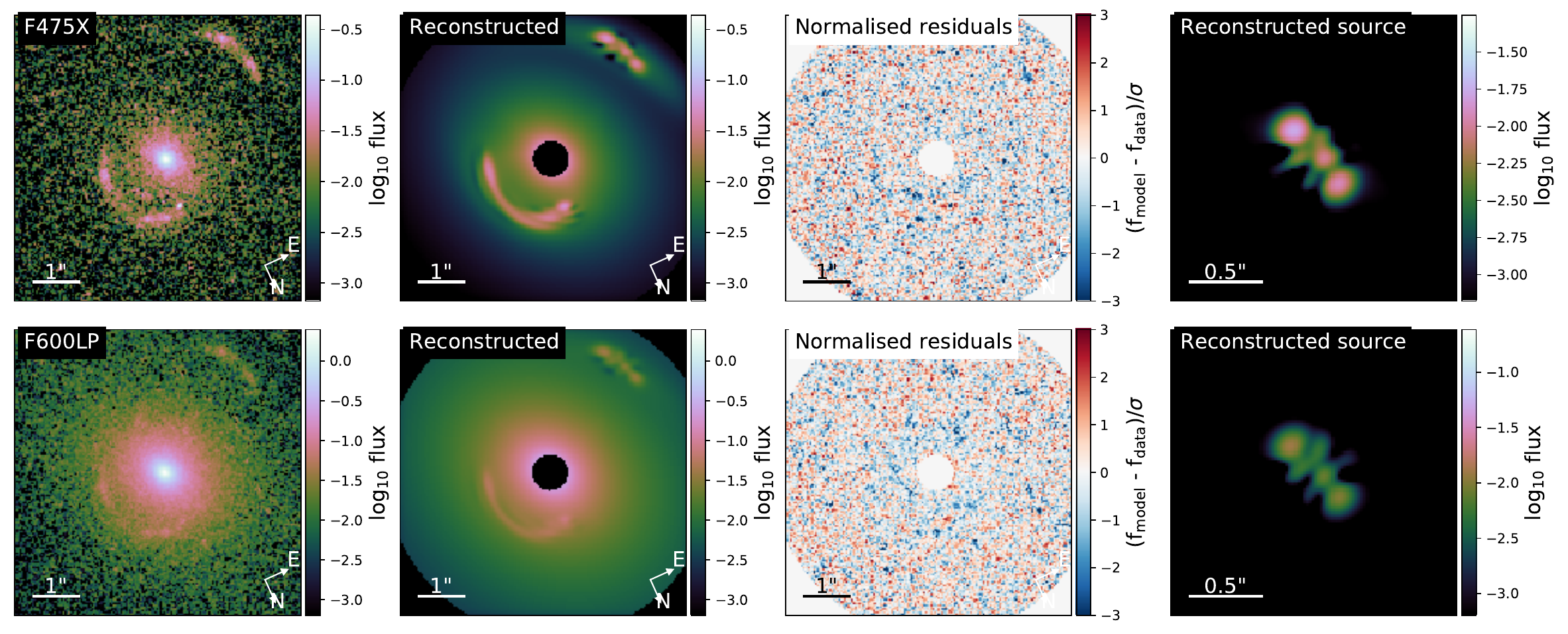}
    \caption{Best-fitting lens model for SL2SJ1401$+$5544.  The columns (from left to right) represent the observed \textit{HST} image, the model-based reconstruction, the normalised residual, and the reconstructed source light distribution. The top row corresponds to the F475X filter, and the bottom row corresponds to the F600LP filter. The flux values are in units of e$^{-}$ s$^{-1}$. All the scale bars are in units of arcsec.}
\label{fig:multi_band+1401}
\end{figure*}

Using the modelling procedure and selection process described above, we successfully modelled 77 lens systems from an initial sample of 105 lens systems. We break down the number of successfully modelled lens systems for each survey in Table \ref{table:models}. Figures \ref{fig:multi_band_0219} and \ref{fig:multi_band+1401} illustrate the best-fitting models for two systems,  SDSSJ1627$-$0053 and SL2SJ1401$-$5544, as examples. All the other models are illustrated in Appendix \ref{Appendix A}.

\begin{table}
\centering
\caption{ \label{table:models}
Number of systems with available \textit{HST} imaging, systems we selected to attempt modelling, and systems successfully modelled for the lens samples considered in this paper.
}
\begin{tabular}{lccc}
\hline
    Survey & \cyone{\textit{HST} data available} & Modelling attempted & Succeeded \\
\hline
SLACS& 85 & 50 &34 \\
SL2S & \cyref{33} & 31&24  \\
BELLS & 36  &25& 19  \\
\hline
Total & 154&103&77 \\
\hline
\end{tabular}
\end{table}

\begingroup
\renewcommand{\arraystretch}{1.25} 
 \begin{table*}
 \caption{Lens model parameters for the SLACS systems. Here, $\Rein$ is the Einstein radius, $\gamma$ is the logarithmic slope of the mass profile, $q_\text{m}$ is the mass axis ratio, $\phi_\text{m}$ is the mass position angle,  $\gamma_\text{shear}$ is the residual shear magnitude, $\phi_\text{shear}$ is the residual shear angle,  $q_\text{L}$ is the light axis ratio, $\phi_\text{L}$ is the light position angle, and $\Reff $ is the effective \ajstwo{or half-light} radius of the light profile. \ajstwo{The angles $\phi_{\rm m}$, $\phi_{\rm shear}$, and $\phi_{\rm L}$ are defined north of east.} The point estimates are the medians of the 1D marginalised posteriors, and the 1$\sigma$ uncertainties represent the 16$^{\rm th}$ and 84$^{\rm th}$ percentiles. \cytwo{ For the logarithmic slope of the mass profile $\gamma$, we provide an additional modelling systematic uncertainty. Additional details about estimating the systematic uncertainty \ajstwo{are provided} in Section \ref{Section_5.2}. The lens light parameters correspond to the \cyref{$V$-band filter (F606W, F600LP, or F555W)}. The effective radii $R_{{\rm eff}, V} $ were obtained from \ajstwo{separately fitting larger \cyref{$V$-band} \textit{HST} image cutouts -- that contain the full extent of the deflector galaxies -- after subtracting the lens-model-predicted arcs} (see Section \ref{Section4:Stellar_Kinematics} for more details). This table can be easily retrieved at \url{https://www.projectdinos.com/dinos-i}.
 }
 }
 \label{table:SLACS_params}
\begin{tabular}{lccccccccc}
\hline
     Lens system &  $\Rein$ &    $\gamma$ &    $q_\text{m}$ &     $\phi_\text{m}$ &  $\gamma_\text{shear}$  &  $\phi_\text{shear}$  &
     $q_\text{L}$ & 
     $\phi_\text{L}$ &
     $R_{{\rm eff}, V}  $  
     \\
     & (arcsec) & &&(deg)   & & (deg) &  & (deg) & (arcsec)\\
\hline
SDSSJ0008-0004 & $1.225_{-0.009}^{+0.008}$ & $1.75_{-0.12}^{+0.14}\pm 0.33$ & $0.60_{-0.02}^{+0.03}$ & $\phantom - 48_{-1}^{+1}$ &$0.104_{-0.023}^{+0.019}$ & $\phantom - 48_{-1}^{+2}$ &$0.856_{-0.005}^{+0.005}$ & $\phantom - 62.8_{-0.8}^{+0.7}$ &$1.93_{-0.02}^{+0.03}$ \\ 
SDSSJ0029-0055 & $0.945_{-0.003}^{+0.003}$ & $2.69_{-0.05}^{+0.04}\pm 0.25$ & $0.66_{-0.04}^{+0.04}$ & $\phantom - 64_{-1}^{+2}$ &$0.009_{-0.005}^{+0.005}$ & $ -26_{-9}^{+10}$  &$0.837_{-0.002}^{+0.001}$ & $\phantom - 63.5_{-0.3}^{+0.2}$ &$2.39_{-0.01}^{+0.01}$ \\ 
SDSSJ0037-0942 & $1.476_{-0.007}^{+0.010}$ & $2.38_{-0.02}^{+0.02}\pm 0.11$ & $0.73_{-0.02}^{+0.02}$ & $\phantom - 74_{-3}^{+4}$ &$0.023_{-0.012}^{+0.009}$ & $\phantom - 17_{-4}^{+4}$ &$0.692_{-0.001}^{+0.001}$ & $\phantom - 82.8_{-0.1}^{+0.1}$ &$3.33_{-0.05}^{+0.05}$ \\ 
SDSSJ0252+0039 & $1.028_{-0.001}^{+0.001}$ & $1.55_{-0.03}^{+0.04}\pm 0.12$ & $0.892_{-0.003}^{+0.005}$ & $ -17_{-1}^{+1}$  &$0.035_{-0.002}^{+0.001}$ & $ -18_{-1}^{+1}$  &$0.923_{-0.002}^{+0.002}$ & $ -17.9_{-0.9}^{+0.9}$  &$1.48_{-0.01}^{+0.01}$ \\ 
SDSSJ0330-0020 & $1.094_{-0.003}^{+0.003}$ & $2.14_{-0.01}^{+0.01}\pm 0.17$ & $0.66_{-0.01}^{+0.01}$ & $ -16_{-1}^{+1}$  &$0.028_{-0.004}^{+0.003}$ & $\phantom - 4_{-3}^{+5}$ &$0.722_{-0.002}^{+0.002}$ & $ -19.2_{-0.2}^{+0.1}$  &$1.24_{-0.01}^{+0.01}$ \\ 
SDSSJ0728+3835 & $1.261_{-0.002}^{+0.003}$ & $1.64_{-0.01}^{+0.47}\pm 0.05$ & $0.78_{-0.18}^{+0.01}$ & $\phantom - 26_{-4}^{+1}$ &$0.068_{-0.003}^{+0.004}$ & $\phantom - 26_{-12}^{+1}$ &$0.757_{-0.008}^{+0.001}$ & $\phantom - 23.8_{-1.9}^{+0.2}$ &$1.90_{-0.01}^{+0.01}$ \\ 
SDSSJ0737+3216 & $0.963_{-0.001}^{+0.001}$ & $2.32_{-0.04}^{+0.03}\pm 0.20$ & $0.85_{-0.01}^{+0.01}$ & $ -11_{-1}^{+1}$  &$0.111_{-0.005}^{+0.004}$ & $\phantom - 80.7_{-0.4}^{+0.4}$ &$0.864_{-0.002}^{+0.002}$ & $ -14.5_{-0.3}^{+0.3}$  &$2.83_{-0.02}^{+0.01}$ \\ 
SDSSJ0819+4534 & $0.784_{-0.007}^{+0.007}$ & $1.84_{-0.06}^{+0.09}\pm 0.20$ & $0.77_{-0.02}^{+0.02}$ & $\phantom - 35_{-1}^{+1}$ &$0.089_{-0.015}^{+0.012}$ & $\phantom - 55_{-3}^{+3}$ &$0.794_{-0.002}^{+0.002}$ & $\phantom - 48.8_{-0.3}^{+0.3}$ &$3.06_{-0.01}^{+0.01}$ \\ 
SDSSJ0903+4116 & $1.275_{-0.001}^{+0.002}$ & $2.17_{-0.01}^{+0.01}\pm 0.04$ & $0.75_{-0.01}^{+0.01}$ & $\phantom - 77_{-1}^{+1}$ &$0.067_{-0.004}^{+0.002}$ & $\phantom - 59_{-1}^{+1}$ &$0.858_{-0.002}^{+0.002}$ & $ -88.1_{-0.4}^{+0.8}$  &$2.98_{-0.07}^{+0.07}$ \\ 
SDSSJ0912+0029 & $1.609_{-0.002}^{+0.001}$ & $1.58_{-0.01}^{+0.01}\pm 0.06$ & $0.76_{-0.01}^{+0.01}$ & $\phantom - 76.4_{-0.5}^{+0.5}$ &$0.032_{-0.003}^{+0.003}$ & $\phantom - 68_{-2}^{+2}$ &$0.636_{-0.001}^{+0.001}$ & $\phantom - 77.15_{-0.05}^{+0.05}$ &$4.21_{-0.01}^{+0.01}$ \\ 
SDSSJ0936+0913 & $1.096_{-0.003}^{+0.003}$ & $2.21_{-0.03}^{+0.05}\pm 0.08$ & $0.74_{-0.02}^{+0.02}$ & $ -56_{-3}^{+3}$  &$0.053_{-0.007}^{+0.006}$ & $ -31_{-4}^{+3}$  &$0.809_{-0.002}^{+0.001}$ & $ -55.9_{-0.2}^{+0.3}$  &$2.11_{-0.01}^{+0.01}$ \\ 
SDSSJ0959+0410 & $0.960_{-0.002}^{+0.002}$ & $2.22_{-0.03}^{+0.02}\pm 0.08$ & $0.73_{-0.01}^{+0.01}$ & $\phantom - 24_{-1}^{+1}$ &$0.020_{-0.003}^{+0.003}$ & $ -29_{-5}^{+6}$  &$0.733_{-0.003}^{+0.003}$ & $\phantom - 32.2_{-0.3}^{+0.3}$ &$0.97_{-0.01}^{+0.01}$ \\ 
SDSSJ1023+4230 & $1.400_{-0.002}^{+0.002}$ & $2.22_{-0.05}^{+0.06}\pm 0.14$ & $0.91_{-0.01}^{+0.01}$ & $ -63.1_{-0.6}^{+0.3}$  &$0.039_{-0.003}^{+0.004}$ & $\phantom - 5_{-2}^{+2}$ &$0.842_{-0.001}^{+0.001}$ & $ -77.8_{-0.2}^{+0.2}$  &$1.97_{-0.01}^{+0.01}$ \\ 
SDSSJ1100+5329 & $1.461_{-0.003}^{+0.003}$ & $1.89_{-0.03}^{+0.03}\pm 0.14$ & $0.61_{-0.01}^{+0.01}$ & $ -10.9_{-0.3}^{+0.3}$  &$0.058_{-0.005}^{+0.005}$ & $\phantom - 50_{-3}^{+2}$ &$0.589_{-0.002}^{+0.002}$ & $ -11.9_{-0.1}^{+0.1}$  &$2.26_{-0.01}^{+0.01}$ \\ 
SDSSJ1112+0826 & $1.497_{-0.012}^{+0.009}$ & $1.92_{-0.03}^{+0.03}\pm 0.18$ & $0.74_{-0.03}^{+0.03}$ & $ -52_{-2}^{+2}$  &$0.020_{-0.017}^{+0.012}$ & $ -41_{-20}^{+12}$  &$0.748_{-0.001}^{+0.001}$ & $ -46.8_{-0.2}^{+0.2}$  &$1.62_{-0.01}^{+0.01}$ \\ 
SDSSJ1134+6027 & $1.243_{-0.005}^{+0.004}$ & $1.75_{-0.01}^{+0.01}\pm 0.20$ & $0.973_{-0.026}^{+0.005}$ & $ -54_{-6}^{+2}$  &$0.002_{-0.001}^{+0.002}$ & $\phantom - 57_{-26}^{+31}$ &$0.799_{-0.001}^{+0.001}$ & $ -65.2_{-0.2}^{+0.2}$  &$1.90_{-0.01}^{+0.01}$ \\ 
SDSSJ1204+0358 & $1.295_{-0.002}^{+0.003}$ & $2.02_{-0.03}^{+0.03}\pm 0.05$ & $0.90_{-0.01}^{+0.01}$ & $\phantom - 19_{-3}^{+2}$ &$0.023_{-0.004}^{+0.004}$ & $ -52_{-4}^{+5}$  &$0.9995_{-0.0006}^{+0.0003}$ & $\phantom - 21.8_{-12.7}^{+9.3}$ &$1.54_{-0.01}^{+0.01}$ \\ 
SDSSJ1213+6708 & $1.451_{-0.005}^{+0.005}$ & $1.71_{-0.01}^{+0.01}\pm 0.22$ & $0.99_{-0.01}^{+0.01}$ & $\phantom - 84_{-4}^{+1}$ &$0.014_{-0.003}^{+0.002}$ & $ -31_{-8}^{+5}$  &$0.838_{-0.001}^{+0.001}$ & $\phantom - 70.2_{-0.3}^{+0.3}$ &$2.77_{-0.01}^{+0.01}$ \\ 
SDSSJ1218+0830 & $1.489_{-0.008}^{+0.007}$ & $1.96_{-0.02}^{+0.02}\pm 0.10$ & $0.64_{-0.02}^{+0.03}$ & $\phantom - 40_{-2}^{+1}$ &$0.042_{-0.007}^{+0.006}$ & $\phantom - 51_{-4}^{+5}$ &$0.698_{-0.001}^{+0.001}$ & $\phantom - 40.0_{-0.1}^{+0.1}$ &$3.51_{-0.02}^{+0.02}$ \\ 
SDSSJ1250+0523 & $1.120_{-0.001}^{+0.001}$ & $2.18_{-0.03}^{+0.03}\pm 0.08$ & $0.93_{-0.01}^{+0.01}$ & $ -22.8_{-0.4}^{+0.5}$  &$0.010_{-0.002}^{+0.002}$ & $ -5_{-4}^{+6}$  &$0.889_{-0.002}^{+0.002}$ & $ -8.1_{-0.3}^{+0.4}$  &$1.75_{-0.01}^{+0.01}$ \\ 
SDSSJ1306+0600 & $1.312_{-0.006}^{+0.006}$ & $1.97_{-0.05}^{+0.06}\pm 0.04$ & $0.87_{-0.01}^{+0.01}$ & $ -46_{-4}^{+5}$  &$0.008_{-0.006}^{+0.008}$ & $ -8_{-24}^{+24}$  &$0.904_{-0.002}^{+0.002}$ & $ -50.3_{-0.7}^{+0.8}$  &$2.37_{-0.01}^{+0.01}$ \\ 
SDSSJ1313+4615 & $1.290_{-0.005}^{+0.006}$ & $2.18_{-0.02}^{+0.02}\pm 0.08$ & $0.76_{-0.02}^{+0.03}$ & $\phantom - 20_{-2}^{+3}$ &$0.032_{-0.006}^{+0.006}$ & $ -6_{-6}^{+7}$  &$0.794_{-0.001}^{+0.002}$ & $\phantom - 32.4_{-0.2}^{+0.2}$ &$2.17_{-0.01}^{+0.01}$ \\ 
SDSSJ1402+6321 & $1.371_{-0.001}^{+0.004}$ & $1.57_{-0.04}^{+0.50}\pm 0.13$ & $0.996_{-0.179}^{+0.002}$ & $\phantom - 28_{-17}^{+6}$ &$0.006_{-0.002}^{+0.001}$ & $ -27_{-4}^{+29}$  &$0.746_{-0.002}^{+0.012}$ & $\phantom - 20.0_{-1.4}^{+0.2}$ &$2.72_{-0.01}^{+0.01}$ \\ 
SDSSJ1531-0105 & $1.712_{-0.009}^{+0.004}$ & $2.07_{-0.10}^{+0.12}\pm 0.23$ & $0.68_{-0.02}^{+0.02}$ & $ -55_{-1}^{+1}$  &$0.029_{-0.023}^{+0.012}$ & $ -58_{-9}^{+7}$  &$0.668_{-0.001}^{+0.001}$ & $ -55.0_{-0.1}^{+0.1}$  &$3.42_{-0.02}^{+0.02}$ \\ 
SDSSJ1621+3931 & $1.266_{-0.004}^{+0.004}$ & $2.02_{-0.06}^{+0.06}\pm 0.04$ & $0.76_{-0.02}^{+0.02}$ & $ -44_{-3}^{+3}$  &$0.014_{-0.008}^{+0.007}$ & $\phantom - 2_{-18}^{+19}$ &$0.729_{-0.002}^{+0.002}$ & $ -52.6_{-0.2}^{+0.2}$  &$2.40_{-0.01}^{+0.01}$ \\ 
SDSSJ1627-0053 & $1.221_{-0.001}^{+0.001}$ & $1.93_{-0.05}^{+0.04}\pm 0.13$ & $0.832_{-0.005}^{+0.005}$ & $\phantom - 84_{-1}^{+1}$ &$0.034_{-0.002}^{+0.003}$ & $\phantom - 89_{-1}^{+1}$ &$0.822_{-0.001}^{+0.001}$ & $\phantom - 86.6_{-0.2}^{+0.2}$ &$2.98_{-0.05}^{+0.06}$ \\ 
SDSSJ1630+4520 & $1.791_{-0.001}^{+0.001}$ & $1.92_{-0.02}^{+0.02}\pm 0.08$ & $0.821_{-0.004}^{+0.004}$ & $\phantom - 16_{-1}^{+1}$ &$0.026_{-0.002}^{+0.002}$ & $\phantom - 18_{-1}^{+2}$ &$0.832_{-0.001}^{+0.001}$ & $\phantom - 19.1_{-0.2}^{+0.2}$ &$2.01_{-0.01}^{+0.01}$ \\ 
SDSSJ1636+4707 & $1.084_{-0.003}^{+0.002}$ & $1.75_{-0.04}^{+0.27}\pm 0.07$ & $0.88_{-0.15}^{+0.03}$ & $ -23_{-7}^{+16}$  &$0.029_{-0.026}^{+0.008}$ & $ -77_{-9}^{+42}$  &$0.823_{-0.011}^{+0.003}$ & $ -15.7_{-0.6}^{+1.9}$  &$1.84_{-0.01}^{+0.01}$ \\ 
SDSSJ2238-0754 & $1.270_{-0.001}^{+0.001}$ & $2.34_{-0.04}^{+0.03}\pm 0.12$ & $0.72_{-0.01}^{+0.02}$ & $ -46_{-1}^{+1}$  &$0.002_{-0.001}^{+0.002}$ & $\phantom - 85_{-26}^{+41}$ &$0.775_{-0.001}^{+0.001}$ & $ -51.1_{-0.2}^{+0.1}$  &$2.52_{-0.01}^{+0.01}$ \\ 
SDSSJ2300+0022 & $1.232_{-0.004}^{+0.004}$ & $1.66_{-0.09}^{+0.12}\pm 0.19$ & $0.79_{-0.02}^{+0.01}$ & $\phantom - 17_{-2}^{+1}$ &$0.044_{-0.005}^{+0.005}$ & $\phantom - 40_{-5}^{+9}$ &$0.797_{-0.002}^{+0.002}$ & $\phantom - 4.5_{-0.2}^{+0.2}$ &$1.82_{-0.01}^{+0.01}$ \\ 
SDSSJ2302-0840 & $1.038_{-0.001}^{+0.001}$ & $2.60_{-0.02}^{+0.02}\pm 0.08$ & $0.66_{-0.02}^{+0.01}$ & $ -74_{-2}^{+1}$  &$0.010_{-0.003}^{+0.003}$ & $\phantom - 47_{-7}^{+9}$ &$0.831_{-0.001}^{+0.001}$ & $ -75.7_{-0.2}^{+0.2}$  &$4.20_{-0.04}^{+0.04}$ \\ 
SDSSJ2303+1422 & $1.622_{-0.002}^{+0.002}$ & $1.94_{-0.02}^{+0.02}\pm 0.08$ & $0.65_{-0.01}^{+0.01}$ & $\phantom - 55.4_{-0.3}^{+0.3}$ &$0.001_{-0.001}^{+0.001}$ & $\phantom - 6_{-10}^{+21}$ &$0.643_{-0.001}^{+0.001}$ & $\phantom - 52.6_{-0.1}^{+0.1}$ &$3.64_{-0.01}^{+0.01}$ \\ 
SDSSJ2343-0030 & $1.506_{-0.002}^{+0.002}$ & $2.16_{-0.05}^{+0.04}\pm 0.12$ & $0.69_{-0.02}^{+0.02}$ & $\phantom - 52_{-1}^{+1}$ &$0.083_{-0.004}^{+0.005}$ & $\phantom - 5_{-1}^{+1}$ &$0.660_{-0.002}^{+0.002}$ & $\phantom - 51.3_{-0.2}^{+0.2}$ &$2.37_{-0.01}^{+0.01}$ \\ 
SDSSJ2347-0005 & $1.077_{-0.004}^{+0.005}$ & $1.98_{-0.03}^{+0.03}\pm 0.07$ & $0.61_{-0.02}^{+0.02}$ & $\phantom - 64_{-2}^{+2}$ &$0.053_{-0.008}^{+0.007}$ & $\phantom - 27_{-3}^{+4}$ &$0.697_{-0.003}^{+0.003}$ & $\phantom - 71.6_{-0.3}^{+0.3}$ &$1.44_{-0.01}^{+0.01}$ \\   
\hline
\end{tabular}
\end{table*}

\begin{table*}
 \caption{Lens model parameters for the SL2S systems. See the caption of Table \ref{table:SLACS_params} for the column descriptions.}
\label{table:SL2S_params}

\begin{tabular}{lccccccccc}
\hline
     Lens system &  $\Rein$ &    $\gamma$ &    $q_\text{m}$ &     $\phi_\text{m}$ &  $\gamma_\text{shear}$  &  $\phi_\text{shear}$  &
     $q_\text{L}$ & 
     $\phi_\text{L}$ &
     $R_{{\rm eff}, V}$  
     \\
     & (arcsec) & &&(deg) & & (deg) &  & (deg) & (arcsec)\\
\hline
SL2SJ0208-0714 & $2.583_{-0.005}^{+0.042}$ & $1.98_{-0.03}^{+0.03}\pm 0.07$ & $0.99_{-0.16}^{+0.01}$ & $\phantom - 11_{-6}^{+14}$ &$0.068_{-0.011}^{+0.003}$ & $ -56_{-1}^{+26}$  &$0.825_{-0.004}^{+0.008}$ & $\phantom - 19.0_{-0.6}^{+0.6}$ &$2.09_{-0.03}^{+0.03}$ \\ 
SL2SJ0214-0405 & $1.221_{-0.006}^{+0.006}$ & $1.86_{-0.04}^{+0.04}\pm 0.33$ & $0.70_{-0.02}^{+0.02}$ & $\phantom - 32_{-1}^{+1}$ &$0.003_{-0.002}^{+0.004}$ & $ -77_{-40}^{+33}$  &$0.941_{-0.011}^{+0.011}$ & $\phantom - 26.0_{-4.8}^{+5.3}$ &$1.18_{-0.04}^{+0.04}$ \\ 
SL2SJ0217-0513 & $1.258_{-0.003}^{+0.003}$ & $2.34_{-0.06}^{+0.06}\pm 0.23$ & $0.85_{-0.02}^{+0.01}$ & $ -25_{-2}^{+2}$  &$0.163_{-0.009}^{+0.009}$ & $\phantom - 89_{-1}^{+1}$ &$0.952_{-0.009}^{+0.010}$ & $ -22.7_{-6.1}^{+5.9}$  &$0.68_{-0.02}^{+0.01}$ \\ 
SL2SJ0219-0829 & $1.285_{-0.006}^{+0.008}$ & $2.62_{-0.18}^{+0.07}\pm 0.08$ & $0.72_{-0.03}^{+0.03}$ & $\phantom - 0_{-1}^{+2}$ &$0.023_{-0.011}^{+0.010}$ & $\phantom - 67_{-17}^{+8}$ &$0.689_{-0.009}^{+0.006}$ & $\phantom - 14.0_{-0.8}^{+0.8}$ &$0.88_{-0.01}^{+0.01}$ \\ 
SL2SJ0225-0454 & $1.716_{-0.017}^{+0.022}$ & $1.96_{-0.03}^{+0.03}\pm 0.11$ & $0.67_{-0.03}^{+0.03}$ & $\phantom - 59_{-2}^{+2}$ &$0.037_{-0.012}^{+0.012}$ & $ -13_{-5}^{+14}$  &$0.690_{-0.003}^{+0.003}$ & $\phantom - 64.7_{-0.3}^{+0.3}$ &$2.55_{-0.01}^{+0.01}$ \\ 
SL2SJ0226-0420 & $1.099_{-0.014}^{+0.014}$ & $2.14_{-0.10}^{+0.10}\pm 0.08$ & $0.82_{-0.11}^{+0.06}$ & $\phantom - 22_{-6}^{+7}$ &$0.013_{-0.009}^{+0.016}$ & $\phantom - 8_{-54}^{+37}$ &$0.832_{-0.024}^{+0.017}$ & $\phantom - 28.7_{-3.4}^{+4.3}$ &$0.89_{-0.01}^{+0.01}$ \\ 
SL2SJ0226-0406 & $1.289_{-0.017}^{+0.015}$ & $2.02_{-0.17}^{+0.22}\pm 0.11$ & $0.76_{-0.08}^{+0.07}$ & $ -30_{-4}^{+2}$  &$0.009_{-0.006}^{+0.015}$ & $ -21_{-36}^{+61}$  &$0.343_{-0.010}^{+0.014}$ & $ -41.6_{-1.0}^{+1.1}$  &$0.88_{-0.04}^{+0.03}$ \\ 
SL2SJ0232-0408 & $1.012_{-0.006}^{+0.007}$ & $2.49_{-0.16}^{+0.15}\pm 0.19$ & $0.77_{-0.08}^{+0.07}$ & $ -29_{-7}^{+8}$  &$0.018_{-0.013}^{+0.020}$ & $ -11_{-22}^{+23}$  &$0.694_{-0.013}^{+0.014}$ & $ -26.8_{-1.9}^{+1.4}$  &$1.13_{-0.01}^{+0.01}$ \\ 
SL2SJ0849-0412 & $1.074_{-0.012}^{+0.012}$ & $2.26_{-0.03}^{+0.03}\pm 0.18$ & $0.61_{-0.03}^{+0.03}$ & $\phantom - 40_{-3}^{+3}$ &$0.032_{-0.015}^{+0.014}$ & $ -29_{-12}^{+11}$  &$0.465_{-0.005}^{+0.006}$ & $\phantom - 49.6_{-0.3}^{+0.3}$ &$0.60_{-0.01}^{+0.01}$ \\ 
SL2SJ0849-0251 & $1.192_{-0.009}^{+0.007}$ & $2.24_{-0.31}^{+0.27}\pm 0.09$ & $0.84_{-0.07}^{+0.05}$ & $ -21_{-4}^{+6}$  &$0.008_{-0.006}^{+0.010}$ & $\phantom - 21_{-37}^{+38}$ &$0.760_{-0.008}^{+0.008}$ & $ -25.7_{-1.1}^{+1.1}$  &$1.54_{-0.01}^{+0.01}$ \\ 
SL2SJ0858-0143 & $0.953_{-0.025}^{+0.024}$ & $2.09_{-0.28}^{+0.45}\pm 0.27$ & $0.84_{-0.08}^{+0.07}$ & $ -61_{-14}^{+11}$  &$0.013_{-0.009}^{+0.013}$ & $\phantom - 13_{-55}^{+41}$ &$0.803_{-0.082}^{+0.089}$ & $ -64.6_{-13.2}^{+14.7}$  &$1.26_{-0.07}^{+0.08}$ \\ 
SL2SJ0901-0259 & $0.997_{-0.064}^{+0.023}$ & $2.11_{-0.07}^{+0.08}\pm 0.05$ & $0.75_{-0.04}^{+0.15}$ & $ -76_{-4}^{+18}$  &$0.004_{-0.003}^{+0.005}$ & $ -7_{-14}^{+12}$  &$0.767_{-0.064}^{+0.091}$ & $ -83.5_{-8.3}^{+23.3}$  &$0.70_{-0.06}^{+0.05}$ \\ 
SL2SJ0904-0059 & $1.381_{-0.013}^{+0.016}$ & $2.24_{-0.11}^{+0.10}\pm 0.19$ & $0.49_{-0.06}^{+0.07}$ & $\phantom - 12_{-1}^{+2}$ &$0.025_{-0.020}^{+0.013}$ & $ -29_{-22}^{+21}$  &$0.754_{-0.035}^{+0.033}$ & $\phantom - 18.1_{-4.3}^{+3.9}$ &$2.26_{-0.10}^{+0.08}$ \\ 
SL2SJ0959+0206 & $0.724_{-0.018}^{+0.002}$ & $2.16_{-0.18}^{+0.27}\pm 0.30$ & $0.94_{-0.05}^{+0.04}$ & $\phantom - 82_{-17}^{+19}$ &$0.075_{-0.031}^{+0.041}$ & $\phantom - 6_{-2}^{+5}$ &$0.793_{-0.039}^{+0.114}$ & $\phantom - 72.7_{-6.0}^{+17.3}$ &$0.46_{-0.01}^{+0.01}$ \\ 
SL2SJ1358+5459 & $1.241_{-0.014}^{+0.013}$ & $2.08_{-0.07}^{+0.08}\pm 0.03$ & $0.81_{-0.02}^{+0.02}$ & $ -47_{-1}^{+3}$  &$0.098_{-0.005}^{+0.005}$ & $ -72_{-2}^{+2}$  &$0.791_{-0.006}^{+0.006}$ & $ -33.4_{-0.9}^{+0.9}$  &$1.02_{-0.02}^{+0.03}$ \\ 
SL2SJ1359+5535 & $1.107_{-0.013}^{+0.014}$ & $1.95_{-0.16}^{+0.15}\pm 0.26$ & $0.69_{-0.12}^{+0.08}$ & $\phantom - 35_{-1}^{+1}$ &$0.004_{-0.002}^{+0.004}$ & $ -82_{-45}^{+37}$  &$0.680_{-0.084}^{+0.075}$ & $\phantom - 46.2_{-4.4}^{+2.9}$ &$2.57_{-0.12}^{+0.18}$ \\ 
SL2SJ1401+5544 & $1.924_{-0.012}^{+0.012}$ & $2.03_{-0.03}^{+0.03}\pm 0.19$ & $0.61_{-0.02}^{+0.02}$ & $\phantom - 59_{-1}^{+2}$ &$0.058_{-0.014}^{+0.011}$ & $ -18_{-2}^{+2}$  &$0.810_{-0.004}^{+0.003}$ & $\phantom - 63.9_{-0.6}^{+0.7}$ &$1.64_{-0.01}^{+0.01}$ \\ 
SL2SJ1402+5505 & $1.233_{-0.009}^{+0.012}$ & $1.76_{-0.15}^{+0.14}\pm 0.21$ & $0.95_{-0.06}^{+0.02}$ & $ -46_{-13}^{+5}$  &$0.022_{-0.018}^{+0.013}$ & $\phantom - 50_{-13}^{+9}$ &$0.868_{-0.008}^{+0.007}$ & $ -52.5_{-1.5}^{+1.5}$  &$1.59_{-0.01}^{+0.01}$ \\ 
SL2SJ1405+5243 & $1.431_{-0.005}^{+0.011}$ & $1.72_{-0.11}^{+0.09}\pm 0.08$ & $0.94_{-0.05}^{+0.02}$ & $ -63_{-8}^{+9}$  &$0.067_{-0.009}^{+0.011}$ & $\phantom - 68_{-2}^{+5}$ &$0.843_{-0.005}^{+0.005}$ & $ -59.6_{-1.0}^{+1.0}$  &$1.02_{-0.01}^{+0.02}$ \\ 
SL2SJ1406+5226 & $1.034_{-0.011}^{+0.010}$ & $1.91_{-0.08}^{+0.06}\pm 0.25$ & $0.60_{-0.03}^{+0.04}$ & $ -3_{-1}^{+2}$  &$0.006_{-0.005}^{+0.007}$ & $\phantom - 75_{-47}^{+72}$ &$0.503_{-0.029}^{+0.035}$ & $\phantom - 0.3_{-3.0}^{+2.8}$ &$0.83_{-0.04}^{+0.04}$ \\ 
SL2SJ1411+5651 & $0.920_{-0.001}^{+0.003}$ & $1.75_{-0.02}^{+0.25}\pm 0.28$ & $0.873_{-0.005}^{+0.005}$ & $\phantom - 70_{-8}^{+1}$ &$0.056_{-0.008}^{+0.002}$ & $\phantom - 86_{-2}^{+1}$ &$0.790_{-0.006}^{+0.053}$ & $\phantom - 84.5_{-14.2}^{+0.7}$ &$0.57_{-0.01}^{+0.01}$ \\ 
SL2SJ1420+5630 & $1.391_{-0.007}^{+0.004}$ & $2.06_{-0.04}^{+0.04}\pm 0.22$ & $0.58_{-0.03}^{+0.04}$ & $ -70.9_{-0.6}^{+0.4}$  &$0.059_{-0.008}^{+0.007}$ & $ -43_{-3}^{+3}$  &$0.575_{-0.004}^{+0.004}$ & $ -75.7_{-0.4}^{+0.3}$  &$1.79_{-0.02}^{+0.01}$ \\ 
SL2SJ1427+5516 & $0.926_{-0.016}^{+0.011}$ & $2.29_{-0.09}^{+0.06}\pm 0.33$ & $0.30_{-0.02}^{+0.04}$ & $ -29_{-1}^{+1}$  &$0.072_{-0.021}^{+0.027}$ & $ -23_{-3}^{+6}$  &$0.255_{-0.005}^{+0.006}$ & $ -28.1_{-0.3}^{+0.3}$  &$0.27_{-0.01}^{+0.01}$ \\ 
SL2SJ2214-1807 & $0.807_{-0.255}^{+0.114}$ & $2.20_{-0.37}^{+0.28}\pm 0.16$ & $0.46_{-0.12}^{+0.20}$ & $\phantom - 49_{-16}^{+9}$ &$0.016_{-0.011}^{+0.018}$ & $\phantom - 10_{-64}^{+49}$ &$0.669_{-0.040}^{+0.038}$ & $\phantom - 46.6_{-4.1}^{+3.5}$ &$0.94_{-0.04}^{+0.04}$ \\  
\hline
\end{tabular}
\end{table*}

\begin{table*}
 \caption{Lens models parameters for the BELLS systems. See the caption of Table \ref{table:SLACS_params} for the column descriptions. \cyref{Due to the lack of $V$-band imaging for the BELLS systems, the lens
 light parameters were obtained from \textit{HST} images in the $I$-band (F814W).}}
\label{table:BELLS_params}
\begin{tabular}{lccccccccc}
\hline
     Lens system &  $\Rein$ &    $\gamma$ &    $q_\text{m}$ &     $\phi_\text{m}$ &  $\gamma_\text{shear}$  &  $\phi_\text{shear}$  &
     $q_\text{L}$ & 
     $\phi_\text{L}$ &
     $R_{{\rm eff}, I}  $  
     \\
     & (arcsec) & &&(deg)   & & (deg) &  & (deg) & (arcsec)\\
\hline
SDSSJ0151+0049 & $0.649_{-0.005}^{+0.004}$ & $2.26_{-0.04}^{+0.05}\pm 0.33$ & $0.54_{-0.02}^{+0.10}$ & $ -13_{-1}^{+3}$  &$0.071_{-0.051}^{+0.013}$ & $ -2_{-4}^{+13}$  &$0.588_{-0.005}^{+0.007}$ & $ -18.8_{-0.6}^{+0.4}$  &$0.72_{-0.01}^{+0.01}$ \\ 
SDSSJ0747+5055 & $0.699_{-0.003}^{+0.002}$ & $2.48_{-0.02}^{+0.02}\pm 0.26$ & $0.61_{-0.03}^{+0.03}$ & $ -80.3_{-0.6}^{+0.4}$  &$0.117_{-0.005}^{+0.004}$ & $ -14_{-1}^{+1}$  &$0.729_{-0.003}^{+0.004}$ & $\phantom - 84.9_{-0.3}^{+0.4}$ &$1.10_{-0.01}^{+0.01}$ \\ 
SDSSJ0801+4727 & $0.504_{-0.009}^{+0.005}$ & $2.68_{-0.19}^{+0.06}\pm 0.31$ & $0.54_{-0.07}^{+0.17}$ & $ -89_{-35}^{+4}$  &$0.020_{-0.015}^{+0.020}$ & $ -84_{-17}^{+61}$  &$0.962_{-0.012}^{+0.024}$ & $\phantom - 78.0_{-19.8}^{+4.9}$ &$0.67_{-0.02}^{+0.02}$ \\ 
SDSSJ0830+5116 & $1.129_{-0.003}^{+0.003}$ & $2.09_{-0.02}^{+0.03}\pm 0.33$ & $0.86_{-0.01}^{+0.02}$ & $ -40_{-2}^{+2}$  &$0.016_{-0.006}^{+0.005}$ & $ -30_{-7}^{+9}$  &$0.723_{-0.004}^{+0.003}$ & $ -30.6_{-0.5}^{+0.5}$  &$0.76_{-0.01}^{+0.01}$ \\ 
SDSSJ0944-0147 & $0.738_{-0.008}^{+0.008}$ & $2.42_{-0.18}^{+0.17}\pm 0.20$ & $0.71_{-0.12}^{+0.07}$ & $\phantom - 9_{-3}^{+3}$ &$0.078_{-0.020}^{+0.019}$ & $\phantom - 9_{-3}^{+3}$ &$0.861_{-0.006}^{+0.006}$ & $ -3.2_{-1.3}^{+1.2}$  &$0.85_{-0.02}^{+0.02}$ \\ 
SDSSJ1159-0007 & $0.674_{-0.002}^{+0.003}$ & $2.67_{-0.06}^{+0.05}\pm 0.33$ & $0.74_{-0.05}^{+0.04}$ & $ -78_{-5}^{+5}$  &$0.068_{-0.007}^{+0.007}$ & $\phantom - 72_{-2}^{+3}$ &$0.934_{-0.007}^{+0.009}$ & $ -74.0_{-3.8}^{+3.3}$  &$0.77_{-0.01}^{+0.01}$ \\ 
SDSSJ1215+0047 & $1.338_{-0.007}^{+0.005}$ & $2.19_{-0.01}^{+0.02}\pm 0.24$ & $0.73_{-0.08}^{+0.02}$ & $ -45_{-1}^{+1}$  &$0.048_{-0.009}^{+0.005}$ & $\phantom - 76_{-4}^{+15}$ &$0.676_{-0.003}^{+0.003}$ & $ -59.5_{-0.3}^{+0.5}$  &$0.79_{-0.01}^{+0.01}$ \\ 
SDSSJ1221+3806 & $0.692_{-0.006}^{+0.006}$ & $2.24_{-0.07}^{+0.10}\pm 0.20$ & $0.68_{-0.04}^{+0.04}$ & $\phantom - 16_{-4}^{+5}$ &$0.067_{-0.016}^{+0.021}$ & $\phantom - 62_{-9}^{+9}$ &$0.844_{-0.005}^{+0.005}$ & $\phantom - 19.3_{-1.2}^{+0.8}$ &$0.64_{-0.01}^{+0.01}$ \\ 
SDSSJ1234-0241 & $0.559_{-0.005}^{+0.005}$ & $1.98_{-0.07}^{+0.07}\pm 0.33$ & $0.51_{-0.03}^{+0.03}$ & $ -5_{-2}^{+2}$  &$0.108_{-0.017}^{+0.014}$ & $ -5_{-3}^{+3}$  &$0.717_{-0.003}^{+0.003}$ & $ -4.4_{-0.4}^{+0.3}$  &$1.12_{-0.01}^{+0.01}$ \\ 
SDSSJ1318-0104 & $0.646_{-0.003}^{+0.002}$ & $2.22_{-0.03}^{+0.03}\pm 0.32$ & $0.91_{-0.03}^{+0.03}$ & $ -32_{-1}^{+2}$  &$0.015_{-0.003}^{+0.005}$ & $ -2_{-11}^{+21}$  &$0.706_{-0.006}^{+0.005}$ & $ -18.5_{-0.6}^{+0.7}$  &$0.74_{-0.01}^{+0.01}$ \\ 
SDSSJ1337+3620 & $1.416_{-0.006}^{+0.003}$ & $2.18_{-0.04}^{+0.08}\pm 0.32$ & $0.86_{-0.02}^{+0.01}$ & $ -15_{-5}^{+3}$  &$0.144_{-0.005}^{+0.006}$ & $\phantom - 32.8_{-0.4}^{+0.5}$ &$0.942_{-0.004}^{+0.003}$ & $ -27.3_{-2.2}^{+2.0}$  &$2.02_{-0.01}^{+0.01}$ \\ 
SDSSJ1349+3612 & $0.673_{-0.004}^{+0.003}$ & $2.36_{-0.04}^{+0.04}\pm 0.23$ & $0.54_{-0.02}^{+0.03}$ & $\phantom - 71_{-2}^{+2}$ &$0.173_{-0.015}^{+0.009}$ & $\phantom - 39_{-2}^{+2}$ &$0.780_{-0.004}^{+0.003}$ & $\phantom - 77.3_{-0.6}^{+0.5}$ &$2.40_{-0.09}^{+0.09}$ \\ 
SDSSJ1352+3216 & $1.693_{-0.006}^{+0.005}$ & $2.21_{-0.01}^{+0.01}\pm 0.33$ & $0.40_{-0.01}^{+0.01}$ & $ -6_{-1}^{+1}$  &$0.202_{-0.004}^{+0.004}$ & $ -2_{-1}^{+1}$  &$0.917_{-0.004}^{+0.004}$ & $ -20.7_{-0.8}^{+0.8}$  &$0.69_{-0.01}^{+0.01}$ \\ 
SDSSJ1542+1629 & $1.029_{-0.006}^{+0.006}$ & $2.16_{-0.04}^{+0.03}\pm 0.33$ & $0.92_{-0.03}^{+0.04}$ & $\phantom - 8_{-10}^{+6}$ &$0.071_{-0.008}^{+0.007}$ & $\phantom - 80_{-3}^{+3}$ &$0.796_{-0.002}^{+0.002}$ & $\phantom - 0.6_{-0.3}^{+0.4}$ &$1.46_{-0.02}^{+0.02}$ \\ 
SDSSJ1545+2748 & $1.122_{-0.009}^{+0.009}$ & $1.94_{-0.03}^{+0.02}\pm 0.24$ & $0.67_{-0.02}^{+0.02}$ & $ -14_{-4}^{+2}$  &$0.102_{-0.012}^{+0.013}$ & $\phantom - 85_{-2}^{+3}$ &$0.645_{-0.004}^{+0.006}$ & $ -15.8_{-0.2}^{+0.2}$  &$2.34_{-0.01}^{+0.01}$ \\ 
SDSSJ1601+2138 & $0.787_{-0.013}^{+0.013}$ & $2.35_{-0.04}^{+0.05}\pm 0.16$ & $0.44_{-0.02}^{+0.33}$ & $ -65_{-6}^{+4}$  &$0.155_{-0.136}^{+0.018}$ & $ -62_{-4}^{+8}$  &$0.924_{-0.004}^{+0.003}$ & $ -73.8_{-1.6}^{+1.2}$  &$0.53_{-0.01}^{+0.01}$ \\ 
SDSSJ1631+1854 & $1.629_{-0.001}^{+0.001}$ & $2.07_{-0.02}^{+0.02}\pm 0.29$ & $0.986_{-0.005}^{+0.005}$ & $\phantom - 84_{-2}^{+2}$ &$0.036_{-0.002}^{+0.001}$ & $ -38_{-1}^{+1}$  &$0.943_{-0.002}^{+0.003}$ & $ -81.3_{-1.3}^{+1.2}$  &$1.87_{-0.03}^{+0.03}$ \\ 
SDSSJ2125+0411 & $1.234_{-0.003}^{+0.003}$ & $1.72_{-0.01}^{+0.01}\pm 0.31$ & $0.81_{-0.01}^{+0.01}$ & $ -36.3_{-0.3}^{+0.4}$  &$0.092_{-0.005}^{+0.005}$ & $ -53_{-1}^{+1}$  &$0.730_{-0.002}^{+0.002}$ & $ -21.5_{-0.3}^{+0.3}$  &$1.66_{-0.03}^{+0.03}$ \\ 
SDSSJ2303+0037 & $0.986_{-0.002}^{+0.005}$ & $2.39_{-0.39}^{+0.08}\pm 0.33$ & $0.69_{-0.04}^{+0.14}$ & $ -4_{-2}^{+1}$  &$0.167_{-0.059}^{+0.015}$ & $ -86.3_{-0.5}^{+1.4}$  &$0.763_{-0.004}^{+0.005}$ & $\phantom - 4.8_{-0.5}^{+0.4}$ &$1.28_{-0.01}^{+0.01}$ \\ 
\hline
\end{tabular}
\end{table*}
\endgroup

\begin{figure*}
    \centering
    \includegraphics[width=0.9\textwidth]{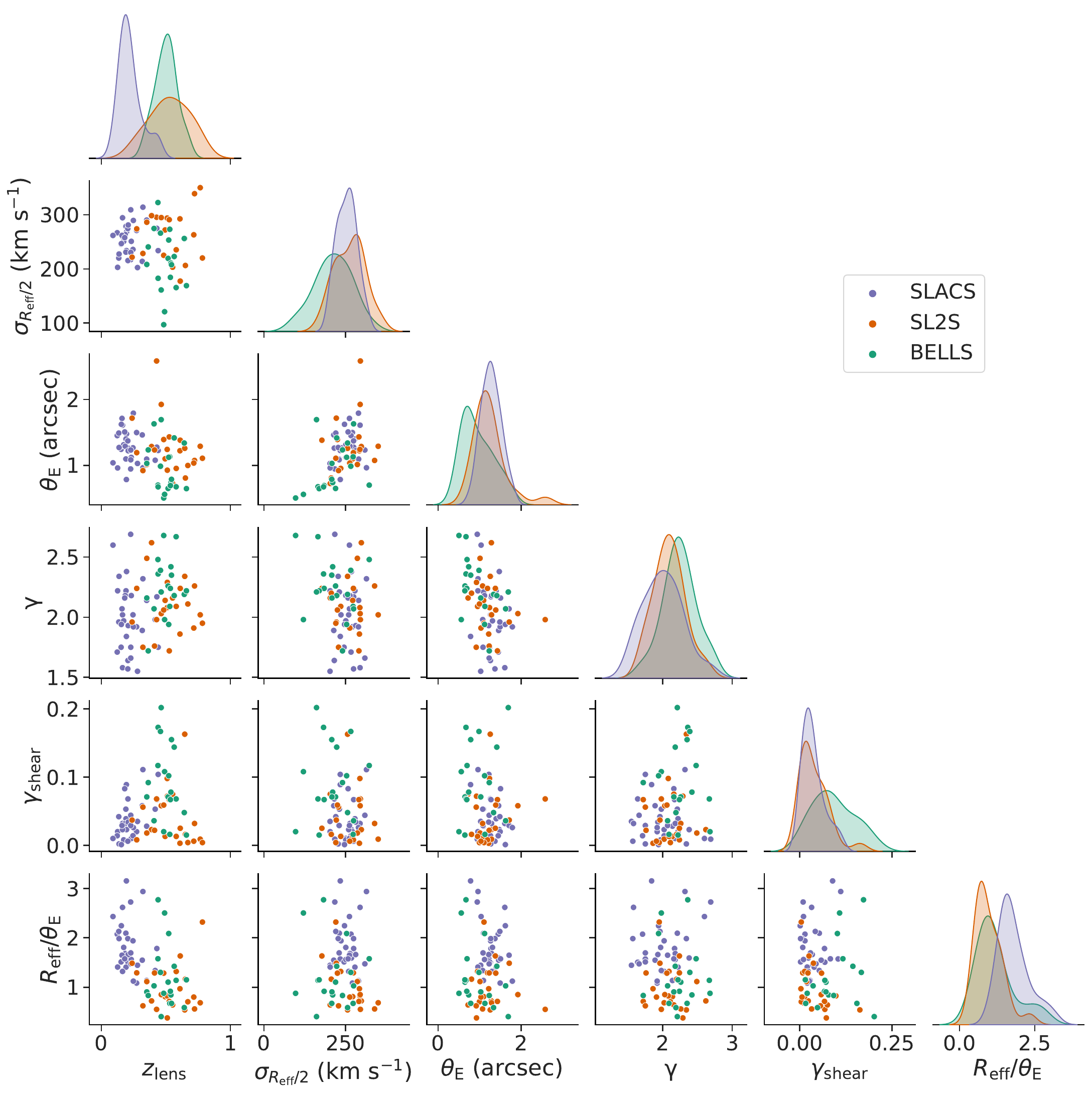}
    \caption{Distributions of deflector redshift $z_{\rm lens}$, central velocity dispersion $\sigma_{R_{\rm eff}/2}$, Einstein radius $\Rein$, logarithmic slope $\gamma$, residual shear magnitude $\gamma_{\rm shear}$, and the ratio of half-light radius to Einstein radius $\Reff/\Rein$ for the SLACS (purple), SL2S (orange), and BELLS (green) systems. The velocity dispersion $\sigma_{R_{\rm eff}/2}$ is adjusted from the measured values \ajstwo{to correspond to} a circular aperture with radius $\Reff/2$ \ajstwo{\citep{Jorgensen95}}. As the distribution of key properties, such as \cyone{$z_{\rm lens}$ and $\Reff/\Rein$,} are different among the samples, selection effects may potentially affect the accuracy of \cyone{our analysis. Therefore, in our hierarchical analysis in Section \ref{Section4:Measure_lambda}, we consider the dependence of \ajstwo{the potential deviation from the power law on} $z_{\rm lens}$ and $\Reff/\Rein$}. The kernel density estimates of the 1D distributions \ajstwo{(in the panels along the diagonal)} are obtained using only the \ajstwo{point estimates} of the corresponding parameters.}
\label{fig:LargeSelectionFunc}
\end{figure*}

We provide the point-estimates of the lens model parameters and 1$\sigma$ uncertainties for the SLACS, SL2S, and BELLS lenses in Tables \ref{table:SLACS_params}, \ref{table:SL2S_params}, and \ref{table:BELLS_params}, respectively. 
In Figure \ref{fig:LargeSelectionFunc}, we show the distribution of the redshift, central velocity dispersion, and other model parameters for the SLACS, SL2S, and BELLS lenses. 
\ajstwo{In the next section, we compare our lens model parameters with those from previous studies and estimate the modelling systematics in the logarithmic slope $\gamma$ based on this comparison.}

\section{\ajs{Model} comparison with previous studies}
\label{Section 5}

In this section, we compare our lens model posteriors with previous analyses.
We compare our parameter posteriors to those previously obtained from imaging data only with lens models similar to ours in Section \ref{Section_5.2} and to those obtained from previous joint lensing--dynamics analysis in Section \ref{Section_5.1}.

\subsection{Comparison with previous lensing-only analyses}
\label{Section_5.2}

We compare our lens model posteriors with those obtained by \citet{Shajib21} and \citet{Etherington22}, who also adopted the PEMD for the lens mass model. \citet{Shajib21} modelled 23 SLACS lenses, all of which are included in our analysis. 
Out of the 43 SLACS lenses modelled by \citet{Etherington22}, 21 overlap with our analysis, and 14 overlap with the analysis by \citet{Shajib21}. 

For our measured Einstein radii, we find a root-mean-square (rms) difference of 1.5 per cent with those from \citet{Shajib21} and 5.6 per cent with those from \citet{Etherington22} (Figure \ref{fig:Comparison_gamma}). \cytwo{ The high rms difference in the Einstein radii between our analysis and \citet{Etherington22} can be attributed to the outliers SDSSJ0912$+$0029, SDSSJ1213$+$6708, SDSSJ1218$+$0830, \ajstwo{for which the lensed arcs have relatively low $S/N$} (see the \cyone{\ajstwo{deflector-light-subtracted images illustrating the arcs} in the fourth columns of Figures \ref{fig:SLACS1} and \ref{fig:SLACS2})}. \cytwo{These three} outliers differ in the Einstein radius by more than 10 per cent between our analysis and \citet{Etherington22}. Interestingly, all of these three outliers were considered failure cases \ajstwo{in the automated modelling done by} \citet{Shajib21}, further highlighting \ajstwo{the low $S/N$ in the lensed arcs as an impediment to robust lens modelling}. Excluding these outliers, the rms difference between our analysis and \citet{Etherington22} falls down to 2 per cent, consistent with \ajstwo{the expected systematic level empirically estimated by} \citet{Bolton08}, \citet{Sonnenfeld13}, and \citet{Etherington22}. We also exclude these three outliers in other comparisons for the remainder of this section, but we include them in our hierarchical analysis in Section \ref{Section4:Model_paramters}.}

We find a significant difference in the logarithmic slope $\gamma$ between different studies. The median absolute difference in  $\gamma$ is 8.2 per cent between our lens models and those from \citet{Shajib21}, and 9.7 per cent between our lens models and those from \cite{Etherington22}. These differences are much larger than the median 3 per cent statistical uncertainty we estimated from the MCMC chains (Figure \ref{fig:Comparison_gamma}).
Our measured logarithmic slopes correlate only weakly with those from \citet{Shajib21} (Pearson correlation coefficient \cytwo{$r=0.30\pm0.11$}) and \citet{Etherington22} (\cytwo{$r=0.17\pm0.16$}). The weak correlation between our $\gamma$ measurements and \citet{Shajib21}'s is noteworthy, as both analyses used the \textsc{dolphin} modelling pipeline with very similar lens model configurations. A major difference, however, is that we use multi-band modelling, whereas \citet{Shajib21} only modelled a single \textit{HST} filter. \cytwo{The apparent differences in the logarithmic slope of the mass profile between the studies suggest that there might be an additional modelling systematic uncertainty not accounted for by the statistical uncertainty.}

\begin{figure*}
    \centering
    \includegraphics[width=0.9\textwidth]{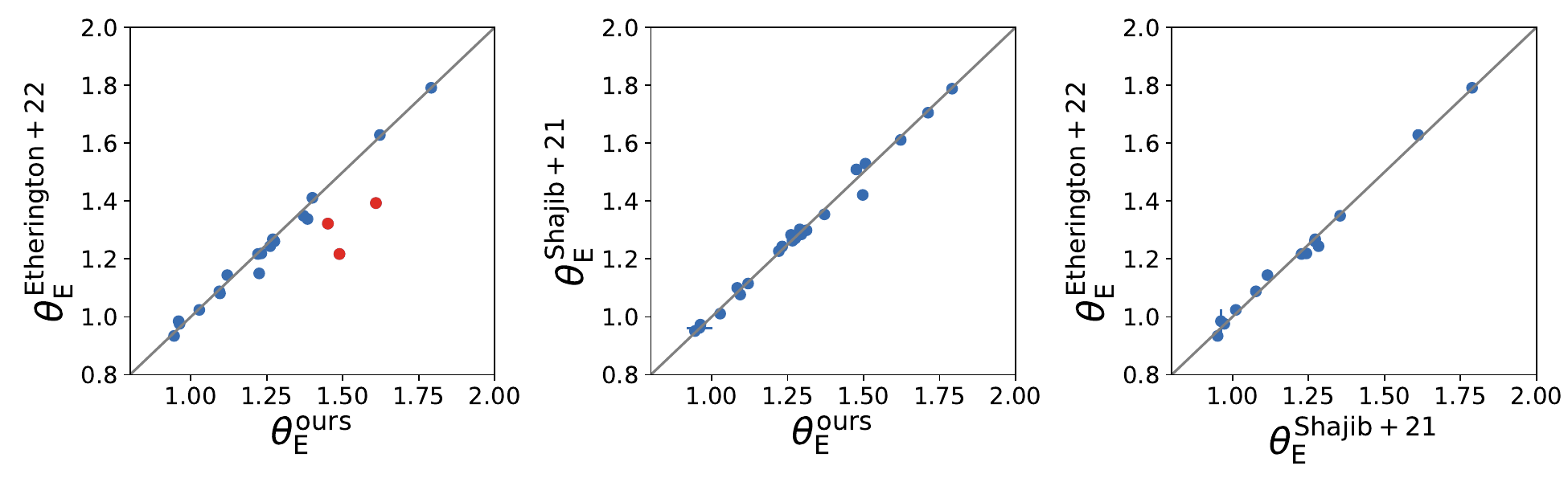}
    \caption{Comparison of the measured Einstein radii between our analysis (denoted by $\Rein^{\rm ours}$) and other lensing-only analyses from \citet{Shajib21} and  \citet{Etherington22} (denoted by $\Rein^{\rm Shajib+21}$ and $\Rein^{\rm Etherington+22}$, respectively). For the three outlier systems in the comparison between this paper and \citet{Etherington22} (marked with red points; SDSSJ0912$+$0029, SDSSJ1213$+$6708, and SDSSJ1218$+$0830), the lensed arcs have low $S/N$ and the Einstein radii differ by more than 10 per cent. If we exclude these outliers, we find the rms difference between the different analyses to be $<$2 per cent.}
\label{fig:Comparison_Rein}
\end{figure*}

\begin{figure*}
    \centering
    \includegraphics[width=0.9\textwidth]{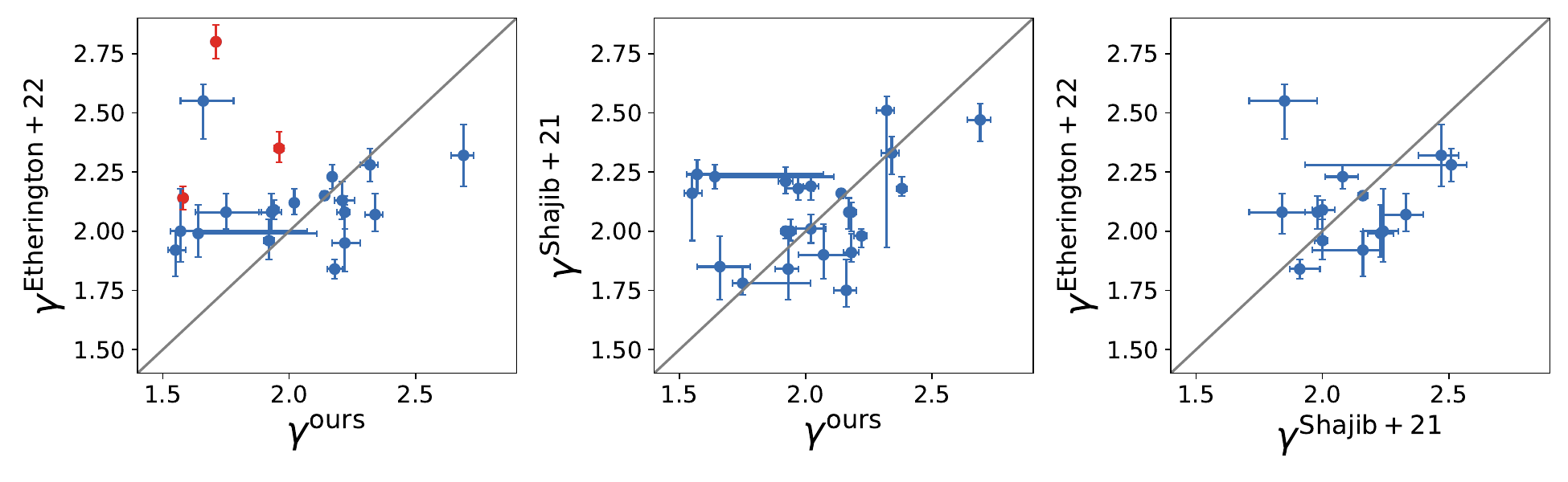}
    \caption{Comparison of the measured logarithmic slopes between our analysis (denoted by $\gamma^{\rm ours}$) and other lensing-only analyses from \citet{Shajib21} and  \citet{Etherington22} (denoted by $\gamma^{\rm Shajib+21}$ and $\gamma^{\rm Etherington+22}$, respectively). The three outlier systems from Figure \ref{fig:Comparison_Rein} are similarly marked with red. We find \ajstwo{only} a \cyone{weak correlation} between the  \ajstwo{measured logarithmic} slope $\gamma$ from different analyses. For example, the Pearson correlation coefficient between the $\gamma$ values from our analysis and \citet{Etherington22} is \cytwo{$r=0.17\pm0.16$}. \ajstwo{We reillustrate the left-hand panel in Figure \ref{fig:modeling_systematic_gamma} after adding the estimated systematic uncertainty levels to the error bars.}}
\label{fig:Comparison_gamma}
\end{figure*}

\ajs{\cytwo{Since \citet{Etherington22} used a different source reconstruction method (adaptive brightness-based pixelisation and regularisation grid) implemented in a different modelling pipeline \textsc{PyAutoLens} \citep{Nightingale21}, we estimated the modelling systematic uncertainty in $\gamma$ by comparing our model predicted values with theirs, excluding the three outlier systems mentioned above.} \ajstwo{We did not use the comparison with the values from \citet{Shajib21} since the difference is smaller. Thus, the comparison with \citet{Etherington22} provides the most conservative estimate of the systematic uncertainty pertaining to lens modelling.}  

We assumed that systems with higher `lensing information' would allow less modelling systematics to manifest in the $\gamma$ measurement. We first quantified this lensing information  using a weighted $S/N$ quantity as}
\begin{equation} \label{eq:lensing_info}
    \mathcal{I} \equiv  \frac{ \sum_{\rm i}^{\rm arc} w_i S_i}{\sqrt{ \sum_{\rm i}^{\rm arc} {N}_{i}^2}},
\end{equation}
\ajs{where the summation index $i$ goes over all the pixels on the lensed arcs, $S_i$ is the flux in the lensed arcs after subtracting off the deflector light, and $N_i$ is the total noise level. We included the pixels on the arcs only with fluxes at least three times the background noise level. Among these pixels, however, we excluded the ones that belong to blobs with less than \ajstwo{five contiguous} pixels to disregard spurious blobs that are less likely to be lensed features. The ad hoc weight $w_i$ for the $i$th pixel is given by}
\begin{equation} \label{eq:lensing_info_weight}
    w_i = \left[1 +  \frac{\left| \theta_i - \Rein \right|}{\Rein} \left(1 + \frac{\left| \phi_i - \phi_{\rm ref} \right| }{\phi_{\rm ref}} \right)^b  \right]^a,
\end{equation}
\ajs{where $\theta_{i}$ is the radial distance of the pixel from the centre of the deflector galaxy, $\phi_i$ is the azimuthal angle of the pixel, and $\phi_{\rm ref}$ is a reference angle which we chose to correspond to the brightest pixel on the arc. The exponents $a$ and $b$ tune the impact of this weighting scheme. For $a=0$, the definition in equation (\ref{eq:lensing_info}) becomes the standard total $S/N$ for the pixels over the lensed arcs. The rationale for the choice of weighting terms in equation (\ref{eq:lensing_info_weight}) is as follows. The constraint on $\gamma$ comes from the radial stretching of the source, which is the next leading-order lensing constraint on the radial mass profile after the Einstein radius. This radial stretching is most effectively constrained by the differential radial thickness of the arcs and then by the tangential stretch for a symmetrical (e.g., elliptical) mass profile \citep{Birrer21}. Therefore, the radial and angular terms in the equation (\ref{eq:lensing_info_weight}) provide more weights to pixels away from the Einstein radius (corresponding to more radial stretch) and away from the reference angle (corresponding to more tangential stretch), respectively. This definition for the lensing information $\mathcal{I}$ will be higher for a system with lensed arcs than a system with only point images with the same total $S/N$. The angular weight term is taken as a multiplicative factor on the fractional radial distance so that the total information is simply the total $S/N$ for the case of a circular lens that creates a perfect Einstein ring from a point source. We obtain the values $a \approx 16$ and $b \approx 0$ by maximizing the anti-correlation between $\log \sigma_{\gamma}$ and $\log \mathcal{I}$ for all the lensing systems in our sample. We maximized the anti-correlation since we expect lower $\gamma$ uncertainty for systems with higher $\mathcal{I}$. We optimized $a$ and $b$ as real numbers but approximated the optimized values to the nearest integers. For the optimized values of $a \approx 16$ and $b \approx 0$, $\log \mathcal{I}$ and $\log \sigma_{\gamma}$ have a Pearson correlation coefficient of $r=-0.71$, whereas the `unweighted' total $S/N$ of the arcs (i.e., corresponding to $a=0$) leads to a lower anti-correlation with $r=-0.65$.}

\begin{figure*}
    \centering
    \includegraphics[width=0.9\textwidth]{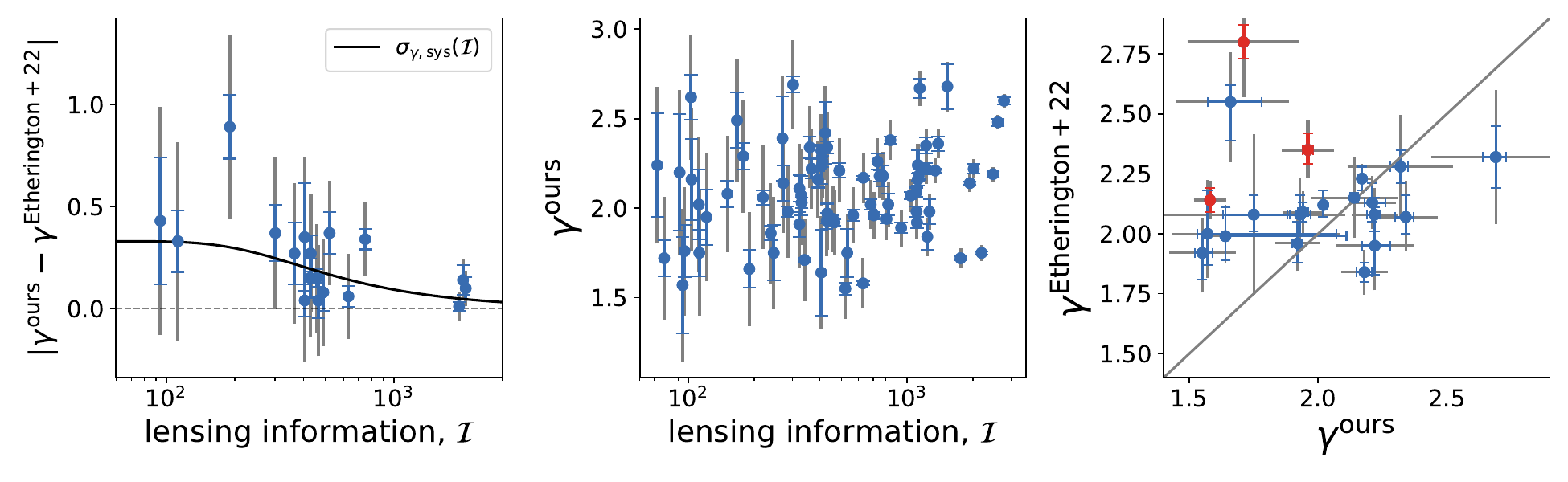}
    \caption{\ajstwo{Estimated modelling systematic uncertainty on the logarithmic slope $\gamma$}. \cytwo{The left-hand \ajstwo{panel} shows the} estimated systematic uncertainty (black line) as a function of the lensing information $\mathcal{I}$ defined in equation (\ref{eq:lensing_info}). The modelling systematic is estimated using the difference in the $\gamma$ values (blue points with blue error bars) of individual systems between the models from this analysis and those from \citet{Etherington22}, excluding the three outliers marked in Figures \ref{fig:Comparison_Rein} and \ref{fig:Comparison_gamma}. The grey error bars show the total uncertainty by adding the estimated systematic uncertainty with the statistical one in quadrature. The \ajstwo{middle panel} shows the total uncertainties (grey error bars) for \ajstwo{the measured $\gamma$ values in our full sample} (blue points, with the blue error bars showing statistical uncertainties). The lens systems with higher lensing information $\mathcal{I}$ have less systematic uncertainty, which was estimated using the model shown by the black line in the left-hand panel. \cytwo{The right-hand \ajstwo{panel} shows a comparison of the measured logarithmic slopes between our analysis \citet{Etherington22} \ajstwo{with the systematic uncertainty of the lenses added to the grey error bars}.} }
    \label{fig:modeling_systematic_gamma}
\end{figure*}

\ajs{We adopted the form of modelling systematics dependent on the lensing information $\mathcal{I}$ to be}
\begin{equation} \label{eq:gamma_sys_uncertainty}
    \sigma_{\gamma, \rm sys} = \sigma_{\gamma, \rm sys}^{\rm max} \tanh \left( {{\mathcal{I}_{\rm scale}} / {\mathcal{I}}} \right),
\end{equation}
\ajs{where we set $\sigma_{\gamma, \rm sys}^{\rm max} = 1/3$ and $\mathcal{I}_{\rm scale}$ is the scaling parameter that determines how fast $\sigma_{\gamma, \rm sys}$ changes with $\mathcal{I}$ in our sample. The total uncertainty on $\gamma$ for an individual lens system is then given by}
\begin{equation} \label{eq:gamma_total_uncertainty}
    \sigma_{\gamma} = \sqrt{\sigma_{\gamma, \rm stat}^2 + \sigma_{\gamma, \rm sys}^2},    
\end{equation}
\ajs{for both our values and those from \citet{Etherington22}.  We obtain \cyref{$\mathcal{I}_{\rm scale} \simeq 290$} by minimizing the following penalty function}
\begin{equation}
    \mathcal{P}(\mathcal{I}_{\rm scale}) \equiv  \left| 1 - \frac{1}{N_{\rm lens}} \sum_{i}^{N_{\rm lens}} \frac{\left(\gamma^{\rm ours}_i - \gamma^{\rm Etherington+23a}_i \right)^2}{(\sigma^{\rm ours}_{\gamma,i})^2 + (\sigma^{\rm Etherington+23a}_{\gamma,i})^2} \right|,
\end{equation}
\ajs{which computes the absolute difference of a reduced $\chi^2$ quantity from 1 for $N_{\rm lens} = 18$ systems that are common between our sample and that from \citet{Etherington22}, excluding the three outliers marked in Figure \ref{fig:Comparison_Rein}. We illustrate the estimated modelling systematic uncertainty $\sigma_{\gamma, \rm sys}$ as a function of $\mathcal{I}$ in Figure \ref{fig:modeling_systematic_gamma}. In the hierarchical analysis done in Section \ref{Section4:Model_paramters}, we considered the total uncertainty for all the $\gamma$ measurements after adding in quadrature the statistical uncertainties with the systematic ones, that is, $\sigma_{\gamma, \rm sys}$ estimated using equation (\ref{eq:gamma_sys_uncertainty}).}

Despite the differences observed between the studies for the individual lens model posteriors, the sample means from these studies agree well. Our estimated sample median of the logarithmic slope $\gamma = 2.04\pm 0.04$ for the SLACS lenses is in agreement within $1\sigma$ with those from \citet[][$\gamma = 2.08\pm 0.03 $]{Shajib21} and \citet[][$\gamma = 2.075\pm 0.024 $]{Etherington22}. \ajsref{Here, 1$\sigma$ difference equals the two uncertainties of the compared values added in quadrature.}  


\subsection{Comparison with previous lensing--dynamics analyses}
\label{Section_5.1}

\begin{figure*}
    \centering
    \includegraphics[width=0.9\textwidth]{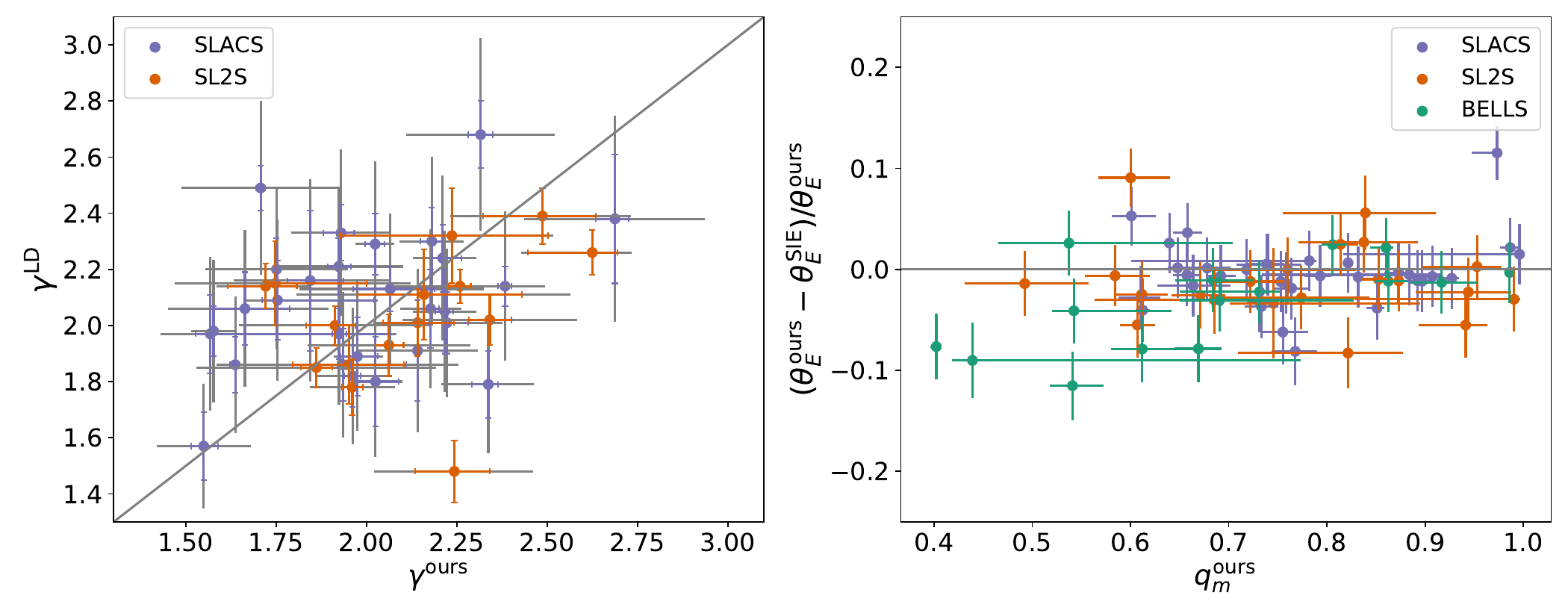}
    \caption{\cytwo{Comparison of the measured logarithmic slope $\gamma$ (left-hand panel) and the Einstein radius $\Rein$ (right-hand panel) between our analysis, which used the PEMD, and previous studies that used the SIE model to obtain $\Rein$ and joint lensing--dynamics (LD) analysis to obtain $\gamma$ \citep{Auger09, Brownstein11, Sonnenfeld13}. The SLACS, SL2S, and BELLS systems are represented with purple, orange, and green colours, respectively. As with Figure \ref{fig:modeling_systematic_gamma}, the grey error bars show the total uncertainty by adding the statistical and systematic uncertainties in quadrature. The 12 per cent systematic uncertainty added to the $\gamma^{\rm LD}$ measurements from the SLACS is based on the 6 per cent systematic uncertainty on SDSS velocity dispersion measurements reported by \citet{Birrer20}. The rms difference between our $\gamma$ and $\Rein$ measurements with those from the literature is 15.9 per cent and 4.5 per cent, respectively. Using the $p$-values of the corresponding $\chi^2$ statistics, we cannot rule out the consistency between our $\gamma$ measurements and those from the previous lensing--dynamics analysis, given the uncertainties. We also do not detect any significant dependence of the $\Rein$ difference on the mass axis ratio $q_{\rm m}$.
    }}
\label{fig:REin_Comparison}
\end{figure*}

The SLACS and SL2S samples were previously modelled with the SIE model \citep{ Bolton08, Auger09, Brownstein11, Sonnenfeld13}. The total density slope of the profile was then obtained using the stellar velocity dispersion. Comparing 
\cyone{the Einstein radius measurements} constrained by our lensing-only analysis to those from these previous joint lensing--dynamics studies, we find a rms difference of 3.3 per cent for the SLACS lenses, 5.6 per cent for the SL2S lenses, \cyone{and  5.5 per cent for BELLS lenses} (Figure \ref{fig:REin_Comparison}).

The \cyone{rms difference} between our measured Einstein radii and those in the literature is higher than the expected uncertainty of 2 per cent that is empirically estimated by \citet{Bolton08} and \citet{Sonnenfeld13}. We note that the Einstein radii obtained by the previous studies are from lens modelling only and do not depend on the stellar kinematics. 
This discrepancy thus can arise from the difference between the SIE model adopted by \citet{Bolton08} and \citet{Auger09} and the PEMD adopted by us. Indeed, by comparing our Einstein radii values to those obtained by \citet{Shajib21} using the same PEMD  for a SLACS subsample, we find a rms difference of 1.5 per cent, consistent within the measurement uncertainty. Other modelling differences could also contribute to the discrepancy mentioned above. \cytwo{For example,} previous lensing--dynamics analyses modelled and subtracted the deflector galaxy's light from the imaging data before lens modelling, where \citet{Bolton08} and \citet{Sonnenfeld13} used the de Vaucouleurs' profile, and \citet{Brownstein11} used B-spline fits. In our analysis, however, we simultaneously fit the deflector light profile (using two S\'ersic profiles) during the lens model optimization.
However, the absolute differences of the sample mean between our Einstein radii and those from the literature \citep{Bolton08, Auger09, Sonnenfeld13, Brownstein11} are 0.01 arcsec, 0.01 arcsec, and 0.02 arcsec for the SLACS, SL2S, and BELLS samples, respectively. Therefore, the systematic bias in the Einstein radius at the sample level is less than 2 per cent between the SIE model and the PEMD.

\cytwo{When comparing our lensing-only analysis to previous joint lensing--dynamics analyses from \citet[][SLACS]{Auger10b}  and \cyref{\citet[][SL2S]{Sonnenfeld13b}}, we find the median \ajstwo{absolute} difference in the \ajstwo{logarithmic} slope $\gamma$ to be 11.3  per cent for the SLACS systems and 6.1 per cent for the SL2S systems (Figure \ref{fig:REin_Comparison}). Alternatively, the rms difference in $\gamma$ is 16.8 per cent for the SLACS systems and 14 per cent for the SL2S systems. We \ajstwo{find only} a weak correlation between the $\gamma$ measurements, where the Pearson correlation coefficient is \cytwo{$r=0.28\pm0.10$} for the SLACS lenses and \cytwo{$r=0.26\pm0.16$} for the SL2S lenses.}
\ajsref{However, our sample mean of $\gamma$ agrees within $1\sigma$ with those from previous lensing--dynamics analyses. For the SLACS lenses, the sample median for $\gamma$ from our analysis is $2.04\pm0.04$ and that from the previous lensing--dynamics analysis is $2.08 \pm 0.03$ \citep{Auger10}. Similarly for the SL2S lenses, the sample median for $\gamma$ from our analysis is $2.08 \pm 0.06$ and that from the previous lensing--dynamics analysis is $2.01\pm 0.04 $ \citep{Sonnenfeld13b}.}

As discussed in Section \ref{Sec2:Spectra}, \citet{Birrer20} find a previously unaccounted systematic uncertainty of 6 per cent in the SDSS velocity dispersion measurements for the SLACS systems, which corresponds to an additional 12 per cent uncertainty on the measured $\gamma$ from their lensing--dynamics analysis. After accounting for this additional systematic uncertainty, we find for the SLACS systems $\Delta_{\gamma, \rm SLACS}^2 \equiv \sum_{i,\rm SLACS}(\gamma^{\rm ours}_i - \gamma^{\rm LD}_i)/({\sigma_{i, \rm ours}}^2 + {\sigma_{i, \rm LD}}^2) = 24.2$. Assuming a $\chi^2$ distribution for $\Delta_{\gamma, \rm SLACS}^2$ with 28 degrees of freedom, we obtain a $p$-value of 0.67. For the SL2S systems -- where we did not include any additional systematic uncertainty to the measured $\gamma$ from the lensing--dynamics analysis -- we find $\Delta_{\gamma, \rm SL2S}^2=27.8$. For a $\chi^2$ distribution with 15 degrees of freedom, this   $\Delta_{\gamma, \rm SL2S}^2$ provides a $p$-value of 0.02. However, the  $\Delta_{\gamma, \rm SL2S}^2$ quantity is dominated by SL2SJ0904$-$0059, which has an atypically low $\gamma^{\rm LD} = 1.48 \pm 0.11$.
When we exclude this system as an outlier, then $\Delta_{\gamma, \rm SL2S}^2 = 18.2$ provides a $p$-value of 0.2 for 14 degrees of freedom. 
Therefore, we cannot rule out the hypothesis that our lensing-only $\gamma$ measurements are consistent with those from the previous lensing--dynamics analyses for both the SLACS and SL2S systems, given the uncertainties. \ajsref{Thus, if the power-law model assumption is correct, then the expected strong correlation in the individual $\gamma$ measurements between our lensing-only and previous lensing--dynamics analyses could have been smeared by the large uncertainties. We can not, therefore, conclude on a deviation from the power law based on our finding of the weak correlation.}



\section{Deviation of the mass profile from a power law}
\label{Section4:Model_paramters}
In this section, we combine the lens model posteriors with the velocity dispersions measured from single-aperture spectra to constrain any deviation from the power law in the total density profile. We parametrize the deviation from the power-law form using the mass-sheet transformation (MST). 
We briefly introduce the MST in Section \ref{Section4:MST}, and describe the formalism and assumptions used to predict the velocity dispersion from our lens models in Section \ref{Section4:Stellar_Kinematics}. Then, in Section \ref{Section4:Measure_lambda}, we describe the hierarchical Bayesian analysis that we used to constrain the deviation from the power law at the population level.

\subsection{Mass-sheet transformation} \label{Section4:MST}
The MST \citep{Falco85} is a multiplicative transformation of the lens equation (i.e., equation \ref{Eqn:Lens_Equation}). This transformation modifies the lens equation as 
\begin{equation}
\boldsymbol{\beta} = \boldsymbol{\theta} - \boldsymbol{\alpha}(\boldsymbol{\theta})  \rightarrow \lambda\boldsymbol{\beta}=\boldsymbol{\theta}-\lambda\boldsymbol{\alpha}(\boldsymbol{\theta}) - (1-\lambda)\boldsymbol{\theta} .
\end{equation}
The convergence field is then transformed as 
\begin{equation}
\kappa(\boldsymbol{\theta}) \rightarrow \kappa'(\boldsymbol{\theta}) =   \lambda\kappa(\boldsymbol{\theta}) + (1-\lambda).
\label{Eqn:MST_convergence}
\end{equation}
 The MST has the property where  image positions remain invariant while the source position scales with $\lambda$ such that
\begin{equation}
\boldsymbol{\theta} \rightarrow  \boldsymbol{\theta} , \boldsymbol{\beta} \rightarrow  \lambda\boldsymbol{\beta}.
\end{equation}
As the MST changes the convergence field and thus the mass distribution of the lensing galaxy, the model-predicted stellar velocity dispersion changes with the transformation as well. The line-of-sight stellar velocity dispersion of the lensing galaxy \ajstwo{integrated} \cyone{within an aperture} scales approximately as
\begin{equation}
\label{Eqn:MST_vel_disp}
\sigma_{\rm  \cyone{ap}} \rightarrow  \sqrt{\lambda} \sigma_\cyone{{\rm ap}}
\end{equation}
\citep{Birrer20}. \cyone{We can therefore estimate the value of the MST parameter \ajstwo{$\lambda$ that transforms the adopted power-law model into the `true' mass distribution} by comparing the measured velocity dispersion $\sigma_{\rm ap, meas }$ and the predicted velocity dispersion $\sigma_{\rm ap, PL}$ by the power-law model as
\begin{equation}
\label{Eqn:MST_lambda}
\lambda \approx \frac{\sigma_{\rm ap, meas }^2}{\sigma_{\rm ap, PL} ^ 2}.
\end{equation}}

In addition to lensing caused by the convergence from the lensing galaxy (denoted with $\kappa_{\rm gal}$), the overdensity or underdensity of the structures along the line of sight with respect to the average matter density of the Universe can also produce additional magnification or demagnification. We denote the `external' convergence from these line-of-sight structures with $\kappa_{\rm ext}$. Therefore, the effective total lens mass distribution of the system, $\kappa$, is then given by 
\begin{equation}
    \kappa = \kappa_{\rm gal} + \kappa_{\rm ext} . 
\end{equation}

The MST can be either associated with transforming the mass profile of the main lensing galaxy with the `internal MST' parameter $\lambda_{\rm int}$, ignoring the line-of-sight structures in the lens modelling, or a combination of both. The total transformation affecting the stellar kinematics is thus given by
\begin{equation}
    \lambda_{\rm total} = (1-\kappa_{\rm ext}) \lambda_{\rm int}
\end{equation}
\citep{Shajib23}. To break this degeneracy, we adopted the estimated $\kappa_{\rm ext}$ from \citet{Birrer20} for the SLACS systems and from Wells et al. (in preparation) for the SL2S systems. For systems without individual $\kappa_{\rm ext}$ measurements in these two samples, we use the overall sample distribution of $\kappa_{\rm ext}$. \ajstwo{The $\kappa_{\rm ext}$ estimates for the BELLS systems are not available from the literature. We leave their $\kappa_{\rm ext}$ estimation for a future study and exclude this sample for the following analysis in this paper.}

\subsection{Modelling of the stellar dynamics} \label{Section4:Stellar_Kinematics}

The \ajstwo{aperture-integrated,} line-of-sight velocity dispersion $\sigma_{\rm \cyone{ap}}$ provides a lensing-independent method to estimate the mass profile of the lensing galaxy. The stellar velocity dispersion can be obtained by solving the Jeans equation, which describes the phase space distribution of a galaxy with a 3D stellar density profile $\rho_*(r)$ located within gravitational potential $\Phi(r)$. In the case of spherical symmetry, the velocity dispersion can be decomposed into its radial component $\sigma_{\rm r}(r)$ and its tangential component $\sigma_{\rm t}(r)$. The stellar anisotropy of the system is parametrized as 
\begin{equation}
    \beta_\text{ani} (r) = 1 - \frac{\sigma^2_{\rm t}(r)}{\sigma^2_{\rm r}(r)} .
\end{equation}
For a relaxed system, the spherical Jeans equation can  be expressed as
\begin{equation}
    \frac{\partial\rho_*(r)\sigma_{\rm r}^2(r)}{\partial r } + \frac{2\beta_\text{ani}(r)\rho_*(r)\sigma_{\rm r}^2(r)}{r} = -\rho_*(r)\frac{\partial\Phi(r)}{\partial r }
\end{equation}
\citep{Binney08, Birrer20}.

We can solve this spherical Jeans equation to obtain
\begin{equation}
    \sigma_{\rm r}^2(r) = \frac{G}{\rho_*(r)}\int^\infty_r\frac{M(s)\rho_*(s)}{s^2}\exp\left[\int^s_r\frac{2\beta_{\rm ani}(r^\prime)}{r^\prime}\,dr^\prime\right] \,ds,
\end{equation}
where $M(s)$ is the 3D total enclosed  mass profile. The line-of-sight velocity dispersion is then given by 
\begin{equation} 
    \sigma_{\rm los}^2(R) = \frac{2}{\Sigma_*(R)} \int ^ \infty _R \left(1-\beta_{\rm ani}(r)\frac{R^2}{r^2}\right)\frac{\rho_*\sigma_{\rm r}^2r \,dr}{\sqrt{r^2-R^2}},
\end{equation}
where $ \Sigma_*(R)$ \cyone{is} the 2D projected stellar density profile,  $R$ is the 2D projected radius \citep{Binney82}. The luminosity-weighted line-of-sight velocity dispersion $\sigma_{\rm ap}$ within an aperture is given by
\begin{equation} \label{eq:los_dispersion}
    \sigma_{\rm ap}^2 = \frac{\int _{\rm ap}[\Sigma_*(R)  \sigma_{\rm los}^2(R) * \mathcal{S} ] \,dA}{\int _{\rm ap}[\Sigma_*(R)  * \mathcal{S}] \,dA} ,
\end{equation}
where  $* \mathcal{S}$ denotes PSF convolution with the seeing \citep{Treu04,Suyu10}.

\ajstwo{We chose a spatially constant anisotropy profile as our baseline model, which is consistent with dynamical observables obtained from both long-slit and integral field spectra of local elliptical galaxies \citep{Gerhard01, Cappellari07, Cappellari08b}. In this model, we treat $\beta_{\rm ani}$ as a free parameter. We also test with an alternative choice of the anisotropy profile given by the Osipkov--Merritt parameterization}
\begin{equation}
\label{eqn:Anisotropy}
    \beta_\text{ani} (r) = \frac{r^2}{(\Reff a_\text{ani})^2+r^2},
\end{equation}
where $a_\text{ani}$ is a unitless scale factor \citep{Osipkov79,Merritt85}.

To compute the model-predicted stellar velocity dispersion, \ajstwo{the 3D mass profile $\rho(r)$ was obtained corresponding to the 2D convergence profile in equation (\ref{Eqn:PEMD}).}
Assuming a constant mass-to-light ratio, we obtained the 2D projected stellar density profile $\Sigma_*(R)$ from the surface brightness profile $I(R)$ constrained from \cytwo{large ($\sim$16$\times$16 \ajstwo{square-}arcsec) cutouts of the \textit{HST} images. We fit the surface brightness profile with either a single or double S\'ersic profile on the \textit{HST} images. These cutouts are larger than those used in lens modelling to \ajstwo{more robustly} estimate the surface brightness profile of the lensing galaxy \ajstwo{up to its full extent}.} The assumed value of the mass-to-light ratio does not matter since it cancels out in equation (\ref{eq:los_dispersion}). To obtain the 3D stellar mass density $\rho_*$, we decomposed the projected stellar density profile $ \Sigma_*(R)$ into multi-Gaussian expansion \citep[MGE;][]{Emsellem94, Cappellari02} and then deprojected the Gaussian profiles into 3D. 

\subsection{Bayesian hierarchical framework}
\label{Section4:Measure_lambda}



\ajstwo{We performed a Bayesian hierarchical analysis to constrain the mean deviation from the power law at the population level while also adopting a prior on the anisotropy profile parameter at the same population level.}
\ajstwo{We performed this using} \textsc{hierArc}\footnote{\url{https://github.com/sibirrer/hierArc}}, a software program for Bayesian hierarchical analysis of strong lensing systems \citep{Birrer20}. 

Here, we provide a brief description of the hierarchical framework. Let $D_i$ be the data for the \cyone{$i^{\rm th}$} individual lens and $D \equiv \{D_i\}_N$ be the dataset \ajstwo{containing} all the data from $N$ number of lenses. \cyone{For this analysis, \ajstwo{the lensing data for each dataset $D_i$ is summarized with the lens model posteriors for the logarithmic slope $\gamma_i$ and the Einstein radius $\theta_{{\rm E}, i}$. The dataset $D_i$ additionally includes the measured velocity dispersion $\sigma_{{\rm ap, meas}, i }$, the estimated external convergence $\kappa_{{\rm ext}, i}$, and the single or double S\'ersic profile parameters fitted to the lens galaxy's surface brightness from Section \ref{Section4:Stellar_Kinematics}.} \ajstwo{We add a 2 per cent systematic uncertainty \citep{Shajib21} in quadrature to the measured $\Reff$ uncertainty provided in Tables \ref{table:SLACS_params} and \ref{table:SL2S_params}. We find from a test that our final results are robust against a more conservative choice of a uniform uncertainty level of 5 per cent in $\Reff$.} For the $i^{\rm th}$ lens system, the individual-level parameter set $\xi_i$ includes the internal MST parameter $\lambda_{{\rm int}, i}$ and the stellar anisotropy parameter -- $\beta_{{\rm ani}, i}$ for the constant anisotropy model or $a_{{\rm ani}, i}$ \cytwo{for the Osipkov--Merritt model}. 
\ajstwo{In the hierarchical framework, the population distribution of individual-level parameters $\xi_i$ are described with population-level parameters $\Xi$.} \ajstwo{Bayes' theorem gives} the posterior $p(\Xi \mid D)$ of the population-level parameters} as 
\begin{equation}
\begin{aligned}
    p(\Xi\mid D)  &\propto p(D\mid \Xi)\, p(\Xi) \\
        & \propto p(\Xi) \prod_{i}^N \int d \xi_i\, p(D_i \mid \xi_i) \, p(\xi_i \mid \Xi),
\label{Eqn:Bayensian}
\end{aligned}
\end{equation}
where $p(D \mid \Xi)$ is the likelihood function and $p(\Xi)$ is the prior of the population-level parameters. While $p(D_i\mid \xi_i)$ is the likelihood \cyone{function} for the \cyone{$i^{\rm th}$} lens system and $ p(\xi_i\mid \Xi)$ describes the distribution of individual lens parameters given the population-level parameters. 

Now, we explain all the population-level model parameters \cyone{$\Xi$} used in this analysis. Following \citet{Birrer20}, we described the population distribution of the \cyone{internal MST parameter} $\lambda_{\rm int}$ using a Gaussian distribution with mean $\mu_{\lambda_{\rm int}}$ and scatter $\sigma_{\lambda_{\rm int}}$. We incorporated linear dependencies of  $\mu_{\lambda_{\rm int}}$ on the redshift $z$ and the scale ratio $\Reff/\Rein$ as
\begin{equation}
    \mu_{\lambda_{\rm int}} (z, \Reff/\Rein) = \mu_{\lambda_{\rm int}}^{\rm ref} + \alpha_{\lambda_{\rm int}}(z-0.3) + \beta_{\lambda_{\rm int}}\left(\frac{\Reff}{\Rein}-1\right).
\label{Eqn:linear_hierarc}
\end{equation}
Here, $\mu_{\lambda_{\rm int}}^{\rm ref}$ is the reference value of $\mu_{\lambda_{\rm int}}$ for $z = 0.3$ and $\Reff/\Rein = 1$, and $\alpha_{\lambda_{\rm int}}$ and $\beta_{\lambda_{\rm int}}$ are linear slope parameters.


\cytwo{
\ajstwo{To describe the population distribution of $\beta_\text{ani}$ for the constant anisotropy model, we used a Gaussian distribution with mean $\mu_{\beta_\text{ani}}$ and scatter  $\sigma_{\beta_{\rm ani}}$.}
For the Osipkov--Merritt model, we described the population distribution of the anisotropy scale factor $a_\text{ani}$ as a Gaussian distribution with mean $\mu_{a_\text{ani}}$ and scatter  $\mu_{a_\text{ani}}\sigma_{a_{\rm ani}}$.}
 

As mentioned in Section \ref{Sec2:Spectra}, \citet{Birrer20} find that the uncertainty in the velocity dispersion measurements for the SLACS lenses from the SDSS spectra was underestimated. Therefore, we included a systematic uncertainty term set by the fractional uncertainty \cyone{$f^{^{\rm SDSS}}_{\cyone{\sigma, {\rm{sys}}}}$}. The total uncertainty $\delta_{\sigma}$ for the measured velocity dispersion $\sigma_{\rm ap}$ of an individual SLACS lens is therefore given by 
\begin{equation}
\label{eqn:SDSS_error}
    \delta_\sigma^2 =  \delta_{\sigma, \rm stat}^2 + (f^{^{\rm SDSS}}_{\cyone{\sigma, {\rm{sys}}}} \sigma_{\rm ap})^2 ,
\end{equation}
where $\delta_{\sigma, \rm stat}$ is the estimated statistical uncertainty for the velocity dispersion measurement. \ajstwo{Thus, for the constant anisotropy model, the population-level parameters sampled by \textsc{hierArc} are $\Xi \equiv \{\mu_{\lambda_{\rm int}}^{\rm ref},\  \alpha_{\lambda_{\rm int}},\ \beta_{\lambda_{\rm int}},\ \sigma_{\lambda_{\rm int}},\  \mu_{\beta_{\rm ani}},\ \sigma_{\beta_\text{ani}},\ f^{^{\rm SDSS}}_{\cyone{\sigma, {\rm{sys}}}}\}$.}
\cytwo{For the Osipkov--Merritt model, the population-level anisotropy parameters are replaced with $\mu_{a_{\rm ani}}$ and $\sigma_{a_\text{ani}}$.} We tabulate the adopted priors on these population-level parameters in Table \ref{table:hiearc_priors}. 

\begingroup
\renewcommand{\arraystretch}{1.13} 
\begin{table*}
\centering
\caption{Description and adopted priors of the population-level parameters in our hierarchical Bayesian analysis.}
\cytwo{
\begin{tabular}{lll}
\hline
    Parameter &{ Description} & {Prior} \\
\hline
    $\mu_{\lambda_{\rm int}}^{\rm ref}$   
    & Population mean of internal MST parameter at $z= 0.3$ and $\Reff/\Rein=1$ (see equation \ref{Eqn:linear_hierarc})
    & \cyone{$ \mu_{\lambda_{\rm int}}^{\rm ref} \sim  \mathcal{U}(0.5, 1.5)$}\\ 
    $\alpha_{\lambda_{\rm int}}$
    & Linear dependency of $\lambda_{\rm int}$ on redshift $z$ (see equation \ref{Eqn:linear_hierarc})
    &$\alpha_{\lambda_{\rm int}} \sim  \mathcal{U}(-1, 1)$\\
    $\beta_{\lambda_{\rm int}}$
    & Linear dependency of $\lambda_{\rm int}$ on  $\Reff/\Rein$ (see equation \ref{Eqn:linear_hierarc})
    &$\beta_{\lambda_{\rm int}} \sim  \mathcal{U}(-1, 1)$\\
    $\sigma_{\lambda_{\rm int}}$
    & 1$\sigma$ Gaussian scatter in $\lambda_{\rm int}$
    &\cyone{$\sigma_{\lambda_{\rm int}} \sim \mathrm{Loguniform}(0, 0.5)$} \\
    $\mu_{\beta_{\rm ani}}$
    & Population mean of the anisotropy parameter $\beta_{\rm ani}$ in the constant anisotropy model
    & $\mu_{\beta_{\rm ani}} \sim \mathcal{U}(-0.49, 1)$ \\
    $\sigma_{\beta_\text{ani}}$
    & 1$\sigma$ Gaussian scatter of $\beta_\text{ani}$ in the constant anisotropy model
    &$ \sigma_{\beta_\text{ani}} \sim \mathrm{Loguniform}(0.01, 0.50)  $   \\
    $\mu_{a_{\rm ani}}$
    & Population mean of anisotropy scale factor in the Osipkov--Merritt model (equation \ref{eqn:Anisotropy})
    & $\mu_{a_{\rm ani}} \sim \mathrm{Loguniform}(0.1, 5)$ \\
    $\sigma_{a_\text{ani}}$
    & 1$\sigma$ Gaussian scatter of $a_\text{ani}$ in the Osipkov--Merritt model is given by $\mu_{a_{\rm ani}}\sigma_{a_\text{ani}}$ 
    &$ \sigma_{a_\text{ani}} \sim \mathrm{Loguniform}(0, 1)  $ \\
    $ f^{^{\rm SDSS}}_{\sigma, {\rm{sys}}}$
    & Systematic uncertainty factor on SDSS velocity dispersion measurements (see equation \ref{eqn:SDSS_error})
    &$f^{^{\rm SDSS}}_{\sigma, {\rm{sys}}} \sim \mathrm{Loguniform}(0.01, 0.5)$  \\
\hline
\end{tabular}}
\label{table:hiearc_priors}
\end{table*}
\endgroup

\begingroup
\renewcommand{\arraystretch}{1.2}
\begin{table*}
    \caption{ \label{tab:hierarichical_posterior}
    Point estimates of the population-level parameters from our hierarchical Bayesian analysis. The parameters are described in Table \ref{table:hiearc_priors}.
    }
    \centering
    \begin{tabular}{lccccccccc}
    \hline
         Anisotropy model
         & $\mu_{\lambda_{\rm int}}^{\rm ref}$ 
         & $\alpha_{\lambda_{\rm int}}$ 
         & $\beta_{\lambda_{\rm int}}$
         & $\sigma_{\lambda_{\rm int}}$
         & $\mu_{\beta_{\rm ani}}$
         & $\sigma_{\beta_{\rm ani}}$
         & $\mu_{a_{\rm ani}}$
         & $\sigma_{a_{\rm ani}}$
         & $f^{^{\rm SDSS}}_{\sigma, {\rm{sys}}}$ \\
    \hline
    Constant & $0.91^{+0.10}_{-0.09}$ &
    $-0.12^{+0.31}_{-0.26}$&
    $-0.04^{+0.04}_{-0.04}$&
    $0.007^{+0.120}_{-0.007}$&
    $0.53^{+0.27}_{-0.45}$&
    $0.021^{+0.350}_{-0.021}$&
    -- & -- &
    $0.05^{+0.03}_{-0.03}$\\
    Osipkov--Merritt & $0.81^{+0.06}_{-0.05}$ &
    $-0.06^{+0.25}_{-0.23}$ &
    $-0.01^{+0.04}_{-0.04}$ &
    $0.006^{+0.090}_{-0.006}$ &
    -- & -- &
    $0.66^{+0.35}_{-0.20}$ & 
    $0.003^{+0.237}_{-0.003}$ &
    $0.05^{+0.03}_{-0.03}$ \\
    \hline 
    \end{tabular}
    \label{tab:my_label}
\end{table*}
\endgroup

\begin{figure*}
    \centering
    \includegraphics[width=0.92\textwidth]{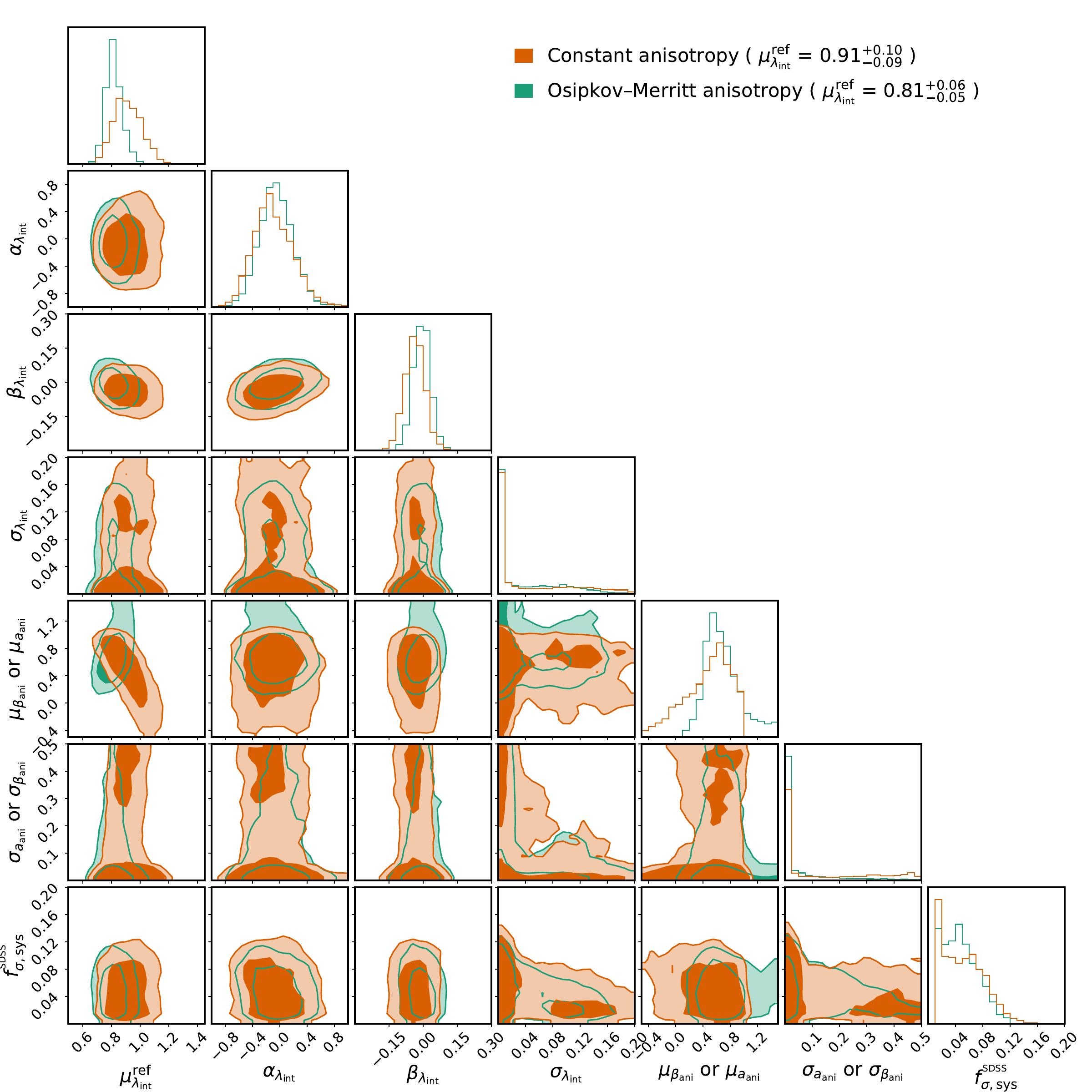}    
    \caption{ \label{fig:hierarc_corner}
    \cytwo{Posteriors of the population-level parameters from our hierarchical Bayesian analysis. The population distribution of the internal MST parameter $\lambda_{\rm int}$ was described with a Gaussian distribution using} \ajstwo{mean $\mu_{\lambda_{\rm int}}(z, \Reff/\Rein) = \mu^{\rm ref}_{{\lambda}_{\rm int}} + \alpha_{\lambda_{\rm int}} (z-0.3) + \beta_{\lambda_{\rm int}} ({\Reff}/\Rein-1)$ and scatter $\sigma_{\lambda_{\rm int}}$. Here, we compare the posterior between the spatially constant anisotropy model (orange, parametrized with $\beta_{\rm ani}$) and the Osipkov--Merritt model (green, parametrized with $a_{\rm ani}$) for the stellar anisotropy. The population distribution of the corresponding anisotropy parameter was modelled using a Gaussian distribution with mean $\mu_{a_{\rm ani}}$ or $\mu_{\beta_{\rm ani}}$ and scatter of $\mu_{ a_\text{ani}} \sigma_{a_\text{ani} }$ or $ \sigma_{\beta_\text{ani} }$ for the constant or Osipkov--Merritt model, respectively. We accounted for a systematic uncertainty in the SDSS velocity dispersion measurements for the SLACS lenses using the factor $f^{^{\rm SDSS}}_{\sigma, {\rm{sys}}}$. The point estimates of these parameters are provided in Table \ref{tab:hierarichical_posterior}. The mean reference parameter $\mu^{\rm ref}_{{\lambda}_{\rm int}}$ is consistent between the two anisotropy models within 0.9$\sigma$. The constant anisotropy model is consistent with the power-law profile (i.e., $\mu^{\rm ref}_{{\lambda}_{\rm int}} = 1$) within 0.9$\sigma$. However, the Osipkov--Merritt model deviates from the power-law model with a larger confidence level of 3.2$\sigma$. We adopted the constant anisotropy model as our baseline choice, as this model is consistent with the dynamical observables in local elliptical galaxies \citep[e.g.,][]{Gerhard01, Cappellari08b}. Our result highlights the impact of the stellar anisotropy model choice in constraining the deviation from the power-law profile \cyref{}{\citep[e.g., as suggested by][]{Etherington23}}, which can be informed by spatially resolved kinematics in the future.}}
\end{figure*}

\begin{figure*}

    \centering
    \includegraphics[width=0.95\textwidth]{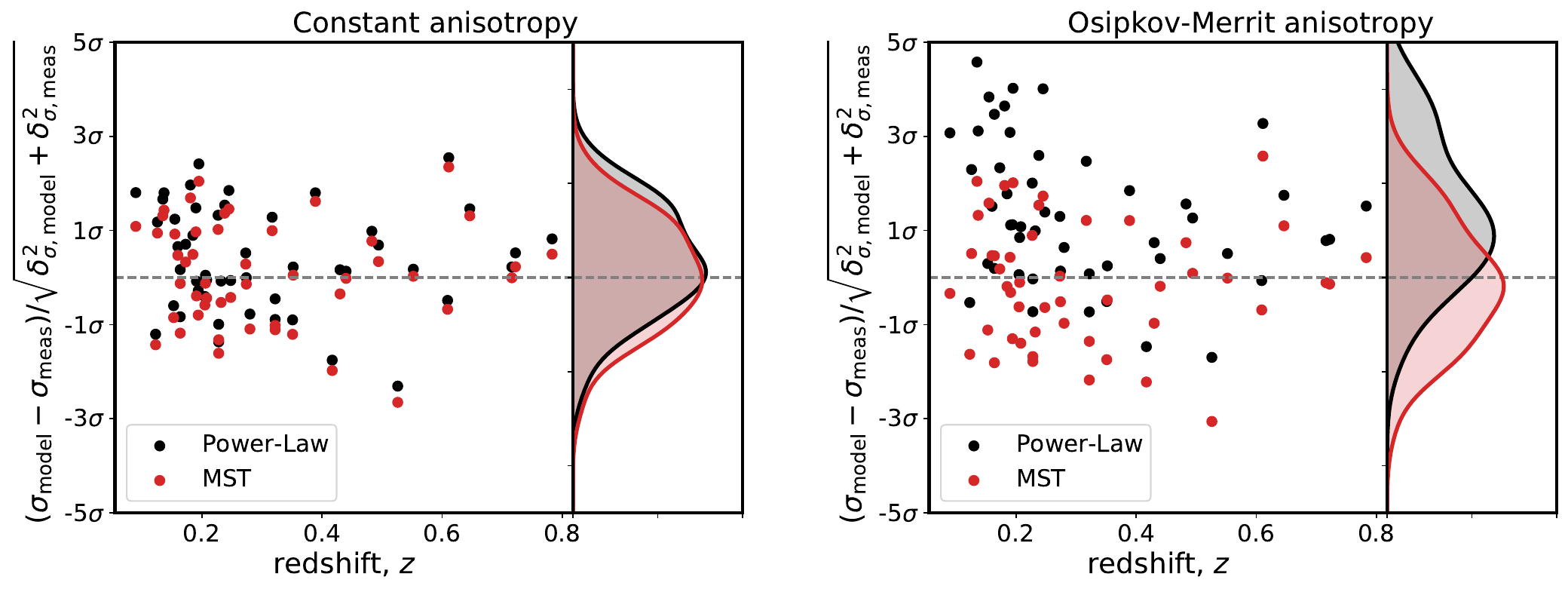}
    \caption{
    \ajsref{Goodness of fit of the dynamical observables between the power-law model and the power-law model with the MST. The goodness of fit is illustrated with the normalised difference between the model-predicted velocity dispersion $\sigma_{\rm model}$ and the measured velocity dispersion $\sigma_{\rm meas}$, which is computed as $(\sigma_{\rm model} - \sigma_{\rm meas}) /{ \sqrt{\delta^2_{\sigma, {\rm model}} + \delta^2_{\sigma, {\rm meas}}}}$. Here, $\delta_{\sigma, {\rm model}}$ and $\delta_{\sigma, {\rm meas}}$ are the uncertainties for $\sigma_{\rm model}$ and $\sigma_{\rm meas}$, respectively.
    The model-predicted velocity dispersion $\sigma_{\rm model}$ for each individual system was computed using the most likely values of the population-level parameters -- except for taking $\mu_{\lambda_{\rm int}}^{\rm ref} = 1$ for the pure power-law case -- given the individual-level parameters $\gamma$, $\Rein$, and $\Reff$ for that system. Thus, the uncertainty term $\delta_{\sigma, {\rm model}}$ incorporates the uncertainties from $\gamma$, $\Rein$, and $\Reff$. The right-hand panel shows the kernel density estimation plot of the normalised difference between $\sigma_{\rm model}$ and $\sigma_{\rm meas}$.  With the constant anisotropy profile, the best-fit mass-sheet transformed profiles predict similar velocity dispersion values as the power-law model, as expected from the consistency of $\mu_{\lambda_{\rm int}}^{\rm ref}$ with the case with no MST (i.e., $\lambda_{\rm int} = 1$) within $1\sigma$ ($\mu_{\lambda_{\rm int}}^{\rm ref} = 0.91_{-0.09}^{+0.10}$). However, using the Osipkov--Merritt anisotropy profile, we find that for a majority of the systems, the model-predicted velocity dispersions are higher than the measured values, and mass-sheet transformed profile is required to fit the measured values better ($\mu_{\lambda_{\rm int}}^{\rm ref} = 0.81_{-0.05}^{+0.06}$). }}
    
\label{fig:hierarc_slits}
\end{figure*}

The posterior distribution of the population-level parameters obtained using \textsc{hierArc} is shown in Figure \ref{fig:hierarc_corner}, and the corresponding point estimates are provided in Table \ref{tab:hierarichical_posterior}. \cyref{Figure \ref{fig:hierarc_slits} shows the goodness of fit for the SLACS and SL2S model-predicted velocity dispersions using the maximum likelihood estimators of the \cyone{population-level parameters}. The model-predicted velocity dispersions are derived for both the power-law profile and the mass-sheet transformed profile assuming the constant or the Osipkov--Merrit anisotropy profile.}

\ajstwo{For the constant anisotropy model, the mean reference MST parameter is $\mu_{\lambda_{\rm int}}^{\rm ref} = 0.91^{+0.10}_{-0.09}$, consistent with the power-law profile (i.e., $\mu_{\lambda_{\rm int}}^{\rm ref} = 1$) within 0.9$\sigma$. The point estimates of the slope parameters $\alpha_{\lambda_{\rm int}} = -0.12^{+0.31}_{-0.26}$ and  $\beta_{\lambda_{\rm int}} = -0.04^{+0.04}_{-0.04}$ are consistent with having no dependence on the redshift and $\Reff/\Rein$, respectively. For the Oskipkov--Merritt model, the mean reference MST parameter $\mu_{\lambda_{\rm int}}^{\rm ref} = 0.81^{+0.06}_{-0.05}$ deviates from the power-law profile with a higher confidence level of $3.2\sigma$. However, $\mu_{\lambda_{\rm int}}^{\rm ref}$ is consistent within 0.9$\sigma$ between the constant anisotropy and Osipkov--Merritt models. Although the likelihood ratio favours the Osipkov--Merritt model by a factor of $\sim$250 (i.e., $\Delta \log \mathcal{L} \approx 5.5$)\footnote{Here, The likelihood ratio is equivalent to the Bayesian information criterion as both models have the same number of free parameters. We extracted the maximum log-likelihood value for each model from the MCMC chain. We checked that the second-highest log-likelihood value in each chain is much closer to the maximum log-likelihood value relative to the difference in the log-likelihood between the two models. Thus, we ensured that the maxima of the likelihood functions were well sampled, giving robustness to our numerical estimation of the maximum-likelihood ratio.}, this is due to the Osipkov--Merritt model being more capable of fitting the population scatter with a Gaussian distribution. Single-aperture velocity dispersions -- as used in this analysis -- have no information to differentiate between different choices of anisotropy model. Therefore, spatially resolved kinematics are necessary to differentiate between the two anisotropy models. In this paper, we adopted the constant anisotropy model as the baseline, given its consistency with dynamical observables of local massive ellipticals, whereas the Osipkov--Merritt model does not allow tangential orbits that are often observed in the local ellipticals \citep[e.g.,][]{Gerhard01, Cappellari07, Cappellari08b}.}

\ajstwo{Accounting for the distribution of redshift and $\Reff/\Rein$ in the corresponding samples, the SLACS galaxies are consistent with the power-law profile (i.e., $\mu_{\lambda_{\rm int}}^{\rm ref} = 1$) within 1.1$\sigma$, and so are the SL2S galaxies within 0.8$\sigma$, for the constant anisotropy model. For the Osipkov--Merritt model, the SLACS and SL2S galaxies are consistent within 2.8$\sigma$ and 2.1$\sigma$, respectively, with the power-law profile.}


\section{Discussion} \label{sec:discussion}

In this section, we first discuss the impact of the anisotropy profile choice on joint lensing--dynamics analysis in Section \ref{Section 5 Anisotropy}. Then, in Section \ref{sec:impact_of_mst}, we discuss the effect of the internal MST on the local logarithmic slope of the density profile. Finally, we discuss the redshift evolution of the logarithmic slope obtained from the different modelling methods in Section \ref{Section_5.3}.

\subsection{Impact of the anisotropy profile on joint lensing--dynamics analysis}
\label{Section 5 Anisotropy}

\ajstwo{Our result highlights the importance of the stellar anisotropy profile choice in the dynamical modelling when constraining the mean deviation from the power law. By comparing our baseline choice of the constant anisotropy model with the Osipkov--Merritt model, we find that the latter provides a higher significant deviation (3.2$\sigma$) from the power-law profile than the former one (1.1$\sigma$), although both choices are consistent with each other at 0.9$\sigma$. 
Spatially resolved kinematics of lens galaxies are essential to differentiate between these anisotropy profile choices to mitigate the associated systematic uncertainty in constraining any potential deviation from the power law, and we leave this investigation for a future study.
}

\ajstwo{In this paper, we directly constrained the deviation from the power law, or the lack thereof, through a joint lensing--dynamics analysis that has several improvements in the dynamical modelling over previous such analyses \citep[e.g.,][]{Auger10b, Sonnenfeld13}. First, we used a stellar anisotropy model with one degree of freedom instead of the isotropic model that has no degree of freedom. The isotropic model is a special case of our constant anisotropy model with $\beta_{\rm ani} = 0$, which is 1.2$\sigma$ away from our population mean for $\beta_{\rm ani}$. Second, we used in the dynamical modelling a more accurate surface brightness profile obtained from the full extent of the lens galaxy instead of the Hernquist or De Vaucouleurs's profile \citep{Hernquist90, deVaucouleurs48}, which can often be inadequate to describe lens galaxy's surface brightness over its full extent and thus potentially induce systematics. 
Our state-of-the-art analysis provides a direct approach to quantify any deviation from the power law, which is an improvement over previous studies investigating such deviation based on inconsistent correlations between lensing-only and lensing--dynamics analyses within the scope of the power-law model.
We find consistency with the power-law profile at 1.1$\sigma$ confidence level for our baseline choice of the constant anisotropy model.
Therefore, the observed discrepancy by \citet{Etherington23} between the correlation of the mass surface density with lensing-only $\gamma$ measurements and that with previous lensing--dynamics $\gamma$ measurements will likely be alleviated with a more flexible and robust model for stellar dynamics, without requiring a significant departure from the power law. The discrepancy may also be alleviated to some further extent if additional modelling systematic uncertainties on $\gamma$ are accounted for, as done in this analysis (Section \ref{Section_5.2}).}

\subsection{Impact of the internal MST on the logarithmic slope} \label{sec:impact_of_mst}

The internal MST changes the shape of the physical mass distribution in the deflector galaxy. Thus, it impacts the local logarithmic slope. In this subsection, we illustrate, with an example, the change in the logarithmic slope when an internal MST is performed on the mass profile. The internal MST does not add mass at a distance far away ($r \gg \Reff$) from the deflector galaxy. Therefore, to ensure that the transformed mass profile goes to zero at a large radius, that is, to have $\lim_{\theta\to\infty}\kappa^\prime (\theta) = 0$, 
we use an `approximate' MST 
\begin{equation}
\label{eqn:approx_mst}
    \kappa(\theta) \rightarrow  \kappa'(\theta) =   \lambda_{\rm int}\kappa(\theta) + (1-\lambda_{\rm int})\frac{\theta_{\rm s}^2}{\theta^2+\theta_{\rm s}^2}
\end{equation}
given by \citet{Blum20}. Here, $\theta_{\rm s}$ is the scale radius where the mass sheet truncates. 
We can deproject this transformed convergence profile from 2D into the 3D density profile
\begin{equation}
    \rho(r) = \Sigma_\text{crit}\left( \lambda_{\rm int}\frac{\rho_0}{r^{\gamma} }  + \frac{(1-\lambda_{\rm int})}{2} \frac{\theta_{\rm s}^2}{(\theta_{\rm s}^2+r^2)^{3/2}}\right) .
\label{eqn:deprojection}
\end{equation}

\begin{figure*}
    \centering
    \includegraphics[width=0.95\textwidth]{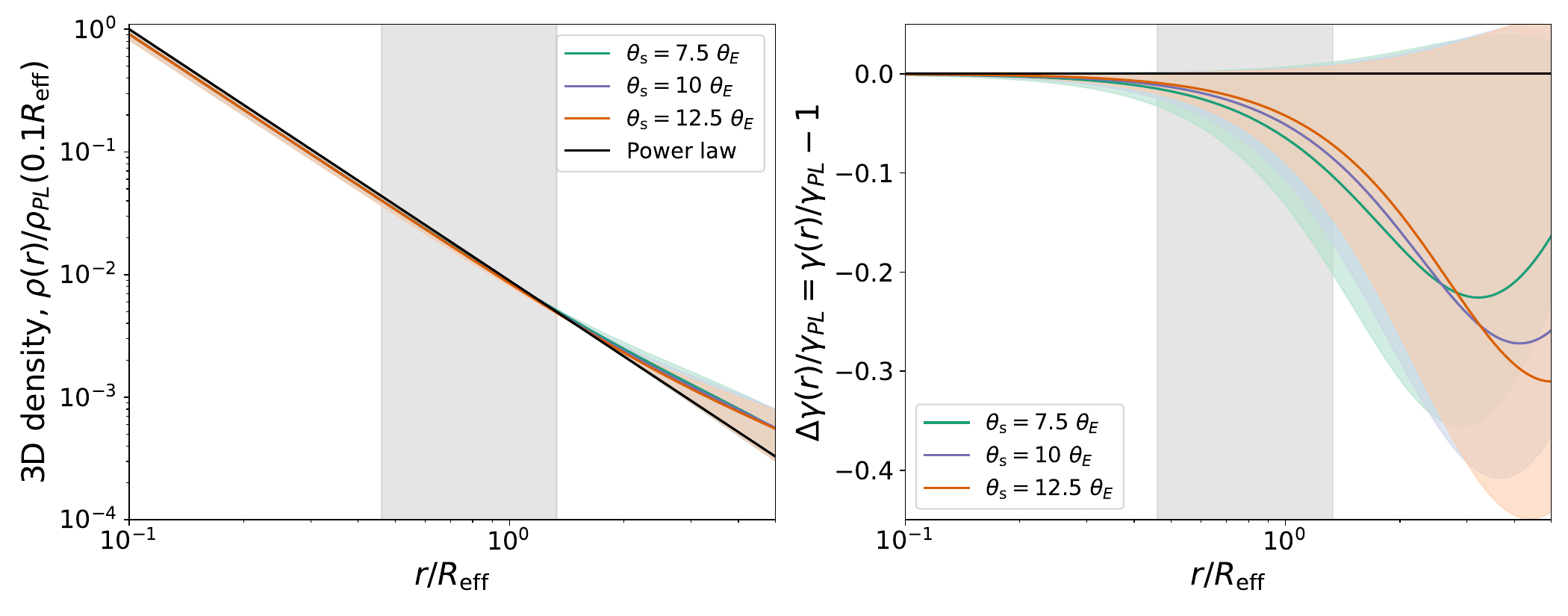}
    \caption{The deviation of the 3D density profile (left-hand panel) and the local logarithmic slope $\gamma (r) \equiv d (\log \rho) / d (\log r)$ (right-hand panel) for the MST of a power-law profile. The 3D density profile is normalised such that $\rho_{\rm PL}(0.1 \Reff) = 1$, where the subscript `PL' denotes the power-law case.
    Different colours represent different core radius $\theta_{\rm s}$ used for the approximate MST (equation \ref{eqn:approx_mst}). The initial mass profile of the system is a power-law profile with $\gamma = 2.05$ and $\Rein = 1.2$ arcsec. The vertical grey band represents the 1$\sigma$ distribution of $\Rein/\Reff$ for our lens sample, illustrating the scale range containing the local lensing information. 
    \cytwo{At $r = \Rein$ with $\theta_{\rm s}$ = 10 $\Rein$, the local logarithmic slope deviates only by \cyone{$\sim$$3\pm3$} per cent from the power-law profile, and the transformed profile is consistent with the power-law profile within 1$\sigma$ over the illustrated range of radius.}}
    \label{Fig:MST}
\end{figure*}

To illustrate an example, we apply the internal MST on a mass profile with $\gamma = 2.05$ and $\Rein = 1.2$ arcsec. These fiducial values are adopted to match the median values from our sample of SLACS and SL2S lenses. We ensure that the approximate MST recreates the same lensing effect observed in the imaging data as the pure MST by setting $\theta_{\rm s} = 10\Rein $ as the lensing image is only sensitive to the inner regions ($\theta \lesssim 2\Rein$) of the mass profile of the lensing galaxy \citep{Schneider13, Birrer20}. For $\lambda_{\rm int} = \cytwo{0.91}$, the mass profile slope at the Einstein radius decreases from $\gamma(\Rein) = 2.05$ to $\gamma_{\rm MST}(\Rein) = 1.99$, which represents a deviation by 3 per cent. The change in the shape of the total mass profile -- with a decrease in the density inside $\Rein$ and an increase outside -- due to the MST is also illustrated in Figure \ref{Fig:MST}. 

We note that different choices of $\theta_{\rm s}$ would lead to slightly different shapes of the mass-sheet transformed profiles, as shown in Figure \ref{Fig:MST}. Thus, we do not have sufficient information to obtain the complete shape of the density profile at all radii. Additional information, such as a strong prior on $\theta_{\rm s}$, or performing the joint lensing--dynamics analysis with a composite model of dark matter and baryonic distributions with empirical priors on their shapes \citep[e.g.,][]{Shajib21}, would be necessary to obtain an accurate density profile shape at all radii. These investigations are left for future papers in the Project Dinos series.

\subsection{Evolution of total density profile in elliptical galaxies}
\label{Section_5.3}


Cosmological hydrodynamical simulations, such as the Magneticum and IllustrisTNG, predict the `global' logarithmic slope averaged over a radius range containing $\Reff$. This global logarithmic slope decreases with decreasing redshift at $z < 1$, as mainly gas-poor mergers make the density profile in massive ellipticals shallower over time \citep[][]{Remus17, Wang19, Wang20}. Although this prediction should be tested using observations in principle, we refrain from doing so for several reasons. 

\ajstwo{First, we do not detect a significant evolution in the mean MST parameter $\mu_{\lambda_{\rm int}}$, as the large uncertainty on the $\alpha_{\lambda_{\rm int}}$ parameter makes the redshift dependence consistent with zero (Table \ref{tab:hierarichical_posterior}). This large $\alpha_{\lambda_{\rm int}}$ uncertainty can be attributed to the large systematic uncertainties added to the measured logarithmic slopes $\gamma$, to a larger extent for the SL2S systems (Table \ref{table:SL2S_params}) owing to their short-exposure images having low $S/N$ on the lensed arcs. A future study will model the SL2S lenses with new, deeper \textit{HST} imaging to constrain the redshift evolution more tightly (Sheu et al., in preparation).}

Second, we cannot uniquely constrain a global (i.e., radially-averaged) logarithmic slope using our analysis framework to make a fair comparison with the predictions from simulations. Although several previous studies have compared the logarithmic slope with those from the simulations \citep[e.g.,][]{Bolton12, Sonnenfeld13b}, those comparisons were performed with the assumption of a power-law mass model, where the local logarithmic slope matches with the global logarithmic slope by definition. A lensing-only analysis provides the local logarithmic slope \citep{Birrer19}. Thus, they cannot be compared to the global logarithmic slope obtained from simulation. Although joint lensing--dynamics analysis can be used to obtain the global logarithmic slope, the adopted model for such analysis requires the flexibility to produce the true mass distribution in galaxies. \ajstwo{Our framework of using the MST to constrain any deviation from the power-law lens model only applies a global scaling on the power-law distribution when fitting dynamical observables without actually changing the shape of the mass profile \citep{Birrer20}. However, applying an approximate internal MST to the power-law lens model to uniquely recover the true mass distribution is also not possible due to the impact of $\theta_{\rm s}$ choice on the transformed profile, as discussed in Section \ref{sec:impact_of_mst}. Future studies will use an empirically motivated composite mass model to constrain the actual mass profile shape from joint lensing--dynamics analysis to perform a fair test of the simulation predictions (Sheu et al., in preparation).}


Third, the logarithmic slope can depend on other galaxy properties, such as the velocity dispersion and the stellar mass density \citep{Auger10, Sonnenfeld13}. Thus, if these properties are not taken as control variables, galaxies at the different redshifts within the same sample (or, `super'-sample as assembled in this study) may be subject to different selection functions. Such selection functions, if unaccounted for, can lead to erroneous detection of a trend in the logarithmic slope across redshift \citep[see][]{Sonnenfeld13b}.

Fourth, the logarithmic slope is only one structural parameter among several other structural properties predicted or assumed by the simulations. For example, the dark matter fraction within a given aperture and the stellar initial mass function (IMF). To comprehensively test the predictions from the simulations to learn about the baryonic astrophysics that has shaped the massive ellipticals, all these structural parameters must be compared simultaneously, not just the logarithmic slope. However, to obtain these additional properties from observations, a composite mass model will be required to individually describe the dark matter and baryonic distributions.

Based on the above discussion, we recommend future studies testing the predictions from simulations with strong lensing to adopt these elements:
\begin{itemize}
    \item perform joint lensing--dynamics analysis with a composite mass profile, where dark matter and baryonic matter distribution in the deflectors are individually described,
    \item control for potential selection effects or covariances with other galaxy properties, such as the central velocity dispersion or the stellar surface density, and
    \item test simulation predictions for multiple structure properties simultaneously, for example, the inner dark matter fraction, the stellar IMF, etc., in addition to the global logarithmic slope.
\end{itemize}

\section{Conclusion}
\label{Section6}

In this paper, we present uniform lens models of 77 galaxy--galaxy lens systems. These 77 systems were collated from the SLACS, SL2S, and BELLS samples. To uniformly model these systems, we used the automated modelling pipeline \textsc{dolphin}, which uses \textsc{lenstronomy} as the modelling engine. This \textsc{dolphin} pipeline successfully modelled these 77 systems out of 103 attempted ones, which is a high success rate for an automated lens modelling pipeline. The remaining systems could be modelled by carefully tuning the models by hand, as previous studies did. However, we excluded those `failure cases' from modelling in this paper, as the minor increase in the statistical power from a larger sample of 103 systems is not worth the additional requirement for investigator time to hand-tune the lens models on a case-by-case basis for the remaining systems.

We combined our lens model posteriors with models to constrain any potential deviation from the power-law profile in the mass distribution. We parametrized the deviation from the power-law mass model using the internal MST parameter $\lambda_{\rm int}$, where $\lambda_{\rm int} = 1$ implies no deviation. We performed a hierarchical Bayesian analysis using the SLACS and SL2S systems from our modelled sample. We incorporated the lensing effect of the line-of-sight structures in this analysis using estimates from the literature, a first for analyses on galaxy--galaxy lenses. We excluded the BELLS systems in this step, as the estimates for their line-of-sight lensing effect are currently unavailable. The key results from this study are as follows:

\begin{itemize}
    \item We constrain the sample mean of the internal MST parameter with dependence on redshift and Einstein radius as $\mu_{\lambda_{\rm int}} =  0.91^{+0.10}_{-0.09} -0.12^{+0.31}_{-0.26}(z-0.3) -0.04^{+0.04}_{-0.04}(\Reff/\Rein -1)$, for the constant anisotropy model. Therefore, at $z = 0.3$ and $\Reff/\Rein = 1$, the population mean for massive ellipticals \cytwo{is consistent with the power-law profile (i.e., $\mu_{\lambda_{\rm int}} = 1$) within 0.9$\sigma$}. 
    \item \ajstwo{For an alternative choice of the stellar anisotropy profile using the Osipkov--Merrit parametrization, the deviation from the power law has a higher significance of 3.2$\sigma$ (i.e., $\mu_{\lambda_{\rm int}}^{\rm ref} =  0.81^{+0.06}_{-0.05}$). However,  $\mu_{\lambda_{\rm int}}^{\rm ref}$ is consistent between the two models within 0.9$\sigma$. This result highlights the potential impact of the stellar anisotropy model in constraining any deviation from the power-law profile by combining lensing and dynamical observables. The constant anisotropy profile is consistent with the dynamical observables in local massive ellipticals, therefore, we chose this model as our baseline. However, spatially resolved kinematics of lens galaxies will be necessary to differentiate between these stellar anisotropy models.}
    \item \ajstwo{We find our logarithmic slope measurements have a median absolute deviation of 8--10 per cent with those from \citet{Etherington22}, who used the same PEMD for the lens model but a different source reconstruction algorithm and software package.} \cytwo{This difference is larger than the median 3 per cent statistical uncertainty obtained from our posteriors and points to potential residual systematics in the lens modelling. We find this difference to be larger for lens systems with lower-$S/N$ arcs. Therefore, we estimated the modelling systematic uncertainty dependent on a `weighted' total $S/N$ quantity of the arc (Section \ref{Section_5.2}).}
    \item We find the Einstein radii constrained by different studies using the PEMD for the lens model agree very well within 2 per cent, which is the expected level of consistency empirically estimated by \citet{Bolton08}. However, the Einstein radii given by the studies using the SIE mass model can differ by 3--6 per cent, falling beyond the level of expected consistency.
\end{itemize}

This study presents the largest sample to date of galaxy--galaxy lenses with uniform, power-law models. However, this decade will see larger samples with similarly exquisite imaging data from the \textit{HST} and velocity dispersion measured using large, ground-based telescopes \citep[e.g.,][]{Tran22}. The sizes of recently discovered samples from the Dark Energy Survey, the DECam Local Volume Exploration Survey, and the Dark Energy Spectroscopic Survey are $\mathcal{O}(10^3)$ \citep{Jacobs19b, Jacobs19, Huang20b, Storfer2022, Zaborowski2022}. In the near future, the sample size of newly discovered systems will rapidly grow to $\mathcal{O}(10^4)$--$\mathcal{O}(10^5)$ thanks to current or future surveys carried out by the Vera Rubin Observatory, the \textit{Nancy Grace Roman Space Telescope}, and the \textit{Euclid} \citep{Oguri10, Collett15}. As employed in this paper, an automated framework will be essential to tackle the computational challenge these very large samples will bring forth. Additionally, these large samples will provide a much higher statistical constraining power to precisely track the evolutionary trends in the structural properties of massive ellipticals and shed light on the precise nature of the baryonic astrophysics at play through comparison with simulations.


\section*{Acknowledgements}
\cyref{We thank the anonymous referee for the many useful comments that helped us improve this manuscript.} \cytwo{We also thank Michele Cappellari, Dominique Sluse, Shawn Knabel, and William Sheu for providing comments and suggestions to improve the analysis and the manuscript. CYT was supported by the National Aeronautics and Space Administration (NASA) through the Space Telescope Science Institute (STScI) grant HST-AR-16149. CYT was also supported by the National Science Foundation (NSF) through the grants AST-2108168 and AST-2307126. This work was also supported by NASA through the NASA Hubble Fellowship grant HST-HF2-51492 awarded to AJS by the Space Telescope Science Institute, which is operated by the Association of Universities for Research in Astronomy, Inc., for NASA, under contract NAS5-26555. This work was completed in part with resources provided by the University of Chicago’s Research Computing Center and the Dark Energy Survey computing resources at Fermilab.}

This analysis has used the following software packages: \textsc{lenstronomy} \citep{Birrer15,Birrer18}, \textsc{dolphin} \citep{Shajib21}, \textsc{hierArc} \citep{Birrer20}, \textsc{numpy} \citep{Oliphant15}, \textsc{scipy} \citep{Jones01}, \textsc{astropy} \citep{AstropyCollaboration13,AstropyCollaboration18}, \textsc{jupyter} \citep{Kluyver16}, \textsc{matplotlib} \citep{Hunter07}, 
\cyref{\textsc{pandas}   \citep{McKinney10}}, 
\textsc{seaborn} \citep{Waskom14} , \textsc{Source Extractor} \citep{Bertin96},  \textsc{deepCR} \citep{Zhang20}, \textsc{astrodrizzle} \citep{Avila15}, \textsc{emcee} \citep{Foreman-Mackey13}, \textsc{fastell}\citep{Barkana99}.

\section*{Data Availability}
 \cyref{The GitHub repository at \url{https://github.com/Project-Dinos/dinos-i} contains the notebooks used for creating \textit{HST} image cutouts and lens modelling. This repository also contains the data products of this analysis, such as the lens models parameters (Table \ref{table:SLACS_params}, \ref{table:SL2S_params}, \ref{table:BELLS_params}) and the posterior of the population-level parameters from the hierarchical Bayesian analysis (Figure \ref{fig:hierarc_corner}).
 The lens model parameters and other model products can also be obtained at the Project Dinos webpage at \url{https://www.projectdinos.com}. The full imaging data are publicly available from the Mikulski Archive for Space Telescopes (MAST). The stellar kinematics measurements from the SLACS and SL2S surveys are made easily retrievable at \url{https://www.projectdinos.com/dinos-i}. The lens modelling pipeline \textsc{dolphin} is publicly available on GitHub at \url{https://github.com/ajshajib/dolphin}. }




\bibliographystyle{mnras}
\bibliography{main} 




\appendix
\section{Lens model settings for individual systems}\label{Appendix B}
\cyone{ Table \ref{table:settings_slacs_sl2s} provides the imaging filter information and model settings for each SLACS and SL2S lens system modelled in our analysis. The imaging filter information and model settings for the BELLS system are provided in Table \ref{table:settings_bells}}.

\begin{table*}
\centering
\caption{\textit{HST} imaging filters and other settings used in modelling the SLACS and SL2S systems. For each system, we manually optimised the shapelet order $n_{\rm max}$ for the source-light model and the radius $R_{\rm inner}$ of the inner mask used to avoid residuals in the deflector galaxy's centre. The `foreground mask' column indicates whether we included an additional mask for the line-of-sight contaminants in the associated filter.
}
\begin{tabular}{lccccc}

\hline
    Lens system & Survey & Filter &Shapelets $n_{\rm max}$  & $R_{\rm inner}$ (arcsec) & Foreground mask \\
\hline
SDSSJ0008-0004 & SLACS & [F606W, F814W] & [6, 6] & [0.4, 0.4] & [No, No]\\ 
SDSSJ0029-0055 & SLACS & [F606W, F814W] & [6, 6] & [0.4, 0.4] & [No, No]\\ 
SDSSJ0037-0942 & SLACS & [F435W, F606W, F814W] & [6, 6, 6] & [0.4, 0.4, 0.4] & [Yes, No, No]\\ 
SDSSJ0252+0039 & SLACS & [F390W, F606W, F814W] & [12, 10, 10] & [0, 0, 0] & [No, No, No]\\ 
SDSSJ0330-0020 & SLACS & [F606W, F814W] & [6, 6] & [0.4, 0.4] & [No, No]\\ 
SDSSJ0728+3835 & SLACS & [F606W, F814W] & [6, 6] & [0.4, 0.4] & [Yes, Yes]\\ 
SDSSJ0737+3216 & SLACS & [F390W, F555W, F814W] & [8, 8, 8] & [0.4, 0.4, 0.4] & [No, No, No]\\ 
SDSSJ0819+4534 & SLACS & [F435W, F606W, F814W] & [6, 6, 6] & [1.5, 1.7, 1.5] & [Yes, Yes, Yes]\\ 
SDSSJ0903+4116 & SLACS & [F390W, F606W, F814W] & [6, 6, 6] & [0.4, 0.4, 0.4] & [Yes, Yes, Yes]\\ 
SDSSJ0912+0029 & SLACS & [F390W, F435W, F555W, F814W] & [6, 6, 6, 6] & [0.4, 0.4, 0.4, 0.4] & [No, No, No, No]\\ 
SDSSJ0936+0913 & SLACS & [F606W, F814W] & [6, 6] & [0.4, 0.4] & [No, No]\\ 
SDSSJ0959+0410 & SLACS & [F390W, F555W] & [12, 12] & [0.4, 0.4] & [No, No]\\ 
SDSSJ1023+4230 & SLACS & [F606W, F814W] & [6, 6] & [0.4, 0.4] & [No, No]\\ 
SDSSJ1100+5329 & SLACS & [F435W, F606W, F814W] & [10, 12, 10] & [0.4, 0.4, 0.4] & [No, No, Yes]\\ 
SDSSJ1112+0826 & SLACS & [F606W, F814W] & [6, 6] & [0.3, 0.3] & [No, No]\\ 
SDSSJ1134+6027 & SLACS & [F606W, F814W] & [10, 10] & [0.4, 0.4] & [No, No]\\ 
SDSSJ1204+0358 & SLACS & [F435W, F606W, F814W] & [6, 6, 6] & [0.4, 0.4, 0.4] & [Yes, No, No]\\ 
SDSSJ1213+6708 & SLACS & [F606W, F814W] & [10, 10] & [0.4, 0.4] & [No, No]\\ 
SDSSJ1218+0830 & SLACS & [F606W, F814W] & [6, 6] & [0.4, 0.4] & [No, No]\\ 
SDSSJ1250+0523 & SLACS & [F435W, F555W, F814W] & [12, 12, 12] & [0.4, 0.4, 0.4] & [No, No, No]\\ 
SDSSJ1306+0600 & SLACS & [F606W, F814W] & [6, 6] & [0.4, 0.4] & [No, No]\\ 
SDSSJ1313+4615 & SLACS & [F606W, F814W] & [10, 10] & [0.3, 0.3] & [Yes, Yes]\\ 
SDSSJ1402+6321 & SLACS & [F435W, F555W, F814W] & [6, 6, 6] & [0.4, 0.4, 0.4] & [No, No, No]\\ 
SDSSJ1531-0105 & SLACS & [F606W, F814W] & [6, 6] & [0.4, 0.4] & [No, Yes]\\  
SDSSJ1621+3931 & SLACS & [F606W, F814W] & [6, 6] & [0.4, 0.4] & [No, No]\\ 
SDSSJ1627-0053 & SLACS & [F390W, F555W, F814W] & [6, 6, 6] & [0.4, 0.4, 0.4] & [No, No, No]\\ 
SDSSJ1630+4520 & SLACS & [F555W, F814W] & [12, 12] & [0.4, 0.4] & [No, No]\\ 
SDSSJ1636+4707 & SLACS & [F435W, F606W, F814W] & [6, 6, 6] & [0.38, 0.35, 0.35] & [Yes, No, Yes]\\ 
SDSSJ2238-0754 & SLACS & [F435W, F555W, F814W] & [6, 6, 6] & [0.4, 0.4, 0.4] & [No, No, No]\\ 
SDSSJ2300+0022 & SLACS & [F435W, F555W, F814W] & [6, 6, 6] & [0.4, 0.4, 0.4] & [No, No, No]\\ 
SDSSJ2302-0840 & SLACS & [F606W, F814W] & [14, 10] & [0.4, 0.4] & [No, No]\\ 
SDSSJ2303+1422 & SLACS & [F435W, F555W, F814W] & [6, 6, 6] & [0.4, 0.4, 0.4] & [No, No, No]\\ 
SDSSJ2343-0030 & SLACS & [F606W] & [6] & [0.4] & [No]\\ 
SDSSJ2347-0005 & SLACS & [F606W, F814W] & [10, 10] & [0.7, 0.7] & [Yes, Yes]\\ 
SL2SJ0208-0714 & SL2S & [F475X, F600LP] & [8, 8] & [0.4, 0.4] & [No, No]\\ 
SL2SJ0214-0405 & SL2S & [F606W, F814W] & [10, 10] & [0.4, 0.4] & [Yes, Yes]\\ 
SL2SJ0217-0513 & SL2S & [F606W, F814W] & [10, 10] & [0.4, 0.4] & [Yes, Yes]\\ 
SL2SJ0219-0829 & SL2S & [F475X, F600LP] & [6, 6] & [0.4, 0.4] & [No, No]\\ 
SL2SJ0225-0454 & SL2S & [F606W] & [6] & [0.4] & [No]\\ 
SL2SJ0226-0420 & SL2S & [F606W] & [6] & [0.4] & [Yes]\\ 
SL2SJ0226-0406 & SL2S & [F606W] & [6] & [0.4] & [Yes]\\ 
SL2SJ0232-0408 & SL2S & [F606W] & [6] & [0.4] & [No]\\ 
SL2SJ0849-0412 & SL2S & [F475X, F600LP] & [6, 6] & [0.3, 0.3] & [No, No]\\ 
SL2SJ0849-0251 & SL2S & [F606W] & [6] & [0.4] & [No]\\ 
SL2SJ0858-0143 & SL2S & [F606W] & [6] & [0.4] & [No]\\ 
SL2SJ0901-0259 & SL2S & [F606W] & [6] & [0.4] & [No]\\ 
SL2SJ0904-0059 & SL2S & [F606W] & [8] & [0.4] & [No]\\ 
SL2SJ0959+0206 & SL2S & [F606W] & [6] & [0.2] & [No]\\ 
SL2SJ1358+5459 & SL2S & [F475X, F600LP, F606W] & [8, 8, 8] & [0.25, 0.25, 0.25] & [Yes, No, Yes]\\ 
SL2SJ1359+5535 & SL2S & [F606W] & [8] & [0.4] & [No]\\ 
SL2SJ1401+5544 & SL2S & [F475X, F600LP] & [6, 6] & [0.4, 0.4] & [No, No]\\ 
SL2SJ1402+5505 & SL2S & [F475X, F600LP] & [6, 6] & [0.4, 0.4] & [No, No]\\ 
SL2SJ1405+5243 & SL2S & [F475X, F600LP, F606W] & [6, 6, 6] & [0.4, 0.4, 0.4] & [No, No, No]\\ 
SL2SJ1406+5226 & SL2S & [F606W] & [10] & [0.4] & [Yes]\\ 
SL2SJ1411+5651 & SL2S & [F475X, F600LP] & [10, 10] & [0.4, 0.4] & [No, No]\\ 
SL2SJ1420+5630 & SL2S & [F475X, F600LP] & [8, 8] & [0.4, 0.4] & [No, No]\\ 
SL2SJ1427+5516 & SL2S & [F606W] & [6] & [0.4] & [No]\\ 
SL2SJ2214-1807 & SL2S & [F606W] & [6] & [0.4] & [No]\\ 
\hline
\end{tabular}
\label{table:settings_slacs_sl2s}
\end{table*}

\begin{table*}
\centering
\caption{\textit{HST} imaging filters and other settings used in modelling the BELLS systems. For each system, we manually optimised the shapelet order $n_{\rm max}$ for the source-light model and the radius $R_{\rm inner}$ of the inner mask used to avoid residuals in the deflector galaxy's centre. The `foreground mask' column indicates whether we included an additional mask for the line-of-sight contaminants in the associated filter.}
\begin{tabular}{lccccc}
\hline
    Lens system & Survey & Filter & Shapelets $n_{\rm max}$  & $R_{\rm inner}$ (arcsec) & Foreground mask \\
\hline
SDSSJ0151+0049 & BELLS & [F814W] & [6] & [0.2] & [No]\\ 
SDSSJ0747+5055 & BELLS & [F814W] & [12] & [0.3] & [Yes]\\ 
SDSSJ0801+4727 & BELLS & [F814W] & [10] & [0.2] & [No]\\ 
SDSSJ0830+5116 & BELLS & [F814W] & [6] & [0.4] & [No]\\ 
SDSSJ0944-0147 & BELLS & [F814W] & [6] & [0.3] & [Yes]\\ 
SDSSJ1159-0007 & BELLS & [F814W] & [6] & [0.25] & [No]\\ 
SDSSJ1215+0047 & BELLS & [F814W] & [12] & [0.25] & [Yes]\\ 
SDSSJ1221+3806 & BELLS & [F814W] & [6] & [0.25] & [No]\\ 
SDSSJ1234-0241 & BELLS & [F814W] & [6] & [0.2] & [No]\\ 
SDSSJ1318-0104 & BELLS & [F814W] & [6] & [0.15] & [Yes]\\ 
SDSSJ1337+3620 & BELLS & [F814W] & [6] & [0.2] & [Yes]\\ 
SDSSJ1349+3612 & BELLS & [F814W] & [6] & [0.225] & [No]\\ 
SDSSJ1352+3216 & BELLS & [F814W] & [10] & [0.4] & [Yes]\\ 
SDSSJ1542+1629 & BELLS & [F814W] & [6] & [0.3] & [Yes]\\ 
SDSSJ1545+2748 & BELLS & [F814W] & [10] & [0.4] & [No]\\ 
SDSSJ1601+2138 & BELLS & [F814W] & [6] & [0.15] & [No]\\ 
SDSSJ1631+1854 & BELLS & [F814W] & [15] & [0.4] & [No]\\ 
SDSSJ2125+0411 & BELLS & [F814W] & [15] & [0.4] & [No]\\ 
\cyref{SDSSJ2303+0037} & BELLS & [F814W] & [10] & [0.3] & [No]\\ 
\hline
\end{tabular}
\label{table:settings_bells}
\end{table*}

\section{Model decomposition plots}\label{Appendix A}
\begin{figure*} 
    \centering
    \includegraphics[width=0.95\textwidth]{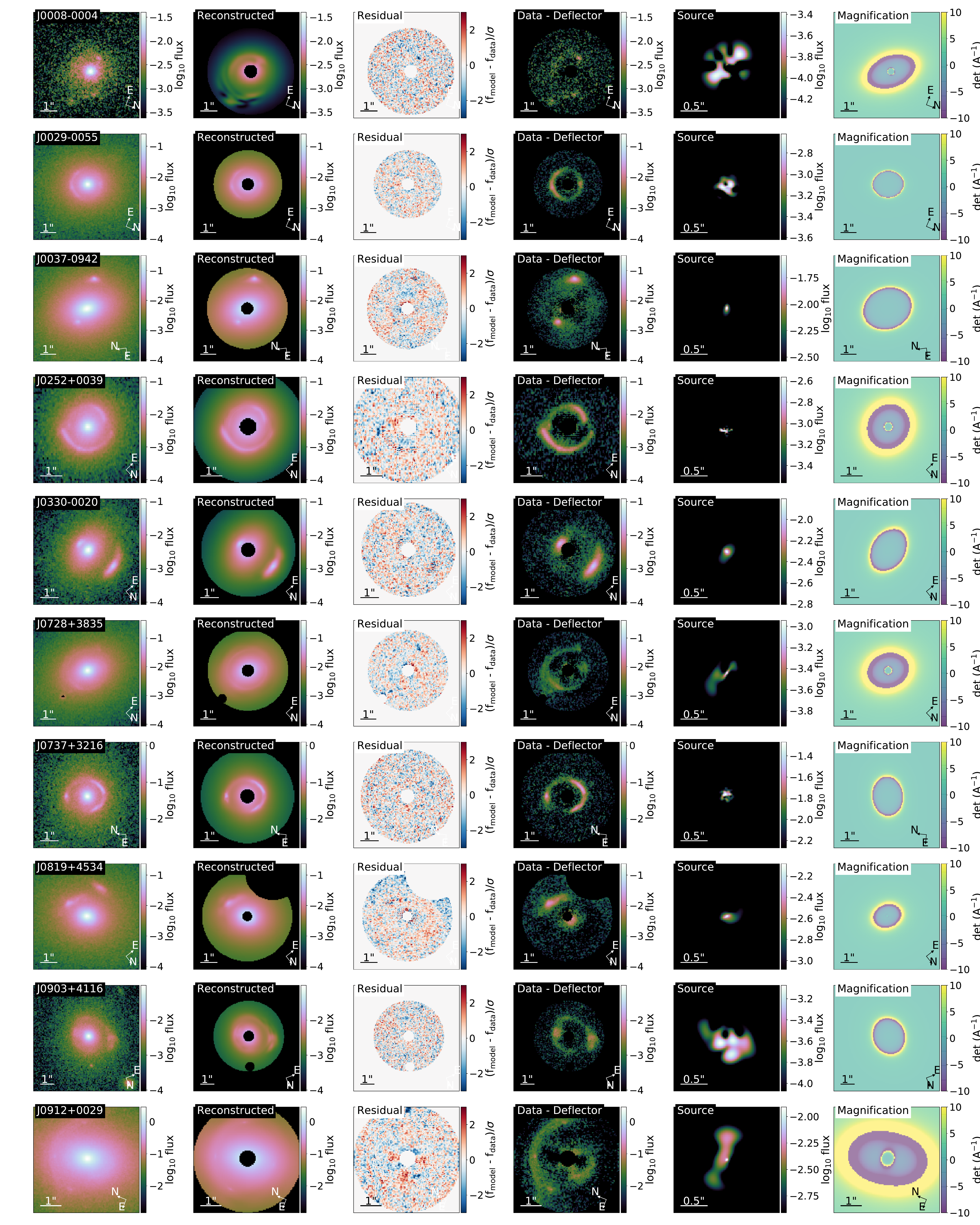}
    \caption{Best-fitting lens models for the first 10 of the 34 SLACS lenses. The first column shows the \textit{HST} images of the lenses in the \cyone{visible band (F555W or F606W)}. The second column shows the model-based reconstruction and the third column shows the normalised residual. We visualise the lensed arcs in the fourth column by subtracting the deflector light profile from the \textit{HST} image. The fifth column shows the reconstructed source light and the sixth column shows the magnification map. The lens models for the other imaging filters are of similar quality (as shown in Figure \ref{fig:multi_band_0219}). The lens models for the remaining SLACS systems are shown in Figures \ref{fig:SLACS2},\ref{fig:SLACS3} and \ref{fig:SLACS4}.}
    \label{fig:SLACS1}
\end{figure*}

\begin{figure*}    
    \centering
    \includegraphics[width=0.95\textwidth]{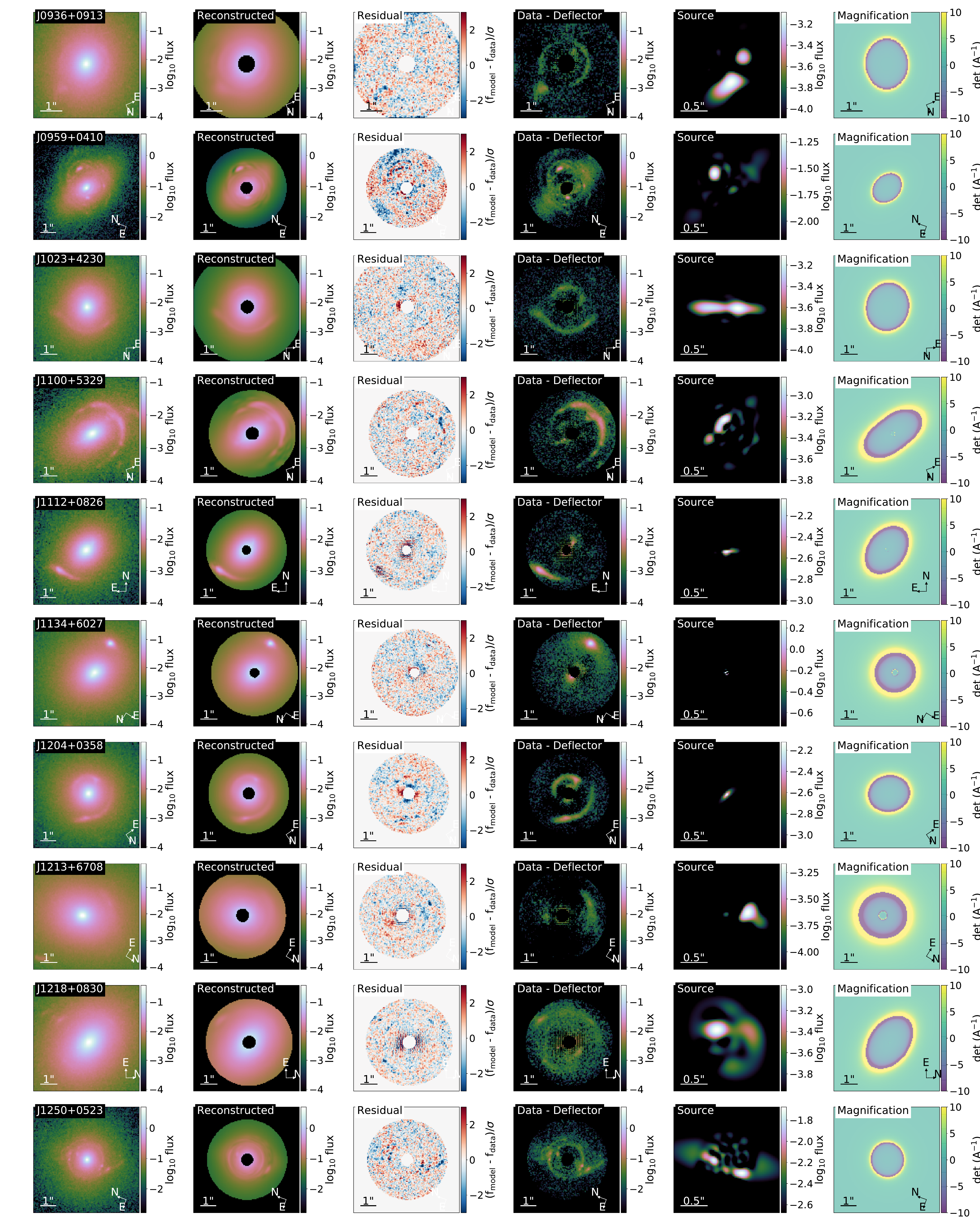}
    \caption{Lens images and models for the next 10 out of the 34 SLACS lenses. This continues from Figure \ref{fig:SLACS1}.}
     \label{fig:SLACS2}
\end{figure*}

\begin{figure*}     
    \centering
    \includegraphics[width=0.95\textwidth]{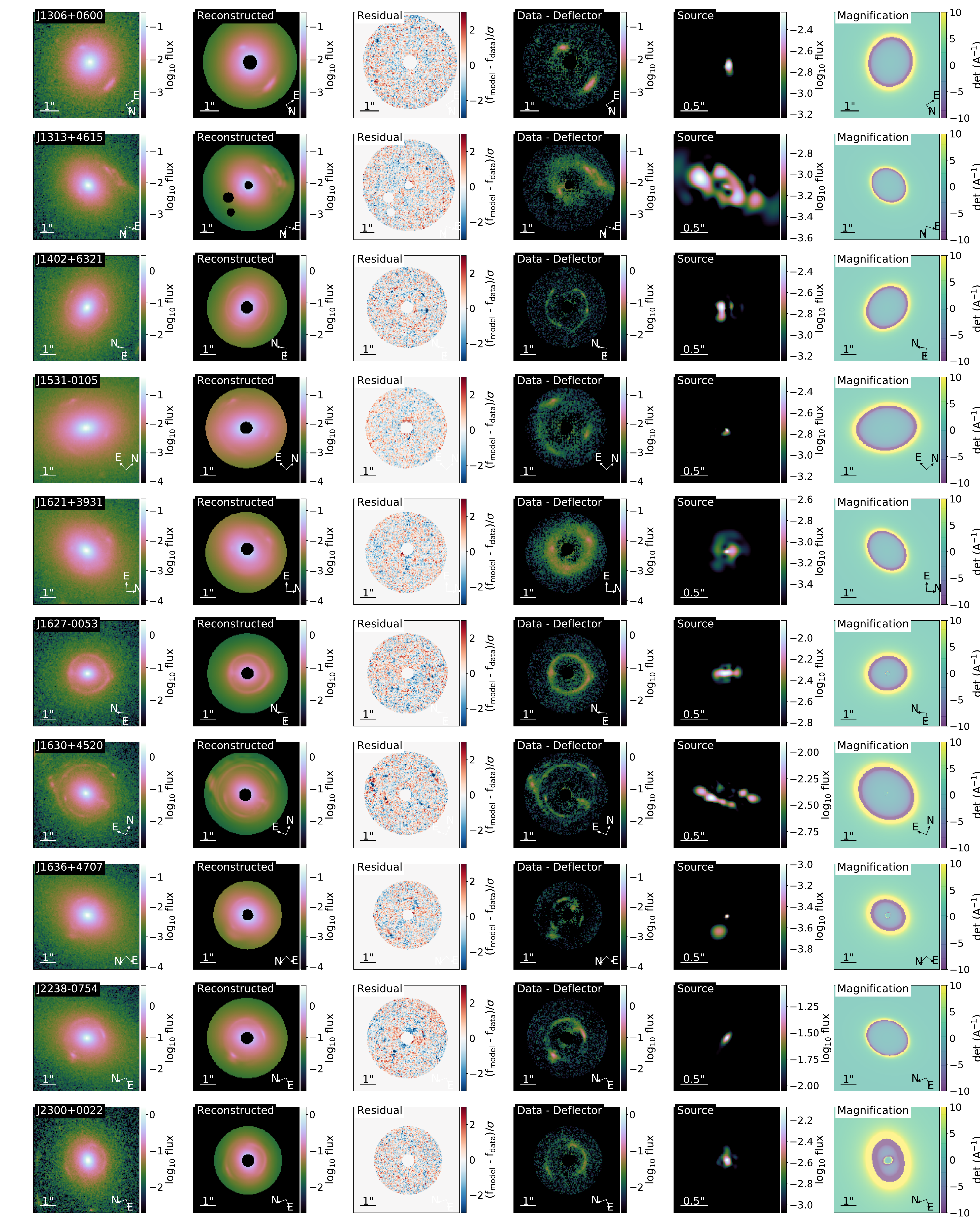}
    \caption{Lens images and models for another 10 out of the 34 SLACS lenses. This continues from Figure \ref{fig:SLACS1} and \ref{fig:SLACS2}.}
    \label{fig:SLACS3}
\end{figure*}

\begin{figure*}  
    \centering
    \includegraphics[width=0.95\textwidth]{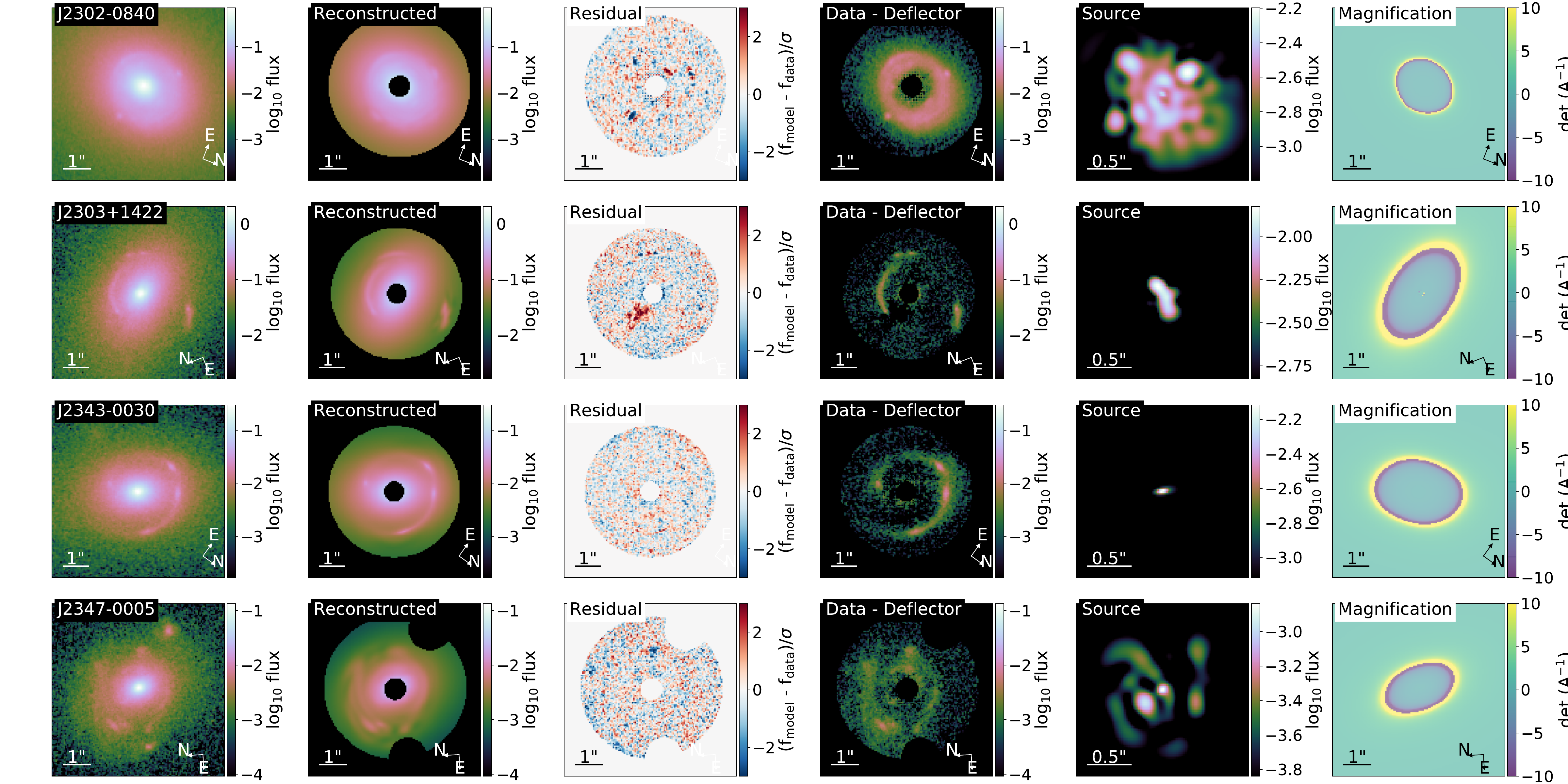}
    \caption{Lens images and models for the last 4 out of the 34 SLACS lenses. This continues from Figure \ref{fig:SLACS1}, \ref{fig:SLACS2} and \ref{fig:SLACS3}.}
    \label{fig:SLACS4}
\end{figure*}

\begin{figure*}
    \centering
    \includegraphics[width=0.95\textwidth]{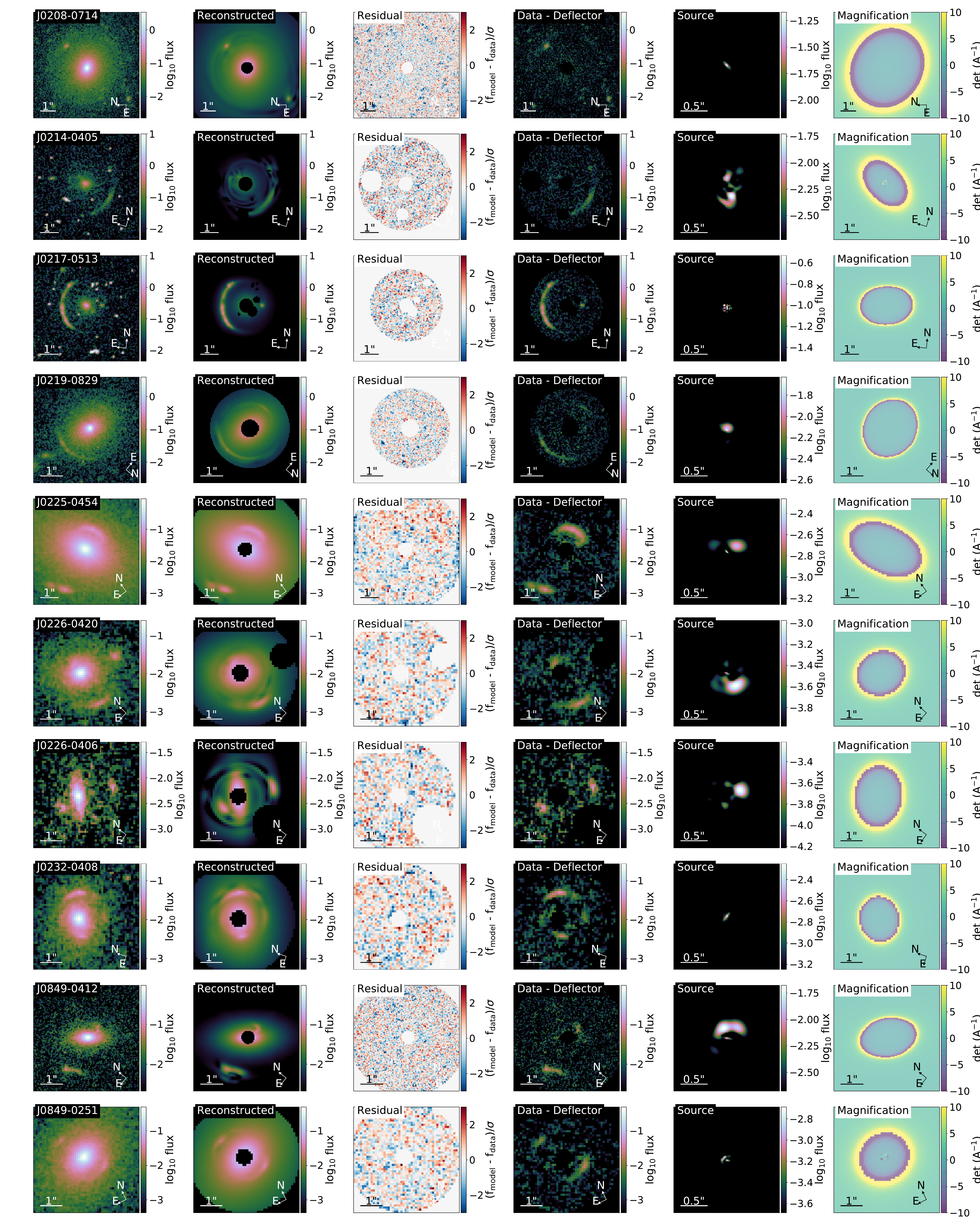}
    \caption{Best-fitting lens models for the first 10 of the 24 SL2S lenses. The first column shows the \textit{HST} images of the lenses in the \cyone{visible band (F600LP or F606W)}. The second column shows the model-based reconstruction and the third column shows the normalised residual. We visualise the lensed arcs in the fourth column by subtracting the deflector light profile from the \textit{HST} image. The fifth column shows the reconstructed source light and the sixth column shows the magnification map. The lens models for the other imaging filters are of similar quality (as shown in Figure \ref{fig:multi_band+1401}). The lens models for the remaining SL2S systems are shown in Figures \ref{fig:SL2S2} and \ref{fig:SL2S3}. \cyone{The lens systems SL2SJ0226-0406 (in Figure \ref{fig:SL2S1})  and SL2SJ1427+5516 (in Figure \ref{fig:SL2S3}) have a disc-like morphology and are not included in the analysis in Section \ref{Section4:Model_paramters}.}  }
    \label{fig:SL2S1}
\end{figure*}

\begin{figure*}
    \centering
    \includegraphics[width=0.95\textwidth]{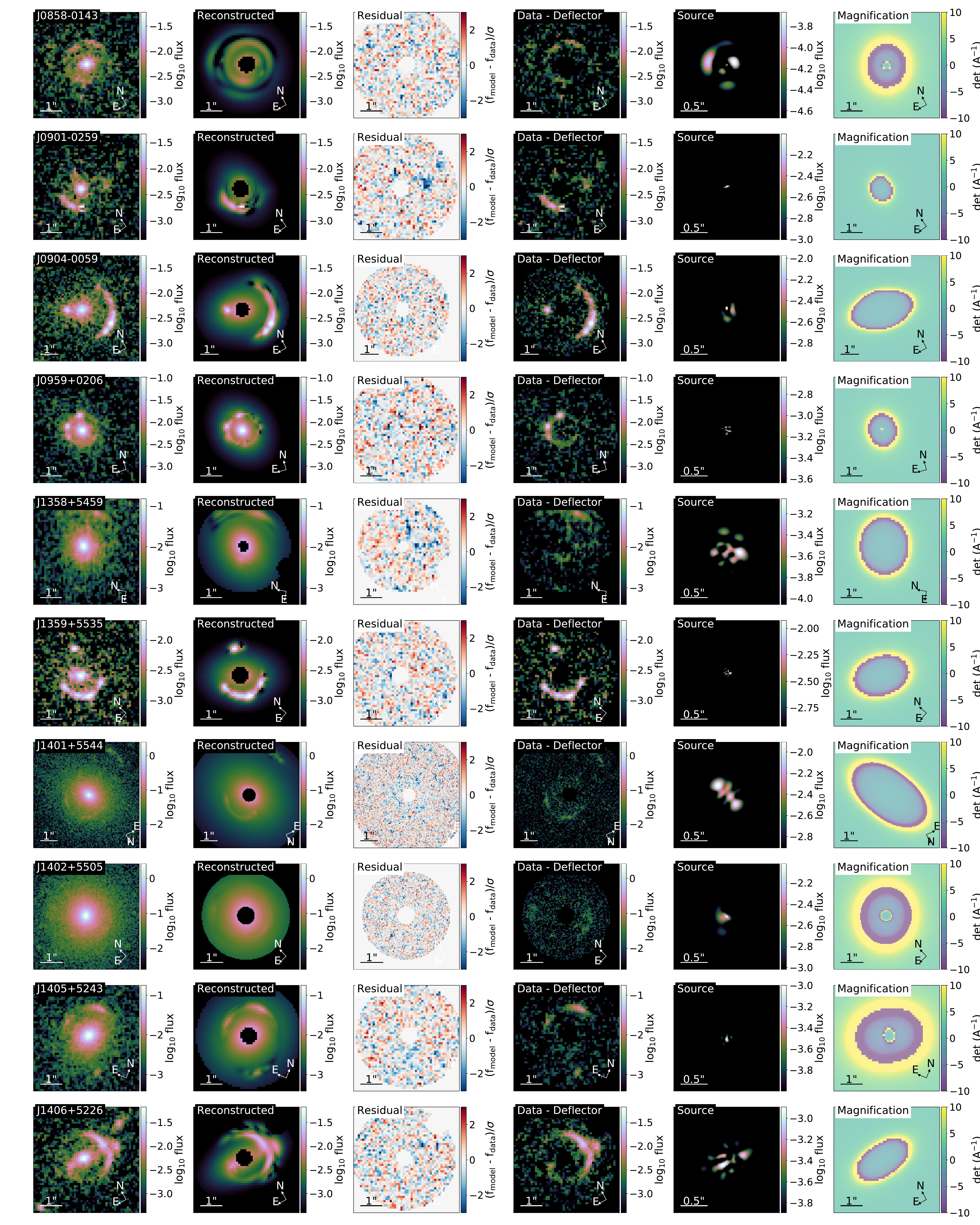}
    \caption{Lens images and models for the next 10 out of the 24 SL2S lenses.This continues from Figure \ref{fig:SL2S1}.}
    \label{fig:SL2S2}
\end{figure*}

\begin{figure*}
    \centering
    \includegraphics[width=0.95\textwidth]{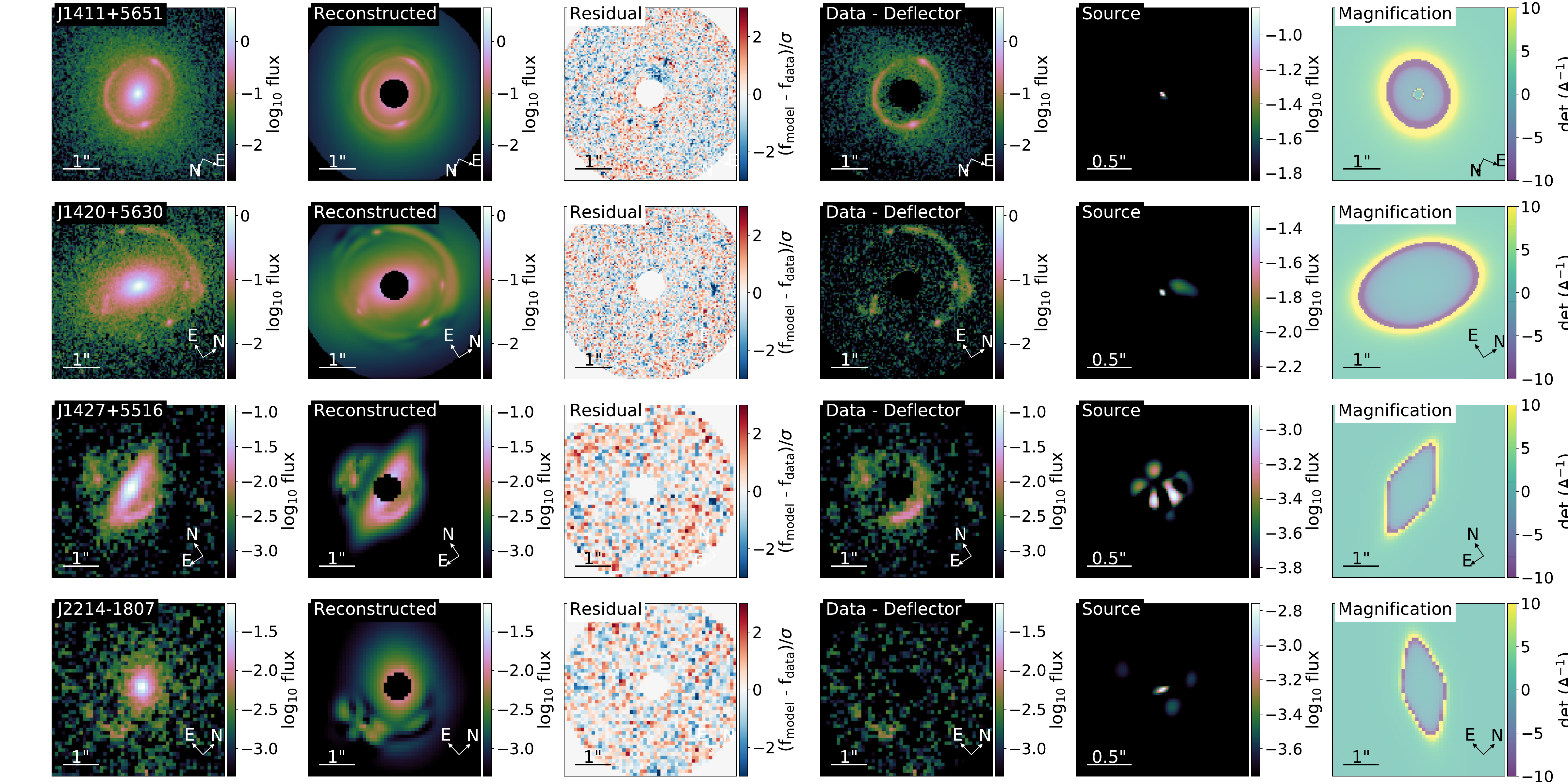}
    \caption{Lens images and models for the last 4 of the 24 SL2S lenses SLACS. This continues from Figure \ref{fig:SL2S1} and  \ref{fig:SL2S2}.}
    \label{fig:SL2S3}
\end{figure*}

\begin{figure*}
    \centering
    \includegraphics[width=0.95\textwidth]{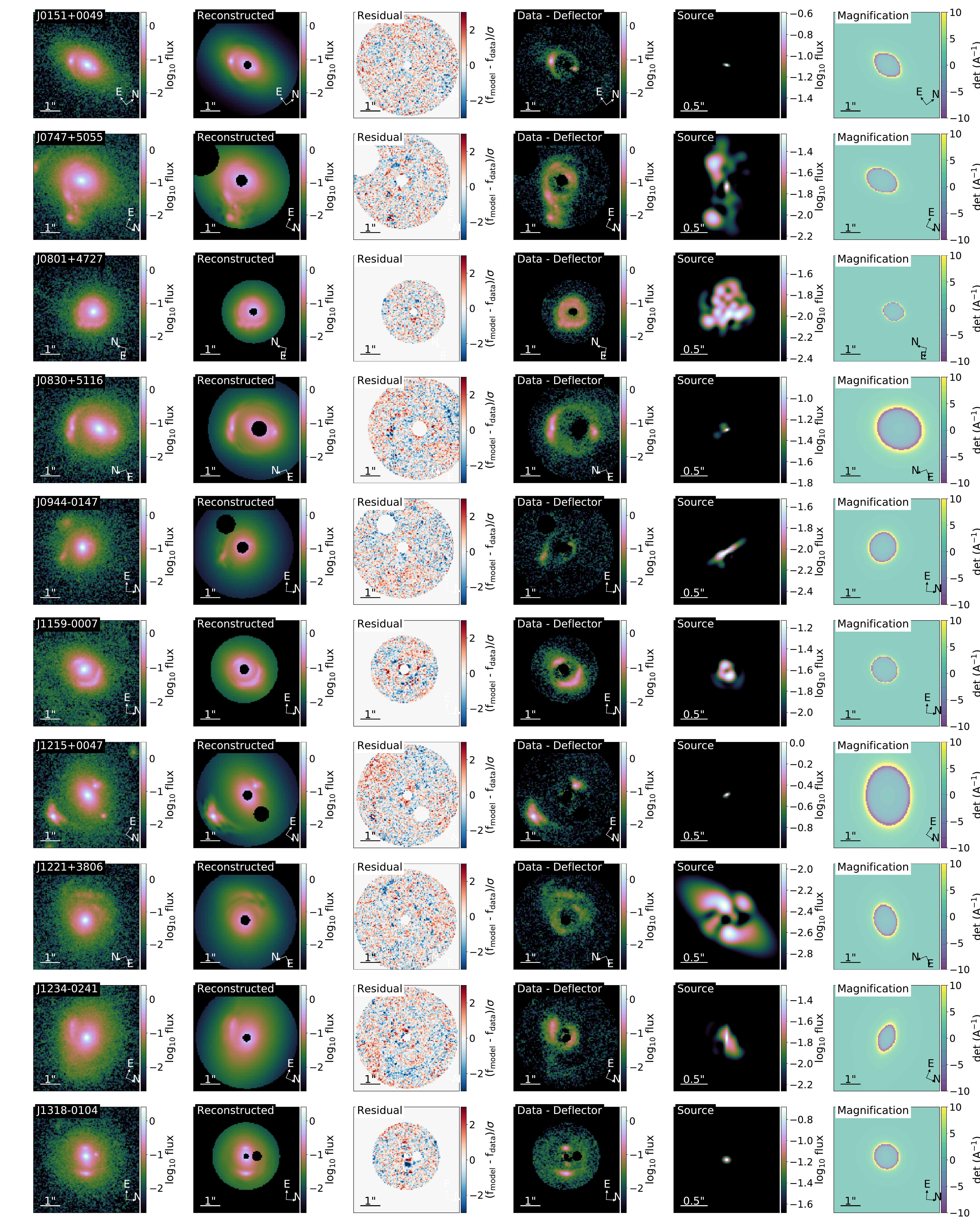}
    \caption{Best-fitting lens models for the first 10 of the 24 BELLS lenses. The first column shows the \textit{HST} images of the lenses in the \cyone{F814W filter}. The second column shows the model-based reconstruction and the third column shows the normalised residual. We visualise the lensed arcs in the fourth column by subtracting the deflector light profile from the \textit{HST} image. The fifth column shows the reconstructed source light and the sixth column shows the magnification map. The lens models for the remaining BELLS systems are shown in Figure \ref{fig:BELLS2}.}
    \label{fig:BELLS1}
\end{figure*}

\begin{figure*}
    \centering
    \includegraphics[width=0.98\textwidth]{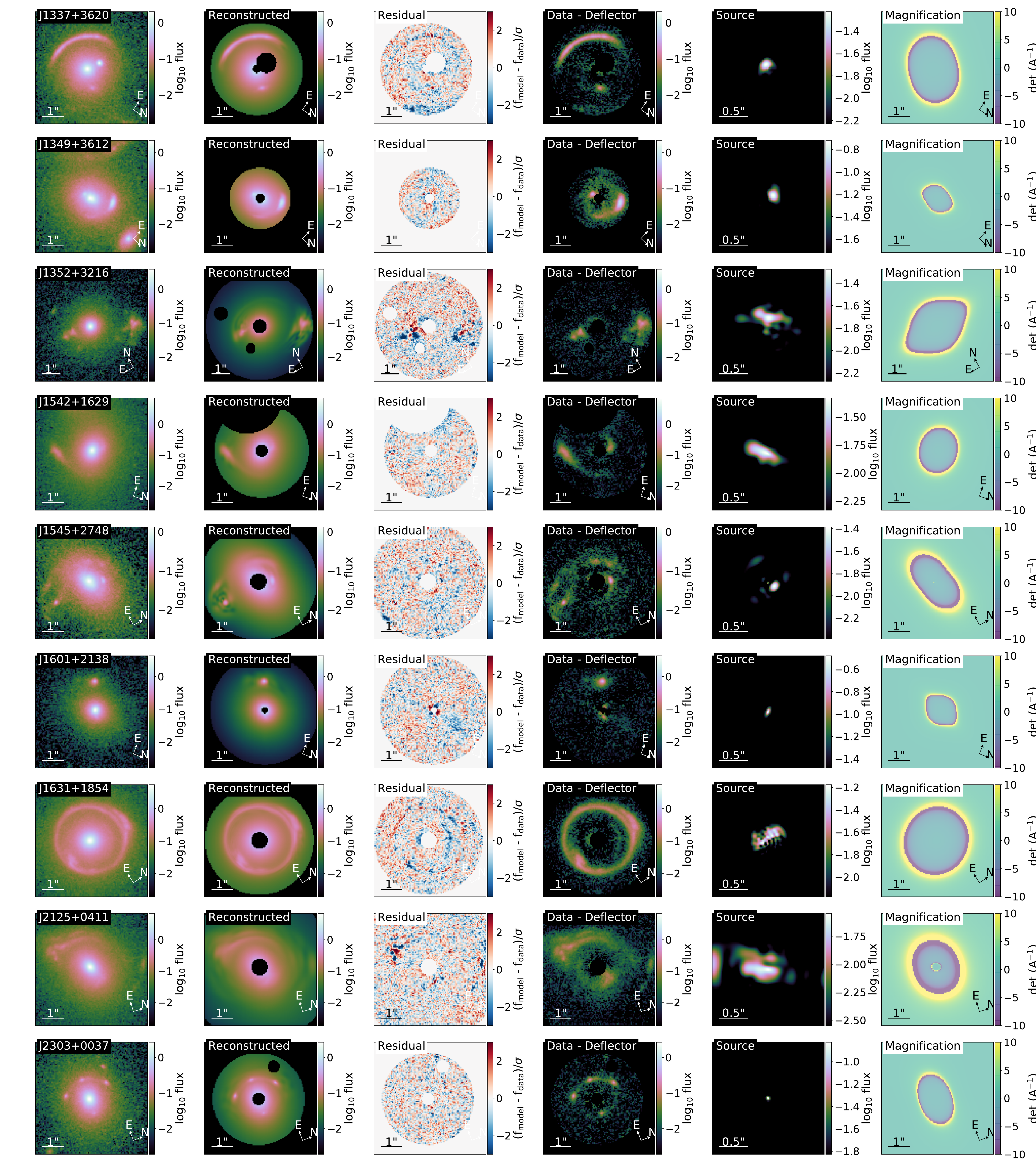}
    \caption{Lens images and models for the next 9 out of the 19 BELLS lenses. This continues from Figure \ref{fig:BELLS1}.}
    \label{fig:BELLS2}
\end{figure*}

We illustrate the best-fitting lens models for the SLACS lenses in Figures \ref{fig:SLACS1}, \ref{fig:SLACS2}, \ref{fig:SLACS3}, \ref{fig:SLACS4}; for the SL2S lenses in Figures \ref{fig:SL2S1}, \ref{fig:SL2S2}, and \ref{fig:SL2S3}; and for the BELLS lenses in Figures \ref{fig:BELLS1}, \ref{fig:BELLS2}. \ajsref{In some cases, the reconstructed sources appear to have artefacts, which are common in lens modelling \citep[e.g.,][]{Shajib21, Shajib22}. Such artifacts are inevitable to some extent since the simplistic nature of the lens mass models that are typical in the literature overlooks additional complex structures in real galaxies, such as dark substructures \citep[e.g.,][]{Vegetti10}, boxyness or discyness \citep{VandeVyvere21, VandeVyvere22}, or ellipticity gradients \citep{Gomer23}. Furthermore, the illustrated reconstructions are created using the best-fit model parameters. Averaging over the model posterior would potentially eliminate some of the artefacts in visualization. Additionally, incorporating a larger number of shapelets to reproduce finer structures in the source galaxy would also alleviate the artefacts. However, our lens model posteriors are robust against increasing the number of shapelets.}

\bsp	
\label{lastpage}

\end{document}